\documentstyle[amscd]{article}

\title{Quantum and Floer cohomology have the same ring structure}

\author{Sergey Piunikhin}

\date{May 25
, 1994}

\begin{document}
\maketitle
\centerline { Department of Mathematics}
\centerline { Massachusetts Institute of Technology}
\centerline { Cambridge, MA 02139  USA}
\centerline { e-mail    serguei@math.mit.edu}
$$  $$

 \begin{abstract}

 	The action of the total cohomology
$  H^*  (M)$  of the almost Kahler manifold  $M$  on its Floer cohomology,
 int roduced originally by Floer, gives a new ring structure on $H^* (M)$ .
 We prove that the  total cohomology space $H^* (M)$, provided with this new
ring structure,
 is  isomorphic to the quantum cohomology ring.
 As a special case, we prove the
 the  formula for the Floer cohomology ring of the complex grassmanians
   conjectured by Vafa and Witten

   \end{abstract}
$$ $$

 \centerline {\bf     CONTENTS}

  		1. Introduction

		2. Moduli spaces of $J$-holomorphic spheres and its 			compactification.

 		3. Various definitions of quantum cup-products

  		4. Review of symplectic Floer cohomology

 		5. Cup products in  Floer cohomology

 		6. Proof of the Main Theorem

	 	7. Floer cohomology ring of Grassmanians

		8. Quantum  cohomology  revisited

	 	9. Discussion

\vskip .2cm

\vskip .2cm

\vskip .2cm

\centerline {\bf     1. INTRODUCTION}
\centerline { }

	Floer cohomology $HF^* (M)$ of the free loop-space of the symplectic
manifold  $M$ is under extensive study by both mathematicians and physicists.

\vskip .2cm

	 Originally the machinery of Floer cohomology was developed in [F1] in order
to prove the classical
Arnold's Conjecture  giving a lower bound on the number of fixed points of a
symplectomorphism in terms of Morse theory. Later  Floer cohomology  appeared
in ``topological sigma-models'' of string theory [Wi1],[BS].

\vskip .2cm

	Quantum cohomology rings of Kahler (or more generally, almost-Kahler)
manifolds were introduced by Witten [Wi2] and Vafa [Va1] using
 moduli spaces of holomorphic curves
[CKM],[Gr],[McD1],[Ru2].

\vskip .2cm

 Recently by Ruan
and Tian [RT] by adjusting
Witten's degeneration argument, proved  that these rings are associative

\vskip .2cm

 	It is difficult to give a rigorous and self-concistent mathematical
definition
of quantum cohomology rings because the moduli spaces of holomorphic curves are
non-compact and may have singularities, the facts
 often ignored by physisists.

\vskip .2cm

The proof of associativity  in [RT] for quantum cohomology  ring
uses a definition of multiplication that involves the moduli spaces of
solutions of inhomogenous Cauchi-Riemann equations [Ru1].

\vskip .2cm

 	The quantum cohomology rings have been computed  for:

a)  complex projective spaces [Wi2]

b)  complex Grassmanians (the formula was conjectured by Vafa [Va1], is
 proved  in
the present paper, and also independently in [AS] and
 [ST] )

c)  toric varieties [Ba]

d)  flag varieties [GK]

e)  more general hermitian symmetric spaces  [AS]

\vskip .2cm

	Quantum cohomology rings of Calabi-Yau 3-fold ``computed'' in several examples
by  ``mirror symmetry '' [COGP]  not considered rigorous. To justify these
``computations'' is a great challenge for algebraic geometers.

\vskip .2cm

	The linear map  $m_F : H^* (M)  \rightarrow End( HF^* (M) )$ or, equivalently,
the action of the classical cohomology of the manifold $M$ on its Floer
cohomology, was defined by Floer himself. He computed this action for the case
$M = {\bf CP^n } $ and noticed the following fact:

\vskip .2cm

	For any two cohomology classes  $A$ and $B$ in $H^* ({\bf CP^n })$  the
product $m_F (A)  m_F (B)$
of the linear operators   $m_F (A)$ and  $m_F (B)$ acting on Floer cohomology
$ HF^* ({\bf CP^n } )$  has the form $m_F (C)$  for some cohomology class  $C$
in $H^* ({\bf CP^n })$. This gives us a new ring structure on
$H^* ({\bf CP^n })$, which is known to be different from the classical
cup-product. We will  call this new multiplication law {\bf Floer
multiplication}.

\vskip .2cm

	 Floer  conjectured that the same phenomenon might be true for all
semi-positive symplectic (or at least Kahler)
manifolds, thus  providing a new ring structure on the total cohomology
$H^* (M)$  our symplectic manifold  $M$.

\vskip .2cm

 Using path-integral arguments, V.Sadov  [S] ``proved'' that
when $M$ is a semi-positive symplectic manifold, then:

\vskip .2cm

1) ``The operator algebra should close'', i.e., for any two cohomology classes
$A$ and $B$ in $H^* (M)$  the product $m_F (A)  m_F (B)$
 always has the form $m_F (C)$  for some cohomology class  $C$ in $H^* (M)$.

and

2)  {\bf Floer  multiplication}  coincides with  {\bf quantum multiplication}.

\vskip .2cm

See also  [Fu1], [CJS], [GK]  and [McD S]
for alternative definitions of cup-products in Floer cohomology and
their conjectural relation with the quantum multiplication.

\vskip .2cm

	The purpose of the present note is to give a rigorous
proof of  Sadov's statements.  We prove

\proclaim  The Main Theorem. For semi-positive symplectic manifold $M$
 the ring structure on $H^* (M)$ inherited from its action $m_F$ on the
  Floer cohomology coincides with the the quantum multiplication on $H^*(M)$
defined as in [RT].

 	Another problem  we will encounter is that some components of the moduli
space of holomorphic curves may not have have its dimension  equal to the
``virtual  dimension'' which can be computed using Atiyah-Singer index theorem.
 This problem was artificially avoided in [RT] paper by introducing the moduli
spaces of solutions of inhomogenous Cauchy-Riemann equations.

\vskip .2cm

	 The definition of ``contribution to the quantum cup-product''
from these ``exceptional'' components of the moduli space  of holomorphic
curves is  different from the definition of  contribution to the quantum
cup-product from the other components.

\vskip .2cm

	When all these irreducible components are smooth , Witten showed in [Wi4] how
to define this contribution.

\vskip .2cm

Witten introduces  ``the bundle of the fermion zero-modes'' over the
``exceptional'' component of the moduli space of holomorphic curves       and
takes  the Euler class of this bundle. He wedges this  Euler class with the
wedge-product of ``geometric'' classes and then ``integrates'' over the whole
component of the moduli space ( taking care with the compactification  of this
component).

\vskip .2cm

 See also [AM] for the sheaf-theoretic computations of contribution from
``multiply-covered curves'' using this definition.

\vskip .2cm

 In the case when these ``exceptional'' components occur, it was not known
whether the  quantum cup-product defined as  in [Wi4] and [AM] was associative.

\vskip .2cm

Assuming that the compactified moduli spaces  admit   smooth desingularization
such that ``the compactification divisor'' of that desingularization has
codimension at least two
we prove associativity by establishing
equivalence of this definition with  the definition used in [RT]

\vskip .2cm

	 The  Floer picture gives  a geometric way to calculate the   Euler class
which encounters in Witten's definition.

$$ $$

\centerline {\bf     2. MODULI  SPACES  OF  J-HOLOMORPHIC  SPHERES  AND
ITS  COMPACTIFICATION}

\centerline { }

\proclaim Definition.
The  manifold $M$  is called is called an almost-Kahler manifold if it admits
an almost-complex     structure $J$ and a symplectic form $\omega$ such that
for any two tangent vectors $x$ and $y$ to $M$,
$ \hskip .4mm \omega (x; y) =  \omega (J(x); J(y))$ and
for any non-zero tangent vector $x$ to $M$ the following inequality holds:

$$ \omega (x; J (x)) \hskip .3mm  > \hskip .3mm  0 \eqno   (2.1)    $$

\proclaim Definition.
An  almost-complex     structure $J$ and a    closed $2$-form $\omega$ on $M$
are called   compatible if for any tangent vector $x$ to $M$
$ \omega (x; J (x)) \hskip .3mm  \geq \hskip .3mm  0 $ and the equality takes
place only if
$ \hskip .2mm \omega (x; y) \hskip .3mm  = \hskip .3mm  0     $ for any tangent
vector
 $\hskip .3mm$   $y$

In particular, the symplectic form $\omega$  is compatible with $J$  iff
$(2.1)$ holds.

\vskip .2cm

	Let $M$ be a compact almost-Kahler
 manifold of  dimension $2n$ which we assume (for simplicity) to be
simply-connected. Let us fix  an almost-complex structure $J_0$ on $M$ and let
us consider the space $\tilde {K}$ of all $J_0$-compatible symplectic forms and
its image $K$ in the   cohomology $H^2 (M, R)$.

\vskip .2cm

If it will not lead to a confusion, we will denote the closed $J_0$-compatible
two-form and the corresponding cohomology class by the same symbol.

\vskip .2cm

	It follows directly from definitions that if $M$ is an almost-Kahler manifold
(which is equivalent to the fact that
$\hskip .2 mm \tilde {K} \hskip .2 mm$
is non-empty) then:

\vskip .2cm

1) $ \hskip .2mm \tilde {K} \hskip .2mm$ is an open convex cone in the space of
all closed 2-forms on $M$. The set   $\tilde {K}$ does not contain any
nontrivial linear subspace (otherwise $\omega$ and  $- \omega$ would be
simultaneously $J_0$-compatible  which is impossible).

\vskip .2cm

2)   $  \hskip .2mm K  \hskip .2mm $ is an open convex cone in
 $  \hskip .2mm H^2 (M, R)  \hskip .2mm $ which does not contain any nontrivial
linear subspace.

\vskip .2cm

 The openness of $K$ follows from the fact that small
perturbations of any given $J_0$-compatible symplectic form, are themserlves
 $J_0$-compatible and symplectic .

\vskip .2cm

Since a  symplectic form compatible with $J_0 \hskip .3mm$ (and in fact any
symplectic form) on a compact
oriented manifold $M$ cannot be cohomologically trivial then  $K$ cannot
contain a nontrivial linear subspace in    $H^2 (M, R)$.

\vskip .2cm

	Let us consider symplectic forms
$\hskip .3mm \{ \omega_1,...,\omega_s \} \hskip .3mm $
such that:

\vskip .2cm

1) they  lie inside  $\hskip .3mm \tilde {K}$.

2)  their cohomology classes form a basis in $H^2 (M, R)$.

3) the elements of this basis are represented by integral cohomology classes

4)  $\hskip .3mm\{ \omega_1,...,\omega_s \}\hskip .3mm$ generate
$\hskip .3mm H^2 (M, Z) \hskip .3mm $ as an abelian group.

\vskip .2cm

 We can always find such a collection of  symplectic forms since any  open
convex cone in $H^2 (M, R)$ contains such a collection.

\vskip .2cm

 	If we fix a homotopy class of $J_0$ in the space of
almost-complex structures on $M$ it will make sence to talk about {\bf the
first Chern class}  $c_1 (TM)$ of the tangent bundle of $M$ as an element in
the second cohomology group of $M$.

\vskip .2cm

\proclaim Definition.
The almost-complex manifold $M$ is called  semi-positive if  $c_1 (TM)$ can be
 represented by a closed 2-form compatible  with the almost-complex
structure $J_0$

The almost-complex manifold $M$ is called  {\bf positive} if  $c_1 (TM)$ can be
  represented by a {\bf symplectic} form compatible  with the almost-complex
structure $J_0$

\vskip .2cm

The above definitions  should be thought as generalizations the notion of
 (semi)-positive simply-connected {\bf Kahler manifold}.

\vskip .2cm

It sometimes
 appears to be useful
 to perturb slightly the complex structure  on $M$ (or on $CP^1 \times M$) and
to work with  non-integrable almost-complex structures. It is much easier to
prove  transversality results if we are allowed to work in this larger
category.

\vskip .2cm

 Let us consider the complex projective line  $CP^1$ with its standard comolex
structure $i$ and Fubiny-Study Kahler form $\Omega$. Let us take the product
$CP^1 \times M$ in the almost-Kahler category. Let ${\cal J}$ be  the space of
all smooth almost-complex structures on
 $CP^1 \times M$ such that the projection on the first factor

$CP^1 \times M \rightarrow CP^1$ is holomorphic.
Let us equip this space with the $W_{5n}^2$-Sobolev norm  topology. This means
that all the partial derivatives up to order $5n$ should be square-integrable
on
 $CP^1 \times M$

\vskip .2cm

	Let ${\cal J}_0$ be a  neighborhood of $i \times J_0$  in    ${\cal J}$
consisting of almost-complex structures
compatible with symplectic forms
 $\{ 1 \otimes \omega_1,..., 1 \otimes \omega_s ,  \Omega \otimes 1 \}$ and
with some
differential form representative of $ 1 \otimes c_1 (TM)$. Since the notion of
compatibility with a 2-form is an open condition in ${\cal J}$, such a
neighborhood always exists.

\vskip .2cm

	Let us consider the vector bundle over the product
 $  \hskip .2mm CP^1 \times M  \hskip .2mm $ consisting of
 $  \hskip .2mm i \times J_0$-antilinear maps from
$  \hskip .2mm T(CP^1)  \hskip .2mm $ to $  \hskip .2mm TM . \hskip .2mm$
``$i \times J_0$-antilinear'' means that for any
 $  \hskip .2mm g \in {\cal G}  \hskip .2mm $ we have
$  \hskip .2mm  J_0     g    =   -  g     i .  \hskip .2mm$  Let
$  \hskip .2mm {\cal G}  \hskip .2mm $ be
 the space  consisting of all
$  \hskip .2mm W_{5n-1}^2$-sections of  the above-defined vector bundle.

\vskip .2cm

 Equivalently, ${\cal G}$  can be thought as a space of all
 $(0,1)$-forms on $CP^1$ with the coefficients in the tangent bundle to $M$.

\vskip .2cm

If $g$ is any such  $(0,1)$-form, we  can construct an almost-complex structure
$J_g$ on $CP^1 \times M$ which is written in coordinates as follows:

$$  J_g =   \left(\matrix{ i & -g  \cr
				0 & J_0 \cr} \right)  \eqno (2.2)  $$

Here we wrote the matrix of $J_g$ acting on $T(CP^1)  \oplus TM$

\vskip .2cm

Thus, we have an embedding $  \hskip .2mm {\cal G}  \subset {\cal J}.  \hskip
.2mm $
Let  $  \hskip .2mm {\cal G}_0  \hskip .2mm $ be the intersection of
$  \hskip .2mm {\cal G}  \hskip .2mm $ and $  \hskip .2mm {\cal J}_0$.

\vskip .2cm

We will assume both
${\cal J}_0$ and ${\cal G}_0$ to be contractible.

\vskip .2cm

Presumably, the introduction of almost-complex structures can be avoided.
We use them to modify  the proofs of some analytic lemmas.

\vskip .2cm

What we really need is to  one fixed almost-complex structure $J_0$ on $M$
(which in all examples will be an actual complex structure) and perturbations
of the product (almost)-complex structure on $CP^1 \times M$  of the form
$(2.2)$.

\vskip .2cm

 So, we desided to use more complicated notations to simplify the proofs.

\vskip .2cm

Let $J \in {\cal J}_0$ be an almost-complex structure on $CP^1 \times M$

\proclaim Definition (Gromov).
A  $J$-holomorphic sphere in $M$ is any almost-complex  submanifold in
$CP^1 \times M$
of real dimension two (or ``complex dimension one'') which projects
isomorphically onto the first factor $CP^1$.

Equivalently, a $J$-holomorphic sphere in $M$ can be defined as a
pseudo-holomorphic section of the the (pseudo-holomorphic) bundle
$M \times CP^1$  over    $CP^1$ where the almost-complex structure $J$ on $M
\times CP^1$ is a  perturbation of the product  almost-complex structure.
Topologically this is the trivial bundle over    $CP^1$  with the fiber $M$ but
(pseudo)-holomorphically it is not trivial.

\vskip .2cm

Any  $J$-holomorphic sphere can be thought as a map $\varphi$ from $CP^1$ to
$M$ which satisfies a non-linear PDE

$$ {\bar \partial}_J \hskip .2mm    \varphi  \hskip .2mm = 0    \eqno (2.3A) $$

If our  almost-complex structure $J_g$ has the form $(2.2)$ than the equation
of
a $J_g$-holomorphic sphere $\varphi$ can be rewritten as

$$ {\bar \partial}_{J_0} \varphi = g    \eqno (2.3B) $$

Here  $ \hskip .2mm {\bar \partial}_{J_0} \hskip .2mm $ is the usual
$ \hskip .2mm{\bar \partial}$-operator on $ \hskip .2mm M \hskip .2mm$
associated with our original (almost)-complex structure $ \hskip .2mm J_0$

\vskip .2cm

 We  assume our manifold $M$ to be {\bf semi-positive} with respect to the
almost-complex structure $J_0$ (and thus, the manifold  $CP^1 \times M$ will
be semi-positive with respect to all  almost-complex structures in ${\cal
J}_0$).
Semi-positivity  implies that
the integrals of the first Chern class   $c_1 (TM)$ over all $J$-holomorphic
curves in $M$ are non-negative if   $ J \in  {\cal J}_0$.

\vskip .2cm

	Let $  \hskip .2mm C \subset H_2 (M, \bf R )  \hskip .2mm $ be the closure of
the convex cone generated by the images of homology classes of $  \hskip .2mm
J$-holomorphic spheres (for all
$  \hskip .2mm J \in    {\cal J}_0$). $ \hskip .2mm$
         Then $C$ will lie in the closure of the convex dual  of the      cone
$  \hskip .2mm K \subset H^2 (M, \bf R )  \hskip .2mm $.

\vskip .2cm

 Following [Ru1], we will call a non-zero cohomology class
${\hskip 2mm} A \in H^2 (M, Z ) {\hskip 2mm}$
 {\bf an effective class }  if  $A$ lies inside the closed cone
 ${\hskip 2mm} C$.

\vskip .2cm

 Let  $q_1,...,q_s$ be the dual to  $w_1,...,w_s$ basis in 	 $H_2 (M)$. We will
write the elements of
$H_2 (M) = H_2 (M,Z)$ in multiplicative notation. The monomial
$q^d = q_1^{d_1}... q_s^{d_s}$ is by definition the sum
$  \hskip .2mm \sum_{i=1}^{s} d_i  \hskip .2mm q_i  \hskip .2mm \in  \hskip
.2mm  H_2 (M)  \hskip .2mm . {\hskip 2mm}$
Here $  \hskip .2mm d  \hskip .2mm$  is a vector of integers
$ {\hskip 2mm} (d_1,...,d_s) {\hskip 2mm} $
 and
$ {\hskip 2mm} q = (q_1,...,q_s) {\hskip 2mm} $
 is a multi-index. Then the group ring
 $ {\hskip 2mm} Z_{[H_2 (M)]} {\hskip 2mm}$ is a commutative ring
generated (as an abelian group) by monomials of the form
$  {\hskip 2mm} q^d = q_1^{d_1}... q_s^{d_s}$.

\vskip .2cm

	The group ring $Z_{[H_2 (M)]}$ which is isomorphic to the ring
$  \hskip .2mm Z_{[q_1^{\pm 1},...,q_s^{\pm 1}]}  \hskip .2mm $ of Laurent
polynomials, has an important
subring $Z_{[C]}$. The semi-positivity of $M$ implies that

$$Z_{[C]} \subset Z_{[q_1,...,q_s]} \subset Z_{[q_1^{\pm 1},...,q_s^{\pm 1}]}$$

i.e., that monomials $q_1^{d_1}... q_s^{d_s}$ may appear in $Z_{[C]}$ only
if all $(d_1,...,d_s)$ are non-negative.

\vskip .2cm

	The ring $Z_{[C]}$ has a natural augmentation
$I: Z_{[C]}  \rightarrow Z$ which sends all non-constant
 monomials in $\{ q_i \}$
to zero. Thus, we can consider its completion  $Z_{<C>}$ with respect to the
$I$-adic topology. This completion lies naturally in the ring
$Z_{<q_1,...,q_s>}$ of formal power series in $\{ q_i \}$.

\vskip .2cm

	For our future purposes let us introduce the following ring

$$N  {\hskip 3mm} =  {\hskip 3mm} Z_{<q_1,...,q_s>} \otimes_{Z_{[q_1,...,q_s]}}
Z_{[H_2 (M)]}$$

 The ring $  {\hskip 2mm} N  {\hskip 2mm} $ is called {\bf Novikov  ring}. The
similiar ring  appeared in Novikov's study of Morse theory
of multivalued functions [No].  Novikov's refinement of Morse theory is almost
exactly the kind of  Morse theory we need in our study of
of Floer homology  (see also [Hs]).

\vskip .2cm

	Let us consider the abelian group  $H^* (M,Z) \otimes Z_{<C>}$. It has
an obvious structure of a $ {\hskip 2mm} Z_+$-graded ring inherited from the
usual grading in cohomology, provided that all the elements of the augmentation
ideal ${\hskip 3mm}I(Z_{<C>})  {\hskip 3mm}  $ have degree zero.

\vskip .2cm

	The same abelian group  $H^* (M) \otimes Z_{<C>}$ has
{\bf another }
$ {\hskip 2mm} Z_{+}$-graded ring structurte
which can be constructed as a $q$-deformation of the classical cohomology
ring $H^* (M)$ with  non-trivial  grading of the
``deformation parameters'' $\{ q_i \} . {\hskip 3mm}$
To be more concrete, let us define a $Z$-grading on
$H^* (M) \otimes Z_{<q_1,...,q_s>}$ as follows: any element
${\hskip 2mm} A{\hskip 2mm} $ from
$H^* (M) \otimes Z_{<q_1,...,q_s>}$ can be obtained as a (possibly infinite)
sum of
``bihomogenous pieces''
$ {\hskip 2mm} A {\hskip 1mm} = {\hskip 1mm}  \sum_{m,d} A^{m,d} \otimes q^d
{\hskip 2mm}$    where

$ {\hskip 2mm} A^{m,d} \in H^m (M,Z)$. Then let us define

$$deg [A^{m,d} \otimes q^d] = m + 2< c_1(TM) ; q^d >   \eqno         (2.4A)$$

where the last term means evaluation of the 2-cocycle $ c_1(TM)$

on the 2-cycle $q^d$.

\vskip .2cm

	The formula $(2.4A)$ can be rewritten in more elegant way:

$$deg [A^{m,d} \otimes q^d] = m + 2 \sum_{i=1}^{s} d_i< c_1(TM) ;  q_i>   \eqno
        (2.4B)$$

	Using the semi-positivity condition

 $$deg [q_i] {\hskip 2mm} =  {\hskip 2mm} 2< c_1(TM) ;  q_i>  {\hskip 2mm}
\geq  0
 \eqno     (2.5)  $$

we see that our  $Z$-grading  is actually a  $Z_+$-grading on $H^* (M) \otimes
Z_{<q_1,...,q_s>}$ and on its subring $H^* (M) \otimes Z_{<C>}$. This
$Z$-grading can be extended to the  $Z$-grading on
${\hskip 2mm} H^* (M) \otimes N {\hskip 2mm} $
(when we ``extend the scalars'').

\proclaim Definition.
For each multi-index  $ {\hskip 2mm} d = (d_1,...,d_s) {\hskip 2mm} $
let  ${\hskip 2mm} Map_d {\hskip 2mm} $ be  the space of all Sobolev maps from
$ {\hskip 2mm} CP^1 {\hskip 2mm} $ to $ {\hskip 2mm} M {\hskip 2mm}$ of a given
homotopy type specified by
``the generalized degree'' $ {\hskip 2mm} d = (d_1,...,d_s) .$

``Sobolev'' means that the the map  $\varphi \in Map_d $ should have the
square-integrable partial derivatives up to order $5n$. (The first derivative
of the map  $\varphi$ from $CP^1$ to $M$ is a one-form on  $CP^1$ with the
values in
 ${\varphi}^*(TM)$).

\vskip .2cm

``Homotopy type specified by
the generalized degree'' $d = (d_1,...,d_s)$
 means that
$ {\hskip 2mm} \int_{\varphi (CP^1)} \omega_i = d_i  {\hskip 2mm} $ for each $
{\hskip 2mm} \varphi \in Map_d {\hskip 2mm} $
 and for each     $ {\hskip 2mm} i = 1,...,s$.

\vskip .2cm

	The space  $Map_d$ thus has a natural structure of a (connected)
Hilbert manifold ${\hskip 2mm}$ (see [McD1] and [McD S] for the proof)
 ${\hskip 2mm}$ which is homotopically equivalent to the space of all smooth
(or all continious) maps  from $ {\hskip 2mm} CP^1 {\hskip 2mm} $ to
$ {\hskip 2mm} M {\hskip 2mm} $ of a given homotopy type. This space is a
connected component of the larger space
 $ {\hskip 2mm} Map = \bigcup_d Map_d {\hskip 2mm} $ of all  Sobolev maps from
$ {\hskip 2mm} CP^1 {\hskip 2mm} $ to $ {\hskip 2mm} M {\hskip 2mm} $
(regardless of homotopy type) which is also a Hilbert manifold.

\vskip .2cm

	Let us introduce an infinite-dimensional Hilbert bundle ${\cal H}$
over
$Map \times {\cal J}_0 $. The fiber ${\cal H}_{J, \varphi}$ of the bundle
${\cal H}$
over the ``point'' $(\varphi, J)  \in Map   \times {\cal J}_0 $ will be the
space of all  $ {\hskip 2mm} (0,1)$-forms on $ {\hskip 2mm} CP^1 {\hskip 2mm}$
with the values in the complex $n$-dimensional vector bundle
$ {\hskip 2mm} {\varphi}^*(TM) . {\hskip 2mm}$
 ``All  $ {\hskip 2mm} (0,1)$-forms on $ {\hskip 2mm} CP^1$'' means all $
{\hskip 2mm} (0,1)$-forms on $ {\hskip 2mm} CP^1 {\hskip 2mm} $ lying in the
Sobolev space $ {\hskip 2mm} W_{5n-1}^2 . {\hskip 2mm}$ The almost-complex
structure $ {\hskip 2mm} J {\hskip 2mm} $ on $ {\hskip 2mm} M {\hskip 2mm}$
provides the tangent bundle $ {\hskip 2mm} TM {\hskip 2mm} $ with the structure
of the complex $ {\hskip 2mm} n$-dimensional vector bundle.

\vskip .2cm

	The bundle  ${\cal H}$ is provided with a  section
 ${\hskip 2mm} {\bar \partial}$ , given by the formula

$$   (\varphi, J) \rightarrow {\bar \partial}_J  (\varphi)  \eqno (2.6)
$$

The above-defined section  ${\hskip 2mm} {\bar \partial} {\hskip 2mm} $ is
actually a nonlinear ${\bar \partial}$-operator

\proclaim Proposition 2.1.
 The zero set ${\bar \partial}^{-1}(0)$ consists of the pairs
 $ {\hskip 2mm} (\varphi, J) {\hskip 2mm} $ where
$ {\hskip 2mm} \varphi {\hskip 2mm} $ is a $J$-holomorphic map.

\proclaim Definition.
For each multi-index  $  {\hskip 2mm} d  {\hskip 2mm} $
let $  {\hskip 2mm} {\cal M}_{J,d} \subset  Map_d    {\hskip 2mm} $ be the
space of all solutions of
$  {\hskip 2mm} (2.3A)  {\hskip 2mm} $ of homotopy type specified by
 $  {\hskip 2mm}d$. Let us  and call
 $  {\hskip 2mm} {\cal M}_{J,d}  {\hskip 2mm} $  the moduli space of  $
{\hskip 2mm} J$-holomorphic maps  from $  {\hskip 2mm} CP^1  {\hskip 2mm} $ to
$  {\hskip 2mm} M  {\hskip 2mm} $   of ``the generalized degree''
 $  {\hskip 2mm}d$.

	The above defined  Hilbert bundle $ {\hskip 2mm} {\cal H} {\hskip 2mm} $
over  $ {\hskip 2mm} Map \times {\cal J}_0 {\hskip 2mm} $ can be (trivially)
extended to a Hilbert bundle
over the
product of $ {\hskip 2mm} Map \times {\cal J}_0 {\hskip 2mm} $ and
$ {\hskip 2mm} {\cal G}_0 . {\hskip 2mm} $
It can also be trivially  extended to a Hilbert bundle
over the
product of $ {\hskip 2mm} Map \times  {\cal J}_0 {\hskip 2mm} $ and
$ {\hskip 2mm} {\cal G}_0  \times  {\cal G}_0 \times [0;1]$.

\vskip .2cm

We will denote these three Hilbert bundles by the same symbol  ${\cal H}$. We
will also denote by  ${\cal H}$ the restriction of these Hilbert bundles to
connected components $ {\hskip 2mm} Map_d   \times$ [auxillary space] $ {\hskip
2mm}$
of their bases.

\vskip .2cm

	Since $  {\hskip 2mm} {\cal G}_0  {\hskip 2mm} $ is an open subset in the
vector space $  {\hskip 2mm} {\cal G}  {\hskip 2mm} $ which has has a
base-point (zero), then it makes sense to speak about extension of smooth
sections of $  {\hskip 2mm} {\cal H}  {\hskip 2mm} $ from
$  {\hskip 2mm} Map \times {\cal J}_0  {\hskip 2mm} $
to the larger spaces
 $  {\hskip 2mm} Map \times {\cal J}_0  \times  {\cal G}_0  {\hskip 2mm} $ and
 $  {\hskip 2mm} Map \times J_0    \times  {\cal G}_0  \times  {\cal G}_0
\times [0;1]  {\hskip 2mm} $

We assume that $Map \times {\cal J}_0 $
 is embedded as
$ {\hskip 2mm} Map \times {\cal J}_0 \times \{ 0 \} {\hskip 2mm} $ into the
product with the auxillarty spaces.

\vskip .2cm

\proclaim Proposition 2.2 (Gromov).
 If restricted to the subspace
 $  {\hskip 2mm} Map \times {\cal G}_0  {\hskip 2mm} $  in
 $  Map \times {\cal J}_0 $
the zero set ${\bar \partial}^{-1}(0)  {\hskip 2mm} $  consists of  the pairs
$  {\hskip 2mm} (\varphi, g)  {\hskip 2mm} $   where
 $  {\hskip 2mm} \varphi  {\hskip 2mm}$
is a solution of inhomogenous
Cauchy-Riemann equation
$  {\hskip 2mm} (2.3B)$

\vskip .2cm

\proclaim Definition. A section
 $ \Phi$  of the Hilbert bundle ${\cal H}$ over
 some base Hilbert manifold $ {\hskip 2mm} B {\hskip 2mm} $
is called $  {\hskip 2mm} regular  {\hskip 2mm} $ if its derivative
${\hskip 2mm} D\Phi {\hskip 2mm}$ at each point    in the zero-locus
 ${\hskip 2mm}  \Phi^{-1}(0) {\hskip 2mm} $
is  a surjective linear map from the tangent space to
$ {\hskip 2mm} B {\hskip 2mm} $ to the tangent space to the fiber of ${\cal H}$

It is obvious that the  section $ {\hskip 2mm} {\bar \partial} {\hskip 2mm}$
is regular since its derivative in
$ {\hskip 2mm} {\cal J}_0$-directions is already
surjective linear map from
$ {\hskip 2mm}  T {\cal G}_0   \subset T {\cal J}_0 {\hskip 2mm} $
 to $ {\hskip 2mm} T{\cal H}$ .

\vskip .2cm

Thus, ${\bar \partial}^{-1}(0)$  is a smooth Hilbert manifold and by the
infinite-dimensional version of Sard Theorem we have that
 for ``generic''
$ {\hskip 2mm} g \in \Omega^{0,1}(TM) {\hskip 2mm} $
the space of solutions of inhomogenous Cauchy-Riemann equation  $(2.3B)$ is
smooth finite-dimensional manifold.

\vskip .2cm

By the same reason, for  ``generic'' almost-complex structure
 $J \in {\cal J}_0 $ the moduli space ${\cal M}_{J,d} $ of $J$-holomorphic
spheres of ``degree d'' is a  smooth finite-dimensional manifold.

\vskip .2cm

Dimension of this manifold is given by the index of the Fredholm linear
operator $ {\hskip 2mm} D {\bar \partial}  {\hskip 2mm} $ which acts
from $  {\hskip 2mm} T(Map_d)  {\hskip 2mm} $ to
$  {\hskip 2mm} T{\cal H}.   {\hskip 2mm} $
The operator $ {\hskip 2mm} D {\bar \partial}  {\hskip 2mm} $
 is defined as a derivative of the section ${\bar \partial}$ in
$  {\hskip 2mm}  Map_d$-directions.

\vskip .2cm

``Generic'' here means the Baire second category set.

\proclaim Proposition 2.3 ( Gromov).
For the ``generic'' choice of $J$   the
moduli space ${\cal M}_{J,d} $   will be  a smooth almost-complex manifold
of dimension

$$ dim { } {\cal M}_{J,d} {\hskip 2mm}  = {\hskip 2mm}  dim M {\hskip 2mm}  +
{\hskip 2mm}  \sum_{i=1}^{s} d_i {\hskip 2mm}  deg [q_i]      \eqno
(2.7)$$

	The idea of the proof of $(2.7)$ is as follows. The operator $ {\hskip 2mm} D
{\bar \partial}  {\hskip 2mm} $
 is
actually a (twisted) $ {\hskip 2mm}  {\bar \partial}$-operator on
 $ {\hskip 2mm} CP^1. {\hskip 2mm} $
Then
 Atiyah-Singer index theorem, applied to any of our  `` ${\bar
\partial}$-operators'', gives us the r.h.s. of $(2.7)$.

\vskip .2cm

To prove that the actual dimension of the moduli space
 $  {\hskip 2mm} {\cal M}_{J,d}  {\hskip 2mm} $  is equal to its ``virtual
dimension'' given by the index calculation in the r.h.s. of $(2.7)$, we need
some analytic lemmas. These lemmas were first proved by Freed and Uhlenbeck
[FU]. We also recommend the reader a book [DK].

\vskip .2cm

\proclaim Lemma 2.4 (Proposition 4.3.11 of [DK]).
If $ \Phi$ is the regular section of the Hilbert bundle ${\cal H}$ over
 $  {\hskip 2mm} Map_d \times$ [auxillary space] $ {\hskip 2mm}$  then for
``generic'' value of the parameter $ {\hskip 2mm} g  {\hskip 2mm}$ in the
auxillary space the zero-set of $  {\hskip 2mm} \Phi  {\hskip 2mm}$ restricted
to
$ {\hskip 2mm} Map_d \times g  {\hskip 2mm}$ will be a  smooth submanifold of
dimension equal  to    ``the virtual dimension''

Here   ``the virtual dimension'' means the index of the derivative of the
section $ {\hskip 2mm}  \Phi {\hskip 2mm} $
 in   $ {\hskip 2mm} Map_d$-directions
(these operators are always Fredholm).

\proclaim Lemma 2.5 (Proposition 4.3.10 of [DK]).
Any finite-dimensional pseudo-manifold  of parameters in the auxillary space
can be perturbed to be made transversal to the projection operator.

Here the projection operator  (by definition) projects

 $ \Phi^{-1}(0) \subset   Map_d \times$ [auxillary space] $ {\hskip 2mm}$  to
the second factor $ {\hskip 2mm}$ (the  auxillary space).

\vskip .2cm The particular case of this lemma will be

\proclaim Lemma 2.6.
For the pair $ {\hskip 2mm} g^1  {\hskip 2mm}$ and
 $ {\hskip 2mm} g^2  {\hskip 2mm}$ of ``the regular values'' of parameters in
the auxillary space any path $\gamma$ joining them can be perturbed to be made
transversal to the projection operator.

\vskip .2cm

The Lemma 2.6 implies that the inverse image of this ``transversal path''
$\gamma$ gives us a smooth cobordism between
$ {\hskip 2mm}  \Phi^{-1}(0) \bigcap Map_d \times  \{ g^1 \} {\hskip 2mm}$

and
 $ {\hskip 2mm} \Phi^{-1}(0) \bigcap Map_d \times \{ g^2 \}$.

\vskip .2cm

 Using Lemmas 2.4 and 2.6 we have that there exists a smooth cobordism
 ${\cal M}^t  {\hskip 2mm}$ inside
 $ {\hskip 2mm} Map_d  \times {\cal J}_0 {\hskip 2mm} $
between
the moduli spaces
 $ {\hskip 2mm} {\cal M}_{J_{g^1},d}  {\hskip 2mm} $
 and
 $ {\hskip 2mm} {\cal M}_{J_{g^2},d}  {\hskip 2mm} $
constructed using different ``regular'' almost-complex structures
$  {\hskip 2mm} J_{g^1}  {\hskip 2mm} $ and
$  {\hskip 2mm} J_{g^2} .  {\hskip 2mm} $

\vskip .2cm

	The moduli spaces ${\cal M}_{J,d}$
 of  $J$-holomorphic spheres are not compact.

\vskip .2cm

	There are only two (closely related) sources of non-compactness of these
moduli spaces

\vskip .2cm

1)  ${\hskip 2mm}$ The sequence of  unparemetrized  $J$-holomorphic spheres may
``split'' into two $J$-holomorphic spheres by contracting of some loop on
$CP^1$. The resulting ``splitted  $J$-holomorphic sphere'' is (formally) not in
our space   which means that the above sequence diverges. This ``degeneration''
may occur only if both spheres which appear after this ``splitting process''
have non-trivial homotopy type (and cannot be contracted to a point).

\vskip .2cm

2) ${\hskip 2mm}$  The sequence of  paremetrized  $J$-holomorphic spheres in  $
{\cal M}_d $ may diverge by ``splitting off'' a $J_0$-holomorphic sphere of
lower (or the same) degree at some point on  $CP^1$. This means that the
curvarure of our sequence of maps ``blows up'' at some point on  $CP^1$.
This phenomenon is the famous  Uhlenbeck's ``bubbling off'' phenomenon [SU]

\vskip .2cm

	The bubbling off may be possible even when the classical splitting is
impossible. For example, let us consider the simplest case when
$  {\hskip 2mm}  M = CP^1  {\hskip 2mm} $
with the standard complex structure
 and $  {\hskip 2mm}  d=1 .  {\hskip 2mm} $ Then the sequence of holomorphic
degree-one maps from  $CP^1$ to itself  may diverge by ``bubbling off''  any
any point on  $  {\hskip 2mm}  CP^1 .  {\hskip 2mm} $ This will compactify the
non-compact space $  {\hskip 2mm}  {\cal M}_{J,1}   {\hskip 2mm}  $ (which is
diffeomorphic to
${\hskip 2mm} PSL(2,C) {\hskip 2mm}$ in this example) $  {\hskip 2mm}  $ to a
compact space
$  {\hskip 2mm}  CP^3$.

\vskip .2cm

\vskip .2cm

\vskip .2cm

\vskip .2cm

\centerline {\bf COMPACTIFICATION   OF  MODULI  SPACES}

\vskip .2cm

\vskip .2cm

	 In order to compactify the moduli space
$ {\hskip 2mm} {\cal M}_{J_g,d}  {\hskip 2mm} $
 in the sense of Gromov, we should, roughly speaking,  add to it
 the spaces of $J$-holomorphic maps of the connected sum of several copies of
 $  {\hskip 2mm} CP^1  {\hskip 2mm} $ to $  {\hskip 2mm} M  {\hskip 2mm} $ of
total degree $  {\hskip 2mm} d$.

\vskip .2cm

 In other words, the space of ``non-degenerate'' $J$-holomorphic spheres in
$ {\hskip 2mm} M  {\hskip 2mm} $ is non-compact but it will be compact
if we add to it
``degenerate $J$-holomorphic spheres''.

\vskip .2cm

Ruan [Ru1] gave an explicit description how to stratify the compactified moduli
spaces $ {\hskip 2mm} \bar{ {\cal M}}_{J_g,d}$.

\vskip .2cm

\proclaim Definition (Ruan). Let us call
 $ {\hskip 2mm} degeneration
{\hskip 2mm} pattern {\hskip 2mm}$
the following set of data   ${\hskip 2mm} $ DP1) - DP3) :

DP1) ${\hskip 2mm} $ The class $ d^0 \in C , {\hskip 2mm} $ the  set
${\hskip 2mm} \{ d^1 ; ... ; d^k \} \subset  {\hskip 2mm} C { \ } \{ 0 \}
\subset H^2(M)  {\hskip 2mm} $
of effective classes , and the set   ${\hskip 2mm} \{ a_1 ; ... ; a_k \} $ of
positive integers, such that the following identity holds:
$ {\hskip 2mm}  d = d^0 + \sum^k_{i=1} a_i d^i {\hskip 2mm} $

\vskip .2cm

DP2) The set ${\hskip 2mm} \{ I_1 ; ... ; I_t \} {\hskip 2mm} $  of subsets in
the set
${\hskip 2mm} \{ d^0 ; d^1 ; ... ; d^k \}. {\hskip 2mm} $ We do not allow one
of   ${\hskip 2mm} \{ I_1 ; ... ; I_t \} {\hskip 2mm} $ to be the proper subset
of another.

\vskip .2cm

Using the set of data   $ {\hskip 2mm}  \{ d^0 ; d^1 ; ... ; d^k  ;   I_1 ; ...
; I_t \} , {\hskip 2mm}   $  we can construct a graph
  ${\hskip 2mm}  T {\hskip 2mm} $ with $ {\hskip 2mm} k + 1 + t {\hskip 2mm} $
vertices ${\hskip 2mm}  \{ d^0 ; d^1 ; ... ; d^k ; {\hskip 2mm}  I_1 ; ... ;
I_t \}   {\hskip 2mm}   $  as follows:

\vskip .2cm

If the class ${\hskip 2mm}  d^i {\hskip 2mm}   $ lies in the set
 $ {\hskip 2mm} I_j {\hskip 2mm} $ then we join the vertices
 ${\hskip 2mm}  d^i {\hskip 2mm}   $ and $ {\hskip 2mm} I_j {\hskip 2mm} $
by an edge.

\vskip .2cm

DP3)  ${\hskip 2mm}$ The graph ${\hskip 2mm}  T {\hskip 2mm} $ obtained by
above prescription
is a tree.

\vskip .2cm

\proclaim Definition. We will call a  $J_0$-holomorphic sphere
 $  {\hskip 2mm} C_i \in   {\cal M}_{J_0,d^i}
 {\hskip 2mm} $
 simple  if $ {\hskip 2mm} C_i  {\hskip 2mm}$
cannot be obtained as a branched cover of any other   $J_0$-holomorphic sphere.

\proclaim Definition.
If the $J_0$-holomorphic sphere  $ {\hskip 2mm} C_i  {\hskip 2mm}$  is not
 simple then  we will call it  multiple-covered.

We will denote $  {\hskip 2mm} {\cal M}^*_{J_0,d^i}  {\hskip 2mm} $ the space
of all simple  $J_0$-holomorphic spheres of ``degree $ {\hskip 2mm} d$ ''
$ {\hskip 2mm} $ in $  {\hskip 2mm} M . {\hskip 2mm}$ According to the theorem
of McDuff $ {\hskip 2mm}$ [McD1] $ {\hskip 2mm}$  if the almost-complex
structure  $ {\hskip 2mm} J_0  {\hskip 2mm} $ on
$  {\hskip 2mm} M  {\hskip 2mm} $   is ``generic'' then
 $  {\hskip 2mm} {\cal M}^*_{J_0,d^i}  {\hskip 2mm} $ is a smooth manifold of
dimension given by the formula $ {\hskip 2mm} (2.7)$

\vskip .2cm

Let  $ {\hskip 2mm} D_d {\hskip 2mm} = {\hskip 2mm}  \{ {\hskip 2mm}  \{ d^0 ;
d^1 ; ... ; d^k \} ; {\hskip 2mm}  \{ a_1 ; ... ; a_k \} ; {\hskip 2mm} \{ I_1
; ... ; I_t \} ;   {\hskip 2mm} T  {\hskip 2mm}  \} {\hskip 2mm}   $
 be some degeneration pattern.
Then let us define
$  {\hskip 2mm} {\cal N}_{J_g,D_d}  {\hskip 2mm} $ as a topological subspace in

$  {\hskip 2mm} {\cal M}_{J_g,d^0} \times \Pi^k_{i=1}
{\hskip 1mm} [  {\cal M}^*_{J_0 , d^i} {\hskip 1mm}  / {\hskip 1mm} PSL(2,C) ]
 {\hskip 2mm} $ as follows:

\vskip .2cm

An element $  {\hskip 2mm} \varphi  {\hskip 2mm} $
 in
$  {\hskip 2mm} {\cal N}_{J_g,D_d}  {\hskip 2mm} $
consists  of one parametrized $  {\hskip 2mm} J_g$-holomorphic sphere
$ {\hskip 2mm} C_0 \in  {\cal M}_{J,d^0}   {\hskip 2mm} $ and $k$
 unparametrized $  {\hskip 2mm} J_0$-holomorphic spheres

$ \{  C_i \in [  {\cal M}^*_{J_0,d^i} / PSL(2,C) ]
{\hskip 2mm} \}.  {\hskip 2mm} $
We require that for any  subset

 $  I_j {\hskip 2mm} =  \{ d^{j_1} ;  ... ; d^{j_{n_j}} \} {\hskip 2mm} $ from
$  {\hskip 2mm} \{ I_1 ; ... ; I_t \}  {\hskip 2mm} $
the spheres
$ {\hskip 2mm}    \{ C_{j_1} ;  ... ; C_{j_{n_j}} \}  {\hskip 2mm} $
have a common intersection point. We do not allow this intersection  point
 to lie on any other sphere
 $  {\hskip 2mm} C_i \subset M {\hskip 2mm} $
in our collection\footnote
{
If
$ {\hskip 2mm}    C_{j_1} \bigcap  ... \bigcap C_{j_{n_j}}  \bigcap C_i \neq
\emptyset
{\hskip 2mm} $
then our ``degenerate $J$-holomorphic sphere''
would lie in the other stratum governed by another ``degeneration pattern''.
}

\proclaim Comment 1.

  We can think about {\bf parametrized} spheres in
$ {\hskip 2mm} M{\hskip 2mm}$ as about
 {\bf unparametrized} spheres in $ {\hskip 2mm} M \times CP^1 {\hskip 2mm}$
which have degree one in
$ {\hskip 2mm} CP^1$-directions.

\proclaim Comment 2.

 Degeneration of parametrized $J_g$-holomorphic sphere of
degree  ${\hskip 2mm} d {\hskip 2mm} $ in $ {\hskip 2mm} M {\hskip 2mm} $ can
be translated in this language as
 splitting of  unparametrized ${\hskip 2mm} J_g$-holomorphic sphere of
degree  $ {\hskip 2mm} d + [CP^1] {\hskip 2mm} $ in
 $ {\hskip 2mm} M \times CP^1 {\hskip 2mm} $ in connected sum of several
 unparametrized ${\hskip 2mm} J_g $-holomorphic spheres of total degree
 $ {\hskip 2mm} d + [CP^1] $.

\vskip .2cm

One of these spheres has degree one in $ {\hskip 2mm} CP^1$-directions  (and
should lie in  ${\hskip 2mm} {\cal M}_{J,d^0}$ ). ${\hskip 2mm}$
All the other spheres
have degree zero  in $ {\hskip 2mm}CP^1$-directions. Each of these spheres maps
to a point
under projection
 $ {\hskip 2mm} M \times CP^1 {\hskip 2mm} \rightarrow M  {\hskip 2mm}  $
 and thus, it should lie in  $ {\hskip 2mm} {\cal M}_{J_0, a_i d^i}/PSL(2,C)$.

\proclaim Comment 3.

 The numbers $ {\hskip 2mm} \{ a_i \} {\hskip 2mm} $ respect the fact
that the some of $ {\hskip 2mm} J_0$-holomorphic spheres which appear in
 ``the degeneration process''  are  $ {\hskip 2mm} \{ a_i \}$-fold branched
covers of other
$ {\hskip 2mm} J_0$-holomorphic spheres  $ {\hskip 2mm} \{ C_i  \in  {\cal
M}^*_{J_0,d^i}/PSL(2,C)  \}$.

\vskip .2cm

The topological space $  {\hskip 2mm} {\cal N}_{J_g,D}  {\hskip 2mm} $
is not a smooth manifold. However, it admits a smooth desingularization
 $  {\hskip 2mm} {\cal M}_{J_g,D}  {\hskip 2mm} $ constructed as follows [Ru1]:

\vskip .2cm

	For each point $ {\hskip 2mm} z \in CP^1 {\hskip 2mm} $ let
$ {\hskip 2mm} ev_z {\hskip 2mm} $ be the evaluation at the point $ {\hskip
2mm} z {\hskip 2mm} $ map from
$ {\hskip 2mm} Map {\hskip 2mm} $ to    $ {\hskip 2mm} M {\hskip 2mm} $
defined as follows:
$ {\hskip 2mm} ev_z (\varphi) = \varphi(z) {\hskip 2mm} $

\vskip .2cm

We also have a more general evaluation  map
from $ {\hskip 2mm} Map {\bar \times} (CP^1)^m  {\hskip 2mm} $ to $M^m$:

$$ev (\varphi , z_1, ... ,z_m) {\hskip 2mm} =  {\hskip 2mm}
\{ \varphi(z_1) , ... ,  \varphi(z_m) \} $$

Here the symbol $ {\hskip 2mm} {\bar \times}  {\hskip 2mm}$ means taking the
product and then moding out by the action of
 $ {\hskip 2mm} {PSL(2,C)} . {\hskip 2mm} $ The  group element
$  {\hskip 2mm} g \in {PSL(2,C)}  {\hskip 2mm} $ acts on

 $  Map  \times (CP^1)^m  {\hskip 2mm} $ by the formula:

$$ g \cdot   (\varphi , z_1, ... ,z_m)    {\hskip 2mm} =  {\hskip 2mm}
 (\varphi \cdot  g^{-1}  {\hskip 2mm} ,   {\hskip 2mm} g \cdot z_1 ,  {\hskip
2mm}  ... ,  {\hskip 2mm}  g \cdot z_m )  \eqno (2.8)$$

 To construct the desired desingularization, we also need ``the product
evaluation map'' , which we will also define by
$ {\hskip 2mm} ev $.   $ {\hskip 2mm} $ This ``product map''

$$ ev: {\hskip 2mm}          Map \times  (CP^1)^{m_0} \times  Map  {\bar
\times}   (CP^1)^{m_1} \times ... \times Map  {\bar \times}   (CP^1)^{m_k}
{\hskip 2mm}
\rightarrow       $$

$$  \rightarrow    M^{m_0 + ... + m_k}  \times  (CP^1)^{m_0}    \eqno (2.9)
$$

 acts as identity from the factor
$ (CP^1)^{m_0} $ in the l.h.s. of $  {\hskip 2mm} (2.9)  {\hskip 2mm} $  to
the factor
$ (CP^1)^{m_0} $ in the  r.h.s. of $  {\hskip 2mm} (2.9) . $

\vskip .2cm

 For any  degeneration pattern   $ {\hskip 2mm} D_d {\hskip 2mm}$  let us
consider the evaluation map

$$ ev:  \bigcup_{g \in {\cal G}_0 }  {\cal M}_{J_g,d^0} \times  (CP^1)^{m_0}
\times  {\cal M}^*_{J_0,d^1} {\bar \times}   (CP^1)^{m_1} \times ... \times
{\cal M}^*_{J_0,d^k}  {\bar \times}   (CP^1)^{m_k}   \rightarrow  $$

$$  \rightarrow   {\hskip 2mm}  M^{m_0 + ... + m_k}  \times  (CP^1)^{m_0}
\times  {\cal G}_0  \eqno (2.10)   $$

Here  $ {\hskip 2mm} m_i {\hskip 2mm} $ is the valency of the vertex
 $ {\hskip 2mm} d^i {\hskip 2mm} $   of the ``degeneration graph''
 $ {\hskip 2mm} T {\hskip 2mm} $ of our degeneration pattern ${\hskip 2mm}$
 (how many other components the given component
 $ {\hskip 2mm} C_i {\hskip 2mm} $ intersects )

\vskip .2cm

Let us observe that the factors of $M$ in the r.h.s. of  $(2.10)$ are in
one-to-one correspondence  with the edges of the ``degeneration graph''
 $ {\hskip 2mm} T {\hskip 2mm} $. The set of these edges can be divided in the
union of groups in two different ways:

\vskip .2cm

The first way is to consider two edges lying in the same group iff they have
the common vertex of the type  $ {\hskip 2mm}  \{ d^0 ; d^1 ; ... ; d^k \}
{\hskip 2mm} $ This corresponds to the grouping the   factors of $M$ as in the
r.h.s. of  $(2.10)$ .

\vskip .2cm

The second way is to consider two edges lying in the same group iff they have
the common vertex of the type
$ {\hskip 2mm}  \{ I_1 ; ... ; I_t \} {\hskip 2mm} $.  Using this way of
grouping the edges, we can regroup the factors of $M$ in
$ {\hskip 2mm}  M^{m_0 + ... + m_k}  \times  (CP^1)^{m_0}     $ and rewrite the
r.h.s. of  $(2.10)$ as

$$  M^{m_0 + ... + m_k}  \times  (CP^1)^{m_0} \times  {\cal G}_0  {\hskip 2mm}
=
 {\hskip 2mm}  M^{n_0 + ... + n_t}  \times  (CP^1)^{m_0} \times {\cal G}_0
\eqno (2.11)    $$

For each index    ${\hskip 2mm} j = 1,...,t {\hskip 2mm}$ let us take the
diagonal
$ {\hskip 2mm} \Delta_j {\hskip 2mm} = {\hskip 2mm} M \subset  M^{n_j} {\hskip
2mm}    $
and take the product
$  {\hskip 2mm}  \Delta {\hskip 2mm} = {\hskip 2mm} \Pi^t_{j=0} {\hskip 1mm}
\Delta_j  {\hskip 2mm}
\subset {\hskip 2mm}  M^{m_0 + ... + m_k} {\hskip 2mm} $ of these diagonals

\vskip .2cm

Let $\pi$ be the projection from

$${\hskip 2mm}  {\cal M}_{J_g,d^0}  \times  (CP^1)^{m_0} \times
 {\cal M}^*_{J_0,d^1} {\bar \times}   (CP^1)^{m_1} \times ... \times
{\cal M}^*_{J_0,d^k}  {\bar \times}   (CP^1)^{m_k}    {\hskip 2mm}$$

to
${\hskip 2mm}  {\cal M}_{J_g,d^0}  \times
 {\cal M}^*_{J_0,d^1}  \times ... \times
{\cal M}^*_{J_0,d^k}     {\hskip 2mm}$

\vskip .2cm

 It follows directly from the definition of
  $  {\hskip 2mm} {\cal N}_{J_g,D}  {\hskip 2mm} $ that
 $  {\hskip 2mm} \pi^{-1} ( {\cal N}_{J_g,D} ) {\hskip 2mm} $
lies inside
$  {\hskip 2mm} ev^{-1}  [\Delta \times  (CP^1)^{m_0} \times \{ g \}] {\hskip
2mm} $
(both topological spaces lie inside the manifold

$$ {\hskip 2mm}  {\cal M}_{J_g,d^0}  \times  (CP^1)^{m_0} \times
 {\cal M}^*_{J_0,d^1} {\bar \times}   (CP^1)^{m_1} \times ... \times
{\cal M}^*_{J_0,d^k}  {\bar \times}   (CP^1)^{m_k}    {\hskip 2mm} ). $$

\vskip .2cm

Moreover, dimension-counting [Ru1] implies that the map
$  {\hskip 2mm} \pi  {\hskip 2mm} $   restricted to $  {\hskip 2mm} ev^{-1}
[\Delta \times  (CP^1)^{m_0}  \times \{ g \} ] {\hskip 2mm} $ is a branched
covering.
Let us denote the topological space  $  {\hskip 2mm} ev^{-1}  [\Delta \times
(CP^1)^{m_0}  \times \{ g \} ] {\hskip 2mm} $ by
$  {\hskip 2mm} {\cal M}_{J_g,D}  {\hskip 2mm} $ .

\vskip .2cm

It follows from the theorem proved in [McDS] that the image of
the evaluation map $  {\hskip 2mm} ev  {\hskip 2mm}$ is transversal to the
product of diagonals $  {\hskip 2mm} \Delta  {\hskip 2mm} . $

\vskip .2cm

McDuff and Salamon stated this theorem in slightly different terms without
working with inhomogenous Cauchy-Riemann equations and without
including an additional factor of $ (CP^1)^{m_0}$. However, the transversality
result stated here can be derived from their result by taking
$M \times CP^1$ instead of $M$ in their considerations.

\vskip .2cm

It follows from the lemma 2.4 that for generic value of
$  {\hskip 2mm}  g \in {\cal G}_0  {\hskip 2mm}  $
the space  $  {\hskip 2mm} {\cal M}_{J_g,D}  {\hskip 2mm} $ is a smooth
manifold
which
gives  the desired smooth desingularization of $  {\hskip 2mm} {\cal N}_{J_g,D}
 {\hskip 2mm} $

\vskip .2cm

Now we can state explicitely the following

\vskip .2cm

\vskip .2cm

\centerline   {\bf LIST  OF  STATEMENTS  ABOUT  THE  COMPACTIFICATION }.

\proclaim Statement 2.7.
For the ``generic'' choice of
 $ {\hskip 2mm} {g \in {\cal G}_0} {\hskip 2mm} $
the moduli space ${\hskip 2mm}  {\cal M}_{J_g,d} {\hskip 2mm} $ can be
compactified as sa stratified space
$ {\hskip 2mm} \bar{ {\cal M}}_{J_g,d} {\hskip 2mm} $
   such that each stratum is a smooth manifold.

\proclaim Statement 2.8. The strata of
 $ {\hskip 2mm} \bar{ {\cal M}}_{J_g,d} {\hskip 2mm} $ are labelled by
degeneration patterns $ {\hskip 2mm} \{  D_d \}  {\hskip 2mm} $
and are diffeomorphic to the manifolds
 $  {\hskip 2mm}  \{  {\cal M}_{J_g,D_d} \}   {\hskip 2mm} $

The stratum  $  {\hskip 2mm} {\cal M}_{J_g,D^{\beta}}  {\hskip 2mm} $
lies inside the closure of another stratum
$  {\hskip 2mm} {\cal M}_{J_g,D^{\alpha}}  {\hskip 2mm} $  if the degeneration
pattern
$D^{\beta}$ is a {\bf subdivision}  of the degeneration pattern
$D^{\alpha}$.

\proclaim Definition.

 A degeneration pattern

  $$ {\hskip 2mm} D^{\beta} {\hskip 2mm} = {\hskip 2mm}  \{ {\hskip 2mm}
 \{ (d^0)^{\beta} ; (d^1)^{\beta} ; ... ; (d^{k^{\beta}})^{\beta}  ;
 I^{\beta}_1 ; ... ;
I^{\beta}_{t^{\beta}}    {\hskip 2mm} T^{\beta}
 {\hskip 2mm}  \} {\hskip 2mm}   $$

   is called a {\bf subdivision}  of a degeneration pattern

 $$ {\hskip 2mm} D^{\alpha} {\hskip 2mm} = {\hskip 2mm}  \{ {\hskip 2mm}
 \{ (d^0)^{\alpha} ; (d^1)^{\alpha} ; ... ; (d^{k^{\alpha}})^{\alpha}  ;
I^{\alpha}_1 ; ... ;
I^{\alpha}_{t^{\alpha}}    {\hskip 2mm} T^{\alpha}  {\hskip 2mm}  \}
{\hskip 2mm}   $$

if there is e system of maps

$$ \psi_d : {\hskip 2mm} \{ (d^0)^{\beta} ; (d^1)^{\beta} ; ... ;
(d^{k^{\beta}})^{\beta} \} \rightarrow
 \{ (d^0)^{\alpha} ; (d^1)^{\alpha} ; ... ; (d^{k^{\alpha}})^{\alpha} \}  $$

$ \psi_I : {\hskip 2mm} \{ I^{\beta}_1 ; ... ; I^{\beta}_{t^{\beta}} \}
\rightarrow
 \{ I^{\alpha}_1 ; ... ; I^{\alpha}_{t^{\alpha}} \}  $     and

$$ \psi_T : {\hskip 2mm}  T^{\beta}     \rightarrow  T^{\alpha} $$

which are consistent in an obvious sense and satisfy an additional property

$$ \sum_{ d_{ i^{\beta} } \in \psi^{-1}_d (d_{i^,} ) }  a_{ i^{\beta}  }
 d_{ i^{\beta} }
{\hskip 2mm} = {\hskip 2mm} a_{ i^{\alpha} }  d_{ i^{\alpha} } $$

\proclaim Statement 2.9. The codimension of the stratum
 $ {\hskip 2mm}  {\cal M}_{J_g,D_d} {\hskip 2mm} $ is always greater or equal
to $  {\hskip 2mm} 2 k  {\hskip 2mm}  $ where
$  {\hskip 2mm}  \{ d^0 ; d^1 ; ... ; d^k \}  {\hskip 2mm} $ is the part of the
degeneration pattern $ {\hskip 2mm} D_d  {\hskip 2mm} $

\proclaim Statement 2.10.  For any two generic  $ {\hskip 2mm} g_1  {\hskip
2mm} $
and  $ {\hskip 2mm} g_2  {\hskip 2mm} $ in
 $ {\hskip 2mm} {\cal G}_0 {\hskip 2mm} $ there exists a smooth path
$ {\hskip 2mm} \gamma: [0;1] \rightarrow   {\cal G}_0   {\hskip 2mm} $
joining them, such that for any degeneration pattern
 $ {\hskip 2mm} D_d  {\hskip 2mm} $ the manifold
  $ {\hskip 2mm} \bigcup_{g \in \gamma}  {\cal M}_{J_g,D_d}  {\hskip 2mm} $
gives a smooth cobordism between
$ {\cal M}_{J_{g_1},D_d}  {\hskip 2mm} $
and $ {\cal M}_{J_{g_2},D_d} . {\hskip 2mm} $
This cobordism has dimension at least one smaller than the moduli space
$ {\cal M}_{J_g,d}  {\hskip 2mm} $ istelf.

\vskip .2cm

\vskip .2cm

The  proof of the  statements 2.7 - 2.10 should appear in the  paper
by  Ruan and Tian [RT] . It can also be derived from the analysis carried out
by  McDuff and Salamon [McD S] as they mentioned at the end of their review.
Aldhough we also have a proof of these statements, we give a credit for them to
[RT] and [McD S].

\vskip .2cm

\vskip .2cm

\vskip .2cm

\vskip .2cm

$$ $$
\centerline {\bf     3.  QUANTUM CUP - PRODUCTS }
\centerline { }

\vskip .2cm

\vskip .2cm

	The total cohomology group $H^* (M)$ has a natural bilinear form given by
Poincare duality .
We will denote this bilinear form  by
$ {\hskip 2mm} <  ;  > {\hskip 2mm} $ i.e.,

 $ \eta^{AB} {\hskip 2mm}  =  {\hskip 2mm} <A ; B> {\hskip 2mm} $
where $A \in H^m(M)$ ; $B \in H^{2n-m}(M)$.

\vskip .2cm

	In order to determine the structure constants
$ {\hskip 2mm} (C_{AC}^D)_q {\hskip 2mm} $ of the quantum cohomology ring it is
sufficient to define ``quantum tri-linear pairings''
$ {\hskip 2mm} <A ; B ; C>_q {\hskip 2mm} $ and then put

$$ (C_{AC}^D)_q = \eta^{BD} <A ; B ; C>_q     \eqno         (3.1)$$

where we use Einstein notation and sum over the repeated index $B$.

\proclaim Definition A (Witten).

Let $A,B,C  \in H^* (M,Z) \otimes Z_{<C>}  {\hskip 2mm} $  Then

$$ <A ; B ; C>_q^{Wi}  =  \sum_d q^d \int_{   {\cal M}_{J,d}   }  ev_0^* (A)
\bigwedge      ev_{\infty}^* (B)  \bigwedge     ev_1^* (C)    \eqno
(3.2)$$

Strictly speaking, the r.h.s. does not make sense because the moduli space
${\cal M}_{J,d}$ is non-compact and the notion of its top-dimensional homology
class is not well-defined.

\vskip .2cm

	In order to make it well-defined, the integral in the r.h.s. of $(3.2)$ should
be considered as an integral over the {\bf compactified} moduli space.

\vskip .2cm

Since the evaluation maps $  {\hskip 2mm} ev_0$, $ev_1  {\hskip 2mm} $ and
$  {\hskip 2mm} ev_{\infty}  {\hskip 2mm} $  do not extend to the
compactification divisor, in order to define  the integral in the r.h.s. of
 $  {\hskip 2mm} (3.2)$, we should make some choices of differential forms on
 $  {\hskip 2mm}M  {\hskip 2mm} $ representing cohomology classes
$  {\hskip 2mm} A ,  {\hskip 2mm}  B  {\hskip 2mm} $ and $  {\hskip 2mm} C$.

In addition we need  $ev_0^* (A)$,  $ev_{\infty}^* (B) $
and $ev_1^* (C)$ to be   differential forms on ${\cal M}_{J,d}$ which
should extend (at least as continious differential forms) to the
compactification divisor.

Taubes [Ta2] proved that in this case the integral
$  {\hskip 2mm} (3.2)  {\hskip 2mm} $
always converges (due to the fact that the compactification of
$   {\hskip 2mm} {\cal M}_{J,d}  {\hskip 2mm} $  is known explicitely and can
be ``blown up'' to a manifold with corners).

\vskip .2cm

In order to show that  the integral  $(3.2)$ over the  compactified moduli
space is well-defined, one must prove that it is independent of the choice of
 differential form representatives of cohomology classes
$ {\hskip 2mm} A , {\hskip 2mm} B {\hskip 2mm} $ and $ {\hskip 2mm} C
{\hskip 2mm} $
and on the  choice of $ {\hskip 2mm}J ,  {\hskip 2mm}$  assuming the latter to
be ``generic''

\vskip .2cm

This analytic problem  has not  been solved $ {\hskip 2mm}$
(see $ {\hskip 2mm}$ [Ta2] $ {\hskip 2mm}$ for the most advanced
treatment of it).

\vskip .2cm

	 It appears that in order to handle analytic problems related to the
non-compactness of  the moduli spaces
$  {\hskip 2mm}  \bar{ {\cal M}}_{J,d} ,  {\hskip 2mm} $    it is  more
convenient to work with cycles on $  {\hskip 2mm}  M  {\hskip 2mm} $ and and
their
 intersections instead of forms on $  {\hskip 2mm}  M  {\hskip 2mm} $ and their
wedge product (if we choose our cycles to be ``generic'').

\vskip .2cm

The  two approaches by Poincare duality
$A \rightarrow {\widehat A}$ where $A \in H^m(M)$ ,

${\widehat A} \in H_{2n-m}(M) .$

\vskip .2cm

Let $M$ be a smooth compact $2n$-dimensional manifold.
A $d$-dimensional {pseudo-cycle} of $M$ is a smooth map
$$f:V \rightarrow M$$

where $V=V_1 \cup \dots \cup V_d$ is a disjoint union
of oriented $\sigma$-compact
manifolds without boundary\footnote
{
A finite dimensional manifold $V$ is called $\sigma$-compact
if it is a countable union of compact sets.
}
such that

 $$     \overline{f_d(V_d)} - f_d(V_d)
      \subset \bigcup_{j=0}^{d-2} f_j(V_j),\qquad
      \dim\,V_j=j,\qquad
      V_{d-1}=\emptyset    $$

Of course, the manifolds $V_j$ are not required to be compact.

\vskip .2cm

Every $d$-dimensional singular homology class
$\alpha$ can be represented
by a pseudo-cycle $f:V\to M$.
To see this represent it by a map $f:P\to M$
defined on a $d$-dimensional
finite oriented simplicial complex
$P$ without boundary.
This condition means that the oriented faces of its
top-dimensional simplices
cancel each other out in pairs.\footnote
{
To avoid some technicalities with jiggling
(i.e. making maps transverse) caused by the fact
that $P$ is not a manifold,
one could equally well work with elements in the
rational homology $H_*(M,Q)$.
Because rational homology is isomorphic to rational
bordism $\Omega_*(M)\otimes Q$, there is a basis of $H_*(M,Q)$
consisting of elements which are represented by smooth  manifolds.
Thus we may suppose that $P$ is a smooth manifold, if we wish.
}

\vskip .2cm

Thus $P$ carries a fundamental homology class $[P]$
of dimension $d$ and $\alpha$ is by definition
the class $\alpha=f_*[P]$.
Now approximate $f$ by a map
which is smooth on each simplex.
Finally, consider the union
of the $d$ and $(d-1)$-dimensional
faces of $P$ as a smooth $d$-dimensional manifold
$V$ and approximate $f$ by a map which is
smooth across the $(d-1)$-dimensional simplices.

\vskip .2cm

Pseudo-cycles of $M$ form an abelian group with
addition given by disjoint union.  The neutral element is
the empty map defined on the empty manifold $V=\emptyset$.
The inverse of $f:V\to M$ is given by reversing the
orientation of $V$.
A $d$-dimensional pseudo-cycle $f:V\to M$
is called {\bf cobordant to the empty set}
if there exists a $(d+1)$-dimensional

{\bf pseudo-cycle with boundary}
$F:W\to M$ with $W=\cup_jW_j$ such that
$$
       \delta W_{j+1} = V_j ,\qquad \left.{F_{j+1}}\right|_{V_j}=f_j
$$
for $j=0,\dots,d$.  Two $d$-dimensional pseudo-cycles
$f:V\to M$ and $f':V'\to M$ are called
{\bf cobordant} if $f\cup f':(-V)\cup V'\to M$
is cobordant to the empty set.

\vskip .2cm

Two pseudo-cycles $e:U\to M$ and $f:V\to M$
are called
{\bf transverse}
 if $e_i:U_i\to M$
is transverse to $f_j:V_j\to M$ for all $i$ and $j$.

\proclaim Lemma 3.1 (McDuff-Salamon).
Let $e:U\to M$ be an $(m-d)$-dimensional singular
submanifold and $f:V\to M$ be a $d$-dimensional
pseudo-cycle.

If $e$ is transverse to $f$ then the
set $\left\{(u,x)\in U\times V\,|\,e(u)=f(x)\right\}$
is finite.  In this case define
$$
      e\cdot f
      = \sum_{u\in U,\,x\in V\atop e(u)=f(x)}
        \nu(u,x)
$$
where $\nu(u,x)$ is the intersection number
of $e_{m-d}(U_{m-d})$ and $f_d(V_d)$
at the point $e_{m-d}(u)=f_d(x)$.

\vskip .2cm

The intersection number $e\cdot f$ depends only
on the cobordism classes of $e$ and $f$.

Every $(2n-d)$-dimensional pseudo-cycle $e:W\to M$
determines a homomorphism
$$
      \Phi_e : H_d(M,Z)\to Z
$$
as follows.  Represent the class $\alpha\in H_d(M,Z)$
by a pseudo-cycle $f:V\to M$ .
Any two such representations are cobordant
and hence, by Lemma 2.5. ,
the intersection number
$$
     \Phi_e(\alpha) = e \cdot f
$$
is independent of the choice of $f$ representing $\alpha$.
The next assertion also follows from Lemma 2.5.

\proclaim Lemma 3.2 (McDuff-Salamon).
 The homomorphism $\Phi_e$ depends
only on the cobordism class of $e$.

Using this isomorphism,  ``$q$-deformed tri-linear pairings''
$ <A ; B ; C>_q$ can be defined as follows:

\proclaim Definition B (Vafa,Ruan).

$$  <A ; B ; C>_q^{VR}  =   \sum_d q^d  \sum_{ [\varphi \in {\cal M}_{J,d}
\bigcap  ev_0^{-1} ({\widehat A}) \bigcap      ev_{\infty}^{-1} ({\widehat B})
\bigcap ev_1^{-1} ({\widehat C})  ] } \pm{1}  \eqno         (3.3)$$

Here the sum in  the r.h.s. of $(3.3)$ is only over those values of $d$ that

$dim A + dim B + dim C = dim {\cal M}_{J,g,d}$ and
only over zero-dimensional cpomponents of
${\cal M}_{J,d} \bigcap  ev_0^{-1} ({\widehat A}) \bigcap      ev_{\infty}^{-1}
({\widehat B}) \bigcap ev_1^{-1} ({\widehat C})$

\vskip .2cm

The sign $\pm{1}$ is taken according to the orientation of intersection

${\cal M}_{J,g,d} \bigcap  ev_0^{-1} ({\widehat A}) \bigcap
ev_{\infty}^{-1} ({\widehat B}) \bigcap ev_1^{-1} ({\widehat C})$. This
intersection index is unambigously defined
since the moduli space ${\cal M}_{J,d}$ is provided with its canonical
orientation using the determinant line bundle of the ${\bar \partial}$-operator
[FH].

\vskip .2cm

The above definition requires several comments:

\vskip .2cm

1) We should make some clever choice of cycles representing the homology
classes $  {\hskip 2mm}  \widehat A  , \widehat B , \widehat C  {\hskip 2mm}  $
in order the r.h.s. of
$ {\hskip 2mm} (3.3)  {\hskip 2mm}  $ to be defined (i.e., the intersection of
the cycles to be transverse)

2) We should prove that  the r.h.s. of $(3.3)$ is independent of this choice

3)  We should prove that  the r.h.s. of $(3.3)$ is independent of the choice
of $  {\hskip 2mm}  J  {\hskip 2mm} $ and $  {\hskip 2mm}  g  {\hskip 2mm} $ as
long as   $  {\hskip 2mm}  J  {\hskip 2mm} $  and
$ {\hskip 2mm}  g  {\hskip 2mm}  $   are ``regular''

\vskip .2cm

``Regular'' means that $J$ is a regular value of the projection map
$ {\hskip 2mm}  \pi_{{\cal J}_0 }       {\hskip 2mm}  $
 from ${\bar \partial}^{-1}(0)  \in Map \times {\cal J}_0   $
to $ {\cal J}_0  $

\vskip .2cm

``The  clever choice of cycles'' means that these cycles should be realized by
``pseudo-manifolds''.

\vskip .2cm

The  proof of ``independence of the choices'' is expected to be given in [RT].
This proof uses cobordism arguments and relies on
the Statements 2.4 - 2.7.

\vskip .2cm

The formula $(3.3)$ for the ``$q$-deformed tri-linear pairings'' was first
written by Vafa.

\vskip .2cm

But in [Va] only ``unperturbed'' holomorphic maps were considered. This makes
the formula
$(3.3)$  incorrect in  when the dimension formula $(2.7)$ does not hold for
some components of the moduli  space  $  {\hskip 2mm}  {\cal M}_{J,d}$.

\vskip .2cm

\proclaim Lemma 3.3 (Taubes). There exist choices of smooth  differential form
representatives of cohomology classes $A , B$ and $C$ such that

$ <A ; B ; C>_q^{VR}  =  <A ; B ; C>_q^{Wi}  {\hskip 2mm} . $

Taubes takes differential forms with support near $\widehat A , \widehat B ,
\widehat C$ respectively. Then the integral in the r.h.s. of $(3.2)$ is
well-defined.

\vskip .2cm

\vskip .2cm

Let $  {\hskip 2mm} A {\hskip 2mm} $ and
$  {\hskip 2mm} B  {\hskip 2mm} $ b
e $ {\hskip 2mm} Z_{<C>}$-valued cohomology classes of
 $  {\hskip 2mm} M  {\hskip 2mm} $ and let
 $  {\hskip 2mm}  A*B  {\hskip 2mm} $ be their quantum cup-pruduct. Then we
have:

\proclaim Lemma 3.4.
$  {\hskip 2mm} deg (A*B)  {\hskip 2mm}  =  {\hskip 2mm}
 deg ( A ) {\hskip 2mm} +  {\hskip 2mm} deg ( B ) $

Thus, we have a new $Z$-graded  ring  structure on
$H^* (M,Z) \otimes Z_{<C>} .  {\hskip 2mm}$
We will call this new ring {\bf the  quantum cohomology ring} of
$  {\hskip 2mm} M   {\hskip 2mm} $  and we will
denote it $HQ^*(M)$.

\vskip .2cm

	Let us define the homomorphism
 $  {\hskip 2mm} l^* : HQ^*(M) \rightarrow H^*(M)  {\hskip 2mm} $ as tensor
multiplication on the ring $  {\hskip 2mm} Z  {\hskip 2mm} $
over the ring $  {\hskip 2mm} Z_{<C>}  {\hskip 2mm} $ which is induced bu the
augmentation
$  {\hskip 2mm} I: Z_{<C>}  \rightarrow Z$.

\proclaim Lemma 3.5.
$l^*$ is a ring homomorphism which preserves the grading.

\vskip .2cm

\vskip .2cm

Before going to the Floer cohomology ring and proving that it is isomorphic to
the quantum cohomology ring let me comment once again about the status of the
definitions of the latter ring.

\vskip .2cm

\proclaim Comment
Only Definition B  of the quantum cup-product has well-defined mathematical
objects in its r.h.s.

\vskip .2cm

\vskip .2cm

\centerline {\bf THE  OPERATION  OF  QUANTUM  MULTIPLICATION }

\vskip .2cm

	In Floer theory which will be discussed in the next paragraph there
is a linear map   $m_F : H^* (M)  \rightarrow End( HF^* (M) )$ or,
equivalently, the action of the classical cohomology of the manifold $M$ on its
Floer cohomology   $HF^* (M)$. Latter module is canonically isomorphic with the
total cohomology
group $H^*(M) \otimes N$ defined as a module over the Novikov ring $N$.

\vskip .2cm

	There is a natural analog of this Floer's map $m_F$ in quantum cohomology:
namely, an operation  $m_Q (C )$ of quantum multiplication (from the left) on
the cohomology class $C \in H^*(M) \otimes  N $

$$m_Q ( C ): H^*(M) \otimes  N  \rightarrow  H^*(M) \otimes  N
 \eqno         (3.4) $$

	In order to obtain the action of  $H^*(M) \otimes N$  on the
homology of $M$ instead of   cohomology of $M$  we should apply Poincare
duality to $(2.13)$. Let us fix some (homogenous) basis  $\{ A,B,...\} $ in
$H^*(M,Z)$  and the Poincare dual basis $\widehat{A},\widehat{B},...$ in
$H_*(M,Z)  \subset   H_*(M,Z)  \otimes  N $. Then we can write matrix elements
$<B| m_Q (C) |A>$ of the operator $m_Q (C )$ in this basis.

\proclaim Lemma 3.6.
$$<B| m_Q (C ) |A>  =  <A ; \eta (B) ; C>_q     \eqno         (3.5) $$

Here $\eta : H_m(M)  \rightarrow  H_{2n-m}(M)$ is a Poincare duality
isomorphism
{\bf in homology} of $M$.

\vskip .2cm

\vskip .2cm

\vskip .2cm

$$ $$
\centerline {\bf     4. REVIEW OF SYMPLECTIC FLOER HOMOLOGY}
\centerline { }
	Let ${\cal L}M$ be the free loop-space of our (compact, sipply-connected
semi-positive) almost-Kahler manifold  $M$ and let $\widehat{{\cal L}M}$ be its
universal cover. The points in  $\widehat{{\cal L}M}$ can be described as pairs
$ {\hskip 2mm} (\gamma ; z)  {\hskip 2mm}$
where $\gamma: S^1 \rightarrow M$ be a free-loop in $M$  and
$z : D^2 \rightarrow M$ be a smooth map from 2-disc $D^2$ which coincides with
 $\gamma$ at the boundary of the disc $\partial D^2 = S^1$. The two maps $z_1$
and $z_2$ of the disc are considered to be equivalent if they are homotopic to
each other and the corresponding homotopy leaves their common boundary loop
$\gamma$ fixed.

\vskip .2cm

\vskip .2cm

	Following Floer [F1-F8] we can define            ``the symplectic action
functional''
$S_{\omega}: \widehat{{\cal L}M}  \rightarrow R$ as follows:

$$S_{\omega} (\gamma ; z)  =   \int_{D^2} z^*(\omega)
\eqno         (4.1) $$

where $\omega$ is the symplectic form on $M$  and $ z^*(\omega) $ is its
pull-back

to the 2-disc $D^2$.

\vskip .2cm

	The tangent vectors to the free loop-space   at the point $\gamma \in {\cal
L}M$  can be described as vector fields $\{ \xi , \eta ,...\}$ on $M$
restricted to the loop $\gamma$. The  free loop-space ${\cal L}M$ (and its
universal cover) has a natural
structure of (infinite-dimensional) almost-Kahler manifold described as
follows:

\vskip .2cm

Let $ {\hskip 2mm} g {\hskip 2mm}$ and  $ {\hskip 2mm} \omega  {\hskip 2mm}$ be
the Riemannian metric and the symplectic form on $M$.
Then we can cefine  the Riemannian metric ${\hskip 2mm} \tilde g {\hskip 2mm}$
and the symplectic form
${\hskip 2mm} \tilde{\omega} {\hskip 2mm}$ on the loop-space ${\cal L}M$ by the
formulas:

$$\tilde g (\xi , \eta) = \int_{S^1} g(\xi(\gamma(\theta));
\eta(\gamma(\theta)) d \theta
\eqno         (4.2A) $$
	and
$$\tilde{\omega} (\xi , \eta) = \int_{S^1} \omega(\xi(\gamma(\theta));
\eta(\gamma(\theta)) d \theta
\eqno         (4.2B) $$

where $\theta$ is the natural length parameter on the circle $S^1$ defined
modulo $2\pi$

\vskip .2cm

The  Riemannian metric $\tilde g$ and the symplectic form
$ \tilde{\omega}$ on the loop-space ${\cal L}M$ are related through the
almost-complex structure operator $\tilde J$. Action of this almost-complex
structure operator $\tilde J$ on the tangent vector $\xi$ to the loop $\gamma$
(which is the vector field restricted to the loop $\gamma$) is defined as the
action of the almost-complex structure operator $J$ on the base manifold $M$ on
our vector field $\xi$.

\proclaim Lemma 4.1. ( Givental).
The following statements hold:

A) $S_{\omega}$ is a Morse-Bott function on  $\widehat{{\cal L}M}$

\vskip .2cm

B) All the critical submanifolds of the ``symplectic action'' $S_{\omega}$ on
the universal cover of ${\cal L}M$ are obtained from each other by the action
of
the group $\pi_1({\cal L}M) = \pi_2(M) = H_2(M)$ of covering transformations.
The image of (any of) these critical submanifolds under the universal covering
map $\pi : \widehat{{\cal L}M} \rightarrow  {\cal L}M$ is the submanifold
$M  \subset  {\cal L}M$     of constant loops.

\vskip .2cm

 If we consider $\widehat{{\cal L}M}$ as a symplectic manifold with the
symplectic form $\tilde{\omega}$
given by $(3.2B)$ then:

\vskip .2cm

C) The hamiltonian flow of the functional ${\hskip 2mm} S_{\omega}{\hskip 2mm}$
generates the circle action on ${\hskip 2mm} \widehat{{\cal L}M}{\hskip 2mm}$
and

\vskip .2cm

D) This circle action is just rotation of the loop $\gamma(\theta) \rightarrow
\gamma(\theta + \theta^0)$

\vskip .2cm

	Let us choose (once and for all) one particular critical submanifold
$M  \subset \widehat{{\cal L}M}$
of the symplectic action ${\hskip 2mm} S_{\omega}$. Then any other critical
submanifold
of  ${\hskip 2mm} S_{\omega}{\hskip 2mm} $ has the form
 ${\hskip 2mm} q^d M {\hskip 2mm}$  which means that it is obtained from
${\hskip 2mm} M {\hskip 2mm}$ by the action of the element
 ${\hskip 2mm}q^d {\hskip 2mm} $ of the group ${\hskip 2mm} H_2(M) {\hskip
2mm}$ of covering transformations.

\proclaim Lemma 4.2.
The gradient flow of the symplectic action functional $S_{\omega}$ on the
universal cover of the loop-space
(which is provided with its canonical Riemannian metric  $\tilde g$ )
depends only on the  almost-complex structure $J$ and does not
depend on the symplectic
form  ${\hskip 2mm} \omega$.

We assume that the metric $g$
and the symplectic form  $\omega$ are related in the standard way through
the almost-complex structure $J$ .

\centerline { }

	Let  $\dot{\gamma}(\theta)$ be unit the tangent vector field to the loop
$\gamma \in {\cal L}M$ (this tangent vector coincides with the generator of the
circle action
rotating the loop). Then we have

\proclaim Lemma 4.3.        $$grad  {\hskip 3mm} S_{\omega} (\gamma (\theta)) =
J (\dot{\gamma}(\theta))  \eqno         (4.3) $$

	Let $\{ H_{\theta} \}: M \rightarrow R {\hskip 2mm}$ be some (smooth) family
of functions on  ${\hskip 2mm} M {\hskip 2mm}$ parametrized by
 $\theta \in     S^1$. This  family of functions on  $M$  is usually called
``periodic time-dependent Hamiltinian'' where ${\hskip 2mm} \theta {\hskip
2mm}$ is ``time''. The fact that ${\hskip 2mm} \theta \in     S^1 {\hskip 2mm}$
reflects the fact
that the time-dependence of our Hamiltonian is periodic.
Let
$S_{\omega,H} : \widehat{{\cal L}M} \rightarrow R$ be a functional on
$\widehat{{\cal L}M}$ defined as follows:

$$S_{\omega,H} (\gamma ; z) {\hskip 2mm} = {\hskip 2mm} S_{\omega} (\gamma ; z)
 - \int_{S^1} H_{\theta}(
\gamma(\theta))d \theta   \eqno         (4.4) $$

\proclaim Theorem 4.4 (Floer).
 For ``generic'' choice of $H$ and $J$ the functional $S_{\omega,H}$ is a Morse
functional on $\widehat{{\cal L}M}$ (wchich is usually called ``the symplectic
action functional perturbed by a Hamiltonian term'')

	``Generic'' here means that the statement is true for the Baire second
category set in the product of the space of all functions on $M \times S^1$
and the space of all almost-complex structures on $M$.

\vskip .2cm

	The gradient flow trajectory of ``the perturbed symplectic action
functional''on the universal cover of the loop-space can be defined as a
solution of the following PDE:

$${{\partial \gamma_{\tau} (\theta)} \over {\partial \tau}} = J {{\partial
\gamma_{\tau } (\theta)} \over {\partial \theta}} - grad   {\hskip 2mm}
H_{\theta}  \eqno         (4.5) $$

where $\tau$ is the parameter on the gradient flow line, varying from
minus infinity to plus infinity, and $\theta$ be the parameter on the loop.

\vskip .2cm

	We will consider only those solutions of $(4.5)$ which are
$L^2$-bounded, i.e.  satisfy the estimate

$$  \int_{R} d \tau  \int_{S^1} d \theta
|| {{\partial \gamma_{\tau} (\theta)} \over {\partial \tau}} {\hskip 2mm}
 || ^2  < \infty
\eqno         (4.6) $$

The $L^2$-boundedness condition $(4.6)$ implies that

$$ \gamma_{\tau} (\theta) \rightarrow   \gamma_- (\theta)  {\hskip 6mm}
\tau \rightarrow - \infty  \eqno         (4.7A) $$

and

$$ \gamma_{\tau} (\theta) \rightarrow   \gamma_+ (\theta)  {\hskip 6mm}
\tau \rightarrow + \infty \eqno         (4.7B)  $$

where $ \gamma_+ (\theta) $ and $ \gamma_+ (\theta) $ are some ``critical
loops'' or, in another words, critical points of the perturbed symplectic
action functional on the universal cover of the loop-space.

\vskip .2cm

This means that any $L^2$-bounded  solution  of $(4.5)$ always extends to some
continious map from ${\hskip 2mm} S^1 \times R {\hskip 2mm} $ to
${\hskip 2mm} M {\hskip 2mm} $ (which is actually a smooth map with finitely
many singular points).

\vskip .2cm

 Here $S^1 \times R$ is identified
with $C^*$ by the map

$$(\theta(mod 2 \pi); \tau) {\hskip 2mm} \rightarrow {\hskip 2mm}
 exp(\tau + i \theta)  \eqno   (4.8) $$

Let  $ \gamma_+ , \gamma_-   \in \widehat{{\cal L}M}$ be  two such critical
points of $S_{\omega,H}$.

\vskip .2cm

Let us define     ${\cal M} (  \gamma_- ,  \gamma_+)$ as the space of all
$L^2$-bounded trajectories of the gradient flow of $S_{\omega,H}$ ,
joining the critical point    $ \gamma_-$
  and the critical point $ \gamma_+$

\vskip .2cm

	In more down-to earth terms, the space  ${\cal M} (\gamma_- ,\gamma_+)$ can be
defined
as the space of all solutions of $(4.5)$ , $2 \pi$-periodic in $\theta$ with
the
assymptotics  given by $(4.7A)$ and $(4.7B)$

\vskip .2cm

 ${\cal M} (  \gamma_- ,  \gamma_+)$ can be thought as union of all loops lying
on the gradient flow trajectories, and thus, as
a tolological subspace in $\widehat{{\cal L}M}$

\proclaim Theorem 4.6 (Floer).
For the any ``generic'' choice of the  function  $H$ on $S^1 \times M$ and for
any pair  $ \{ \gamma_+ , \gamma_- \}$ of the critical points of $S_{\omega,H}$
   in $\widehat{{\cal L}M}$ the following statements hold:

A) The space ${\cal M} (  \gamma_- ,  \gamma_+)$ is a smooth submanifold in
$\widehat{{\cal L}M}$

\vskip .2cm

B) The dimension of this submanifold is equal to the spectral flow of the
family
${\hskip 2mm} \{ D_{\tau} = {\bar \partial} - grad (H_{\theta}) \} {\hskip 2mm}
(- \infty < \tau < \infty) {\hskip 2mm} $ of  ${\bar \partial}$-operators

acting from the space $W_{5n}^2(S^1, \gamma_{\tau}^*(TM)$
 to the space $W_{5n-1}^2(S^1, \gamma_{\tau}^*(TM)$

\vskip .2cm

C) For any element $q^d \in \pi_2(M)$ we have

$$dim ({\cal M} ( \gamma_- , q^d \gamma_+) ) = {\hskip 2mm} dim ({\cal M} (
\gamma_- ,  \gamma_+) ) +  2< c_1(TM) ; q^d  > \eqno         (4.9) $$

(this formula follows from the computation of the spectral flow)

\vskip .2cm

Since the Hessian of $S_{\omega,H}$ at any of its critical points has
infinitely many positive and infinitely many negative eigenvalues, the usual
Morse index of the critical point is not well-defined.

\vskip .2cm

But the relative Morse index of the pair $ \gamma_- $ and $\gamma_+$  of the
critical points is well-defined as
 ${\hskip 2mm} vdim ({\cal M} ( \gamma_- ,  \gamma_+) ) $

\vskip .2cm

Here by  ${\hskip 2mm} vdim ({\cal M} ( \gamma_- ,  \gamma_+) ){\hskip 2mm} $
 we denote ``the virtual dimension'' of the manifold ${\hskip 2mm}{\cal M} (
\gamma_- ,  \gamma_+){\hskip 2mm}$ which is defined as a spectral flow of the
family

$ \{ D_{\tau} \}(- \infty < \tau < \infty) {\hskip 2mm} $ of
 ${\bar \partial}$-operators

\vskip .2cm

In the case when $J$ and $H$ are ``generic'' (or ``regular'' in the sence of
the previous section),
 this  virtual dimension ${\hskip 2mm} vdim {\hskip 2mm}$ is equal to actual
dimension  $ {\hskip 2mm} dim ({\cal M} ( \gamma_- ,  \gamma_+) ) {\hskip 2mm}$
 of this manifold.

\vskip .2cm

But for some choices of $H$ which will be of interest to us this might  not be
true. In these cases ${\cal M} ( \gamma_- ,  \gamma_+) $ is no longer smooth.
Different components of ${\cal M} ( \gamma_- ,  \gamma_+) $
are allowed to have different dimensions and to meet each other
nontransversally.

\proclaim Lemma 4.7.
 Let ${\hskip 2mm} \gamma_1 , \gamma_2 , \gamma_3 {\hskip 2mm} $ be three
``critical loops'' in $\widehat{{\cal L}M}$ Then

$$ vdim ({\cal M} ( \gamma_1 ,  \gamma_3) ) {\hskip 2mm} = {\hskip 2mm} vdim
({\cal M} ( \gamma_1 ,  \gamma_2) ) {\hskip 2mm} +  {\hskip 2mm} vdim ({\cal M}
( \gamma_2 ,  \gamma_3) ) {\hskip 2mm} \eqno (3.10)$$

This formula follows from the spectral flow calculations and from the fact that
we are working on the simply-connected space  $\widehat{{\cal L}M}$.

\vskip .2cm

It is worth mentioning that the formula $(4.10)$ is not true if we do not go
from   ${\cal L}M$ to its universal cover  $\widehat{{\cal L}M}$.
Without going to the universal cover  the formula $(4.10)$ is only true modulo
$ {\hskip 2mm} 2 \Gamma {\hskip 2mm}$ where ${\hskip 2mm} \Gamma {\hskip 2mm}$
is the least common multiple of the numbers
${\hskip 2mm} \{ < c_1(TM) ;  q_i> \}$

\proclaim Lemma 4.8.

$$ {\cal M} (q^d \gamma_-  , q^d \gamma_+  ) {\hskip 2mm}  = {\hskip 2mm} q^d
[{\cal M} (  \gamma_- ,  \gamma_+)]  \eqno         (4.11) $$

\vskip .2cm

	Although  Morse index of the critical points $\{ \gamma_i \}$ of
$S_{\omega,H}$ is not defined in the usual sense, the formulas $(4.9)$ and
$(4.10)$ allow us to define it by hands.

\vskip .2cm

Let us fix some ``basic critical point''  $\gamma_0  \in \widehat{{\cal L}M}$

\vskip .2cm

For any other critical point  $\gamma  \in \widehat{{\cal L}M}$ we can always
find $q^d \in H_2(M)$ such that either the manifold
$ {\cal M} (\gamma_0 , q^d \gamma  )$
or the manifold
$ {\cal M} (\gamma , q^d \gamma_0  )$
is non-empty. Then we can define

$$ deg  \gamma {\hskip 2mm} = {\hskip 2mm}  deg \gamma_0 {\hskip 2mm}  +
{\hskip 2mm} vdim ({\cal M} ( \gamma_0 , q^d \gamma) ) {\hskip 2mm}  - {\hskip
2mm} deg [q^d]  \eqno         (4.12A)  $$

$$ deg  \gamma {\hskip 2mm} = {\hskip 2mm} deg \gamma_0 {\hskip 2mm} - {\hskip
2mm} vdim ({\cal M} ( \gamma , q^d \gamma_0) ) {\hskip 2mm} + {\hskip 2mm}  deg
[q^d]  \eqno       (4.12B)  $$

Here ${\hskip 2mm} deg [q^d] {\hskip 2mm} $ is defined by $(2.4)$

\vskip .2cm

The formulas $(4.12A)$ and $(4.12B)$ for different
${\hskip 2mm} \{ d \} {\hskip 2mm}$ are consistent with each other.

\vskip .2cm

	So,  our grading on the set of critical points of $S_{\omega,H}$ is defined
uniquely up to an additive constant ${\hskip 2mm} deg \gamma_0$

\vskip .2cm

	The manifolds $\{ {\cal M} ( \gamma_-  ,  \gamma_+  ) \}$ of the gradient flow
trajectories are non-compact. There are two basic reasons of their
non-compactness:

\vskip .2cm

A) ${\hskip 2mm}$ The gradient flow trajectory may goes through the
intermediate critical point, i.e., it may ``split'' into the union of two
trajectories

\vskip .2cm

B) ${\hskip 2mm}$  The sequence of the gradient flow trajectories in
 $ {\hskip 2mm}{\cal M} ( \gamma_-  ,  \gamma_+  ) {\hskip 2mm}$ may diverge by
``bubbling off'' a $J$-holomorphic sphere of degree $d$. The formal limit of
this diverging sequence will be  of a gradient flow trajectory from
 $ {\cal M} ( q^d \gamma_-  ,  \gamma_+  ) $ (which can be thought as a
pseudo-holomorphic cylinder in $M$ in the sence which will be explained in the
next section) and a $J$-holomorphic sphere of degree $d$ attached to this
cylinder at some point.

\vskip .2cm

In order to have a good intersection theory on manifolds of gradient flow
trajectories (which is the main ingredient in the definition of cup-product in
Floer cohomology) we should {\bf compactify} them

\vskip .2cm

The compactification of the manifold $ {\cal M} ( \gamma_-  ,  \gamma_+  ) $
includes:

\vskip .2cm

A) ${\hskip 2mm}$ The loops lying inside the product

$${\cal M} ( \gamma_-  ,  \gamma_1  ) \times
{\cal M} ( \gamma_1 ,  \gamma_2  ) \times ... \times
{\cal M} ( \gamma_{k-1}  ,  \gamma_k  ) \times
{\cal M} ( \gamma_k ,  \gamma_+  ) $$

B) ${\hskip 2mm}$  Those trajectories in
 $ {\hskip 2mm} {\cal M} ( q^d \gamma_-  ,  \gamma_+  ) {\hskip 2mm} $
 which can be obtained by bubbling off from some sequences of trajectories in
$ {\hskip 2mm} {\cal M} ( \gamma_-  ,  \gamma_+  ) . $

\vskip .2cm

The part A) of the compactification is easy to handle. We just add this part to
$ {\cal M} ( \gamma_-  ,  \gamma_+  ) $ to obtain a smooth manifold with
corners.

\vskip .2cm

The above constructed  manifold with corners is desingularized by a canonical
Morse-theoretic  procedure of ``gluing trajectories'' (see [CJS1],[AuBr] for a
precise constructiontion)   to obtain  smooth manifold with boundary. The
boundary of
this  ``desingularized''  manifold consists of the gradient flow trajectories
going through the intermediate critical points together with ``the gluing
data'' which corresponds to
 ``blowing up''
the corners.

\vskip .2cm

The part B) of the compactification is much more complicated object to work
with. It was proved by Floer himself using dimension-counting argument ${\hskip
2mm} (4.9) {\hskip 2mm}$ that if we bubble off the sphere of degree $d$ such
that ${\hskip 2mm}  < c_1(TM) ;  q_i>  {\hskip 1mm}    >  0$ then the
corresponding part of the compactification has codimension at least two.

\vskip .2cm

For the case when
 $< c_1(TM) ;  q_i>  {\hskip 1mm}    \geq  0 {\hskip 2mm}$
  this was proved by Hofer and Salamon [HS]  (assuming that  the almost-complex
structure
$ {\hskip 2mm} J_0 {\hskip 2mm} $ on $ {\hskip 2mm} M {\hskip 2mm} $ is
``generic'').

\vskip .2cm

	Let us consider the free abelian group  $CF_*(M)$ genetated by the critical
points
 of the perturbed symplectic action $S_{\omega,H}$ in $\widehat{{\cal L}M}$.
This abelian group has a structure
of $ {\hskip 2mm} Z_{[H_2 (M)]}$-module since the group
${\hskip 2mm} H_2 (M) {\hskip 2mm}$ of the covering transformations acts on the
set of critical points.

\vskip .2cm

 Since the action of the group of covering transformations is free, the module
$CF_*(M)$ is a free module, generated by the finite set  of the critical points
of the multivalued functional ${\hskip 2mm} S_{\omega,H} {\hskip 2mm}$ on the
loop-space  ${\cal L} M$ (before going to the
universal cover)

\centerline { }

	Let us take a completion of this abelian group  $CF_*(M)$ by allowing
certain infinite linear combinations of the critical points
 of  $S_{\omega,H}$ to appear in $CF_*(M)$. More precisely, let us tensor our
$Z_{[H_2 (M)]}$-module  $CF_*(M)$ on the Novikov ring $N$ over the ring
 $Z_{[H_2 (M)]}$. We will denote this extended abelian group by the same symbol
$CF_*(M)$
(which is actually an $N$-module) and call it  a {\bf Floer chain complex}
corresponding to ``perturbed symplectic action'' $S_{\omega,H}$.

\centerline { }

	A Floer chain complex $CF_*(M)$ has a natural $Z$-grading {\bf deg}
induced from the above-defined grading of the critical points

\centerline { }

Let $\{ x,y,...\}$ be some  set of critical points of $S_{\omega,H}$ on
$\widehat{{\cal L}M}$. We assume that this set  maps isomorphically onto the
set of all
critical points of  $S_{\omega,H}$ on
${\cal L} M$. In other words, we choose one point in the fiber of the universal
cover over each critical point.

\vskip .2cm

	Now it is time to define a boundary operator ${\hskip 2mm} \delta : CF_*(M)
\rightarrow CF_*(M)$ which will:

\vskip .2cm

A) $ {\hskip 2mm}$ commute with $N$-action ${\hskip 2mm}$ (i.e. $\delta$ will
be $N$-module homomorphism) ;

\vskip .2cm

B) ${\hskip 2mm}$ decrease the $Z$-grading {\bf deg} by one.

\vskip .2cm

	Let us define

$$\delta x =  \sum_y  \sum_d < \delta x ; q^d y >  q^d y    \eqno  (4.13) $$

where the sum in the r.h.s. of $(3.13)$ is taken only over such values of
 ${\hskip 2mm} y {\hskip 2mm}$ and of ${\hskip 2mm} d {\hskip 2mm}$  that the
critical points ${\hskip 2mm} x {\hskip 2mm} $ and
$ {\hskip 2mm} q^d y {\hskip 2mm}$ have relative Morse index one.

\vskip .2cm

Let  ${\hskip 2mm} < \delta x ; q^d y > {\hskip 2mm} $ be the number of
connected components
 of ${\cal M}_ (x ,q^d y){\hskip 2mm} $ (all of them are one-dimensional)
counted with ${\hskip 2mm} \pm 1$-signs depending on orientations of
these components relative to their ends  ${\hskip 2mm} x {\hskip 2mm} $ and
${\hskip 2mm} q^d y$

\proclaim Lemma 4.9.
The boundary operator $\delta$ is defined over the Novikov ring $N$

This means that for any index $i = 1,...,s$ there exists an integer $N_i$ such
that only those values of $(d_1,...,d_s)$ could contribute to the r.h.s.
 of  $(4.13)$ that $d_i > - N_i {\hskip 2mm} $ for all $i$.

\proclaim Proof.

\vskip .2cm

By definition of the gradient flow, if the manifold
 $ {\hskip 2mm} {\cal M} ( x  ,  q^d y  ) {\hskip 2mm}$
is non-empty, then
$ {\hskip 2mm} S_{\omega,H}(x) {\hskip 1mm}  >   {\hskip 1mm}  S_{\omega,H}(q^d
y) {\hskip 2mm}$
for any $ {\hskip 2mm} J_0$-compatible symplectic form ${\hskip 2mm}  \omega
{\hskip 2mm}$
(and in particular for our basic forms
 $  {\hskip 2mm} \{  \omega_1,...,  \omega_s \}$ )
This means that for any positive real number
 $  {\hskip 2mm} t  {\hskip 2mm} $ and for any trajectory
 $  {\hskip 2mm} \gamma_(\tau, \theta) \in  {\cal M} ( x  ,  q^d y  )
 {\hskip 2mm}$ we have

$$S_{ \omega_i ,H}(x) {\hskip 3mm} - {\hskip 3mm} S_{ \omega_i ,H}(q^d y)
{\hskip 3mm}  =   $$

$$  = {\hskip 3mm}
 \int_{S^1 \times R} \gamma^*(\omega_i) {\hskip 3mm} + {\hskip 3mm}
\int_{S^1} H_{\theta}(y(\theta))d \theta   {\hskip 3mm}
- {\hskip 3mm}   \int_{S^1} H_{\theta}(
x(\theta))d \theta {\hskip 3mm}  > {\hskip 3mm} 0  \eqno (4.14)   $$

Since the values of the integrals
$  {\hskip 2mm}  \int_{S^1} H_{\theta}(y(\theta))d \theta  {\hskip 2mm}  $
and $  {\hskip 2mm} \int_{S^1} H_{\theta}(
x(\theta))d \theta  {\hskip 2mm} $    are independent of the symplectic form,
and
$  {\hskip 2mm}  \int_{S^1 \times R} \gamma^*(\omega_i)  {\hskip 2mm} $ is a
homotopy invariant which depends only on the limit values of
$  {\hskip 2mm}  \gamma  {\hskip 2mm}  $ as $  {\hskip 2mm}  \tau \rightarrow
\pm \infty  , {\hskip 2mm}  $ then
we can conclude that  in our case $  {\hskip 2mm}  \int_{S^1 \times R}
\gamma^*(\omega_i)  {\hskip 2mm}  $
depends only on the value of $  {\hskip 2mm}  d$.

\vskip .2cm

It follows directly from the fact that $  {\hskip 2mm} \{  \omega_1,...,
\omega_s \}  {\hskip 2mm} $
form a basis dual to

$\{  q_1,...,  q_s \}  {\hskip 2mm} $ that
 if the value of $d_i$ decreases by one then the value of the integral
$  {\hskip 2mm} \int_{S^1 \times R} \gamma^*(\omega_i)  {\hskip 2mm} $  also
decreases by one.

\vskip .2cm

 This observation implies existence of the lower bound $ {\hskip 2mm} - N_i
{\hskip 2mm} $ on the value of
$ {\hskip 2mm} d_i  {\hskip 2mm} $ in order  the inequality $ (4.14)$ to hold.
   This is equivalent to the statement of the Lemma 4.9.

\vskip .2cm

\proclaim Theorem 4.10 (Floer). $\delta^2 = 0$ .

The proof of this statement is highly non-trivial and relies heavily on the way
how we compactify the manifolds   ${\hskip 2mm} \{ {\cal M} ( \gamma_-  ,
\gamma_+  ) \} {\hskip 2mm}$   of the gradient flow
trajectories. This allows one to prove that the contributions to
 $ {\hskip 2mm}  \delta^2 {\hskip 2mm}$ ``from the boundary'' of the
appropriate manifold  of the gradient flow
trajectories will cancel each other.

\proclaim Lemma 4.11.
Homology  $HF_*(M)$ of the Floer chain complex inherit both the $N$-module
structure
 and the $Z$-grading {\bf deg} from $ CF_*(M)$.

\proclaim Theorem 4.12 (Floer).       $HF_*(M) = H_*(M) \otimes N$

The idea of the proof of this theorem is as follows:

\vskip .2cm

First, Floer proved that the graded module $HF_*(M)$ is well-defined and
independent of the choice of ``hamiltonian perturbation'' $H$ involved in its
definition.

\vskip .2cm

Floer constructed an explicit chain homotopy between Floer chain complexes
$ CF_*(M, H_1) {\hskip 2mm} $ and
$ {\hskip 2mm} CF_*(M, H_2) {\hskip 2mm} $ constructed from two different
hamiltonians ${\hskip 2mm}  H_1 {\hskip 2mm} $ and
 $ {\hskip 2mm} H_2 {\hskip 2mm} $
 (which are functions from $ S^1 \times M$ to $R {\hskip 2mm}$ )

\vskip .2cm

	Second, if we consider $\theta$-independent Hamiltonian
 $H: M \rightarrow R$  which is small in $C^2$-norm, then all the critical
points of perturbed symplectic action functional  $S_{\omega,H}$  on
$ {\hskip 2mm} \widehat{{\cal L}M} {\hskip 2mm} $  can be obtained from the
critical points of $ {\hskip 2mm} H {\hskip 2mm}$ on
${\hskip 2mm} M {\hskip 2mm}$  by covering transformations. Here $M$ is
embedded in $\widehat{{\cal L}M}$  as a submanifold of constant loops  as
specified above.

\vskip .2cm

	Saying the same thing in another words, only constant loops can be critical
points of  $S_{\omega,H}$ . These ``critical loops''can take values  in the
critical points of $H  {\hskip 2mm} $ on the manifold
$  {\hskip 2mm} M  {\hskip 2mm}$ and only in those points.

\vskip .2cm

The gradient flow trajectories joining these critical points can be of two
types:

\vskip .2cm

A) Lying inside submanifold $M \subset {\cal L}M$ of constant loops

\vskip .2cm

B) Not lying inside any submanifold  of constant loops

\vskip .2cm

	The trajectories of type B) cannot be isolated due to non-triviality of

$S^1$-action (which rotates the loop) on the space of those trajectories.

\vskip .2cm

Thus, only trajectories of type A) can contribute to the Floer boundary
operator  $,  {\hskip 2mm} \delta .  {\hskip 2mm} $ But the chain complex
generated by these trajectories is exactly the Morse complex of $M$.

Thus, the homology of the Floer complex will be the same as homology of $M$
(tensored by the appropriate coefficient ring due to the action of the group of
covering transformations)

\vskip .2cm

	Before starting to explain cup-product structure, let us define
{\bf Floer cohomology}  $HF^*(M)$ and {\bf  Floer cochain complex} $CF^*(M)$
for both perturbed and unperturbed symplectic action. To define those objects
we should define:

\vskip .2cm

A) Floer cochain complex $CF^*(M) = Hom_N (CF^*(M),N) $

\vskip .2cm

B) Coboundaty operator $ \delta^*$ in  the Floer cochain complex

\vskip .2cm

C) Floer cohomology $HF^*(M)$ as homology of the complex
$(CF^*(M) ; \delta^* )$

\proclaim Lemma 4.13. The following statements hold:

A) $  {\hskip 2mm} HF^*(M) = Hom_N (HF^*(M),N) $

B)   $ {\hskip 2mm} HF^*(M) = H^*(M) \otimes N = H^*(M,N)  {\hskip 2mm}$
i.e. Floer cohomology are isomorphic to ordinary cohomology with the
appropriate coefficient ring.

\centerline { }

	 The Floer cochain complex of the perturbed symplectic action functional
$S_{\omega,H}$ has a canonical basis corresponding to the critical points
$\{q^{d^1} x , q^{d^2} y , ... \}$ of $S_{\omega,H}$. This basis is dual to the
basis of  the critical points $\{q^{-d^1} x , q^{-d^2} y, ... \}$  in the Floer
chain complex $CF_*(M)$.

\centerline { }

	Proceeding as above, we can develop the  Morse-Bott-Witten theory for the
Morse-Bott  functional  $S_{\omega}$  on the universal cover of the loop-space
$\widehat{{\cal L}M}$
in the same way as Floer developed his theory for Morse functional
$S_{\omega,H}$ on the same space.

\vskip .2cm

	The main ingredient of such a theory is a {\bf Floer chain complex}
corresponding to the ``unperturbed symplectic action'' $S_{\omega}$ .
Algebraically this chain complex is
 defined
 as $H_*(M) \otimes N$.

\centerline { }

Geometrically, this Floer chain complex
is generated (as an abelian group) by the total homology of all the critical
submanifolds $ {\hskip 2mm} \{ q^d M \} {\hskip 2mm} $ of the symplectic action
functional.

\vskip .2cm

Here, as above, we
allow certail infinite linear combinations to appear. The appearance of these
infinite linear combinations stands for the fact that we are
 working over the Novikov ring $ {\hskip 2mm} N$.

\vskip .2cm

	This new  Floer chain complex
(we will again denote it $ {\hskip 2mm} CF_*(M) {\hskip 2mm} $)     also has a
 $ {\hskip 2mm} N$-module structure
 and the $ {\hskip 2mm} Z$-grading {\bf deg}. The latter is defined as follows:

$$deg [q^d \widehat{A}]  {\hskip 2mm} =  {\hskip 2mm} deg [\widehat{A}]
{\hskip 2mm} -  {\hskip 2mm} \sum_{i=1}^{s} d_i  {\hskip 2mm} deg [q_i]
 \eqno    (4.15) $$

where $ {\hskip 2mm} \widehat{A}  {\hskip 2mm}$ be some homology class of
degree
 $  {\hskip 2mm} deg [\widehat{A}]$.

	 Here, as usual, $ A \rightarrow \widehat{A}  {\hskip 2mm} $ stands for
Poincare duality isomorphism between cohomology class
$  {\hskip 2mm} A \in H^{2n - deg [\widehat{A}]}(M)  {\hskip 2mm} $ and
homology class $ {\hskip 2mm} \widehat{A} \in H_{deg [\widehat{A}]}(M)  {\hskip
2mm}$.

	Let  $ {\hskip 2mm} q^{d^1} \widehat{A}  {\hskip 2mm} $ and
$  {\hskip 2mm}q^{d^2}  \widehat{B}  {\hskip 2mm} $ be two (bihomogenous)
elements of the Floer chain complex $ CF_*(M)$. Proceeding as above, we can
define:

\centerline { }

A) $  {\hskip 2mm} $ The manifold
$  {\hskip 2mm}  {\cal M} (q^{d^1}  \widehat{A} , q^{d^2} \eta(\widehat{B}))
{\hskip 2mm}  $
 of the gradient trajectories of
  $  {\hskip 2mm}  S_{\omega}  {\hskip 2mm}  $
which flow from the cycle
$ {\hskip 2mm} q^{d^1}  \widehat{A}  {\hskip 2mm} $ in $ {\hskip 2mm}  q^{d^1}
M  {\hskip 2mm} $  as
$ {\hskip 2mm} \tau \rightarrow - \infty  {\hskip 2mm} $  to the cycle
 $  {\hskip 2mm} q^{d^2} \eta(\widehat{B})  {\hskip 2mm} $  in
$ {\hskip 2mm}  q^{d^2}  M  {\hskip 2mm} $ as
 $  {\hskip 2mm} \tau \rightarrow + \infty  {\hskip 2mm} $

\centerline { }

We compactify this manifold  by the gradient flow trajectories passing through
the intermediate critical submanifolds and by trajectories in
$  {\hskip 2mm} {\cal M} (q^{d^1+d}  \widehat{A} , q^{d^2} \eta(\widehat{B}))
{\hskip 2mm}$ obtained by bubbling off.

\centerline { }

 (Here, as above, $ \widehat{B} \rightarrow \eta(\widehat{B})$ stands for
Poincare duality {\bf in homology} of $M$.)
\centerline { }

B) ${\hskip 2mm}$ Relative Morse index of $q^{d^1} \widehat{A}$ and $q^{d^2}
\widehat{B}$ as the virtual dimension of ${\cal M} (q^{d^1}  \widehat{A} ,
q^{d^2} \eta(\widehat{B}))  {\hskip 2mm} $ defined as
$  {\hskip 2mm} deg [q^{d^2} \eta( {\widehat B})]  - deg [q^{d^1} { \widehat
A}]$

\centerline { }

C) ${\hskip 2mm} Z$-grading   {\bf deg} on the  Floer chain complex $  {\hskip
2mm} CF_*(M) {\hskip 2mm} $
(defined by the formula $(4.15)$)
$ {\hskip 2mm}$ such that the relative Morse index of
 $ {\hskip 2mm} q^{d^1} \widehat{A} {\hskip 2mm} $ and
 ${\hskip 2mm} q^{d^2}  \widehat{B} {\hskip 2mm}$
is equal to the difference of their degrees

\centerline { }

D) Floer boundary operator $\delta : CF_*(M) \rightarrow CF_*(M)$ which
commutes with $N$-action and decreases the $Z$-grading {\bf deg} by one.

\centerline { }

	This  Floer boundary operator is defined as

$$\delta \widehat{A} =  \sum_{\widehat{B}}  \sum_d < \delta \widehat{A} ; q^d
\widehat{B} >  q^d \widehat{B}    \eqno  (4.16) $$

Here (as usual)
 ${\hskip 2mm} < \delta \widehat{A} ; q^d \widehat{B} > {\hskip 2mm}$
counts the number (weighted with $\pm 1$-signs depending on orientation)
of isolated gradient flow trajectoties inside the manifold
$  {\hskip 2mm} {\cal M}_d (  \widehat{A} ,  \eta(\widehat{B}))  {\hskip 2mm}$
defined as

 $${\cal M}_d (  \widehat{A} ,  \eta(\widehat{B}))  = {\cal M} ( \widehat{A} ,
q^{d} \eta(\widehat{B}))   \eqno  (4.17)  $$

Here the r.h.s. of $(4.17)$ gives the definition to its l.h.s.

\proclaim Lemma 4.14 (Givental).     $\delta = 0$ .

	The proof of this lemma relies on the fact that any Morse-Bott function
which is a hamiltonian of an  $S^1$-action has this property.

\centerline { }

	Thus, we have

\proclaim Lemma 4.15.
Floer homology   $HF_*(M)$  coincide with the Floer chain complex  $CF_*(M)$
of the unperturbrd symplectic action functional.

	The Floer cochain complex of the unperturbed symplectic action functional $
{\hskip 2mm} S_{\omega}  {\hskip 2mm} $ also has a canonical basis   $  {\hskip
2mm} \{q^{d^1} A , q^{d^2} B \}  {\hskip 2mm} $ where
$  {\hskip 2mm} \{A,B,...\}  {\hskip 2mm} $ is some (homogenous) basis in the
cohomology of $  {\hskip 2mm} M .  {\hskip 2mm}$  This basis is dual to the
basis $  {\hskip 2mm} \{q^{-d^1}  \widehat{A} , q^{-d^2} (\widehat{B}) , ...
   \}  {\hskip 2mm} $  in the Floer chain complex $  {\hskip 2mm}CF_*(M)$.

\centerline { }

	Later on we will use these two bases in these two Floer cochain complexes when
we will work with  Floer cohomology instead of  Floer homology.

\vskip .2cm

\vskip .2cm

\vskip .2cm

$$  $$

 \centerline {\bf       {5. CUP-PRODUCTS IN FLOER COHOMOLOGY}}
\centerline {  }

	Original Floer's motivation for introducing the object which is now known as
``symplectic Floer cohomology'' was to give an interpretation of fixed points
of the symplectomorphism of $M$  in terms of Morse theory.

\vskip .2cm

	In order to have such an interpretation, one has to develop some Morse theory
on the loop space  $  {\hskip 2mm} {\cal L}M  {\hskip 2mm} $  instead of the
usual Morse theory on $  {\hskip 2mm} M  {\hskip 2mm}$.
By identifying  the  fixed points of our  symplectomorphism
(constructed canonically from  ``the periodic time-dependent Hamiltonian''
$H_{\theta}:S^1 \times M \rightarrow R$)
with the critical points of Floer's  ``perturbed symplectic action functional''
on the loop-space,
we have such a Morse-theoretic interpretation.

\vskip .2cm

	If we assume all  the  fixed points of our  symplectomorphism to be
non-degenerate (which is the case only if ``the Hamiltonian'' $H$ is
``generic'' in the sense of Lemma 4.4), and  use the fact that homology of our
Morse-Floer complex
$CF_*(M)$ are isomorphic to the classical homology of $M$ , then the lower
bound
on the number of  the  fixed points of our  symplectomorphism will be given by
usual Morse inequalities. This was one part of the Arnond's Conjecture which
Floer proved.

\vskip .2cm

\vskip .2cm

	The other part of the same Arnond's Conjecture was:
 what will be if we drop the non-degeneracy assumption on the Jacobian at the
fixed points? Classical Morse theory gives us the lower bound on the number of
(not necessarily  non-degenerate) critical points of the function
 $ {\hskip 2mm} H {\hskip 2mm} $ on the compact manifold $  {\hskip 2mm} M
{\hskip 2mm} $ in terms of the so-called  cohomological length of $M$.

\proclaim Definition.
The cohomological length of the topological space $M$ is an integer $k \in
Z_{+}$ such that:

A) $ {\hskip 2mm}$  There exist $  {\hskip 2mm} k-1  {\hskip 2mm} $ cohomology
classes  $  {\hskip 2mm} \alpha_1,...,\alpha_{k-1}  {\hskip 2mm} $ on
$  {\hskip 2mm} M  {\hskip 2mm} $ of positive degrees such that
$  {\hskip 2mm} \alpha_1 \bigwedge...\bigwedge   \alpha_{k-1} \neq 0  {\hskip
2mm} $ in $  {\hskip 2mm} H^*(M)  {\hskip 2mm} $   and

B) $ {\hskip 2mm}$  There are no $  {\hskip 2mm} k  {\hskip 2mm} $ cohomology
classes on $  {\hskip 2mm} M  {\hskip 2mm} $ with this property.

\vskip .2cm

	Thus we see that in order to try to prove this part of the  Arnond's
Conjecture in the framework of Floer's Morse theory, {\bf one needs to invent
some multiplicative structure in Floer cohomology}. A kind of such a
multiplicative structure was also constructed by Floer [F1] and successfully
applied to this part of  Arnond's Conjecture in another Floer's paper [F2].

\vskip .2cm

However, Hofer [Ho2] have found a proof of  this part of  Arnond's Conjecture
without using Floer homology.

\centerline { }

	Using the fact that Floer cohomology  $HF^*(M)$ are canonically isomorphic (as
an abelian group) to the ordinary cohomology  $H^*(M) \otimes N$ the following
five statements are equivalent:

\vskip .2cm

A) we have a multiplication in Floer cohomology

$$HF^*(M)  \otimes   HF^*(M)  {\hskip 2mm}  \rightarrow  {\hskip 2mm}  HF^*(M)
\eqno  (5.1A)    $$

 which is $N$-module homomorphism and which preserves the $Z$-grading;

\vskip .2cm

B) we have an action

$$HF^*(M) \rightarrow  End( HF^*(M) )  \eqno  (5.1B)   $$

of Floer cohomology on itself (by left multiplication)
 which is $N$-module homomorphism and which preserves the $Z$-grading;

\vskip .2cm

C) we have an action

$$H^*(M) \rightarrow  End( HF^*(M) )    \eqno  (5.1C)    $$
of classical cohomology of the manifold $M$ on its Floer cohomology which
preserves the $Z$-grading;

\vskip .2cm

D) we have an action

$$H_*(M) \rightarrow  End( HF^*(M) )    \eqno  (5.1D)    $$
of classical homology of the manifold $M$ (related by Poincare duality with the
cohomology of  $M$) on its Floer cohomology which preserves the $Z$-grading;
\centerline { }

E) we have an action

$$\Omega_*(M) \rightarrow  End( CF^*(M) )    \eqno  (5.1E)     $$

of the space of singular chains in $M$ which can be realized by pseudo-cycles
on the Floer cochain complex of $M$. This action  commutes with the boundary
operator and  preserves the $Z$-grading.

\vskip .2cm

Later on we will denote all these four maps $(5.1A)-(5.1E)$ by the same symbol
$m_F$ and call them {\bf the Floer multiplication}.

\vskip .2cm

	In order to define the Floer multiplication
$  {\hskip 2mm} m_F  {\hskip 2mm} $ in the form $  {\hskip 2mm} (5.1E)  {\hskip
2mm}$
it is enough to define its matrix elements
$  {\hskip 2mm} < y | m_F({\widehat C}) | x >  {\hskip 2mm}  $
where  $  {\hskip 2mm} x, y \in  CF^*(M)  {\hskip 2mm} $ ,
$  {\hskip 2mm} {\widehat C} \in \Omega_*(M)  {\hskip 2mm}$
and then put

$$m_F({\widehat C}) ( x)  {\hskip 2mm}  =  {\hskip 2mm} \sum_{y,d} < q^d y |
m_F({\widehat C}) | x> q^d y \eqno  (5.2)     $$

where the sum in the r.h.s. of $(5.2)$ is taken over the basis $\{ x,y,...\}$
in
the $N$-module $CF^*(M)$.

\vskip .2cm

	Let $x$ , $y$ be two ``basic'' critical point of the perturbed symplectic
action $S_{\omega,H}$
on $\widehat{{\cal L}M}$, and let ${\hskip 2mm} q^{d^1} x {\hskip 2mm}$ and $
{\hskip 2mm} q^{d^2} y {\hskip 2mm}$ be the corresponding elements of the Floer
cochain complex $CF^*(M)$. Then let us put

$$ <  q^{d^2} y | m_F( {\widehat C} ) | q^{d^1} x  > {\hskip 2mm} =
{\hskip 2mm} {\cal M} (q^{d^1} x ; q^{d^2} y) \bigcap
\tilde{ev}_1^{-1}({\widehat C} )    \eqno  (5.3)     $$

The r.h.s. of $ {\hskip 2mm} (5.3)$ is defined here as an intersection index.

\vskip .2cm

Here $\tilde{ev}: S^1 \times  \widehat{{\cal L}M} \rightarrow M $  is the
standard
``evaluation map'' where the circle $S^1$ is assumed to be embedded as a unit
circle $|z|=1$ in the complex plane $C$. The map $\tilde{ev}_1$ means
evaluation of the loop at the point $z=1$

\vskip .2cm

\proclaim Theorem 5.1. The following two statements hold:

A) ${\hskip 2mm}$ The action $ {\hskip 2mm} m_F {\hskip 2mm} $ of `
$ {\hskip 2mm} \Omega_*(M) {\hskip 2mm} $ on
$ {\hskip 2mm} CF^*(M) {\hskip 2mm} $ defined by $ {\hskip 2mm} (5.2)$  and
$(5.3) {\hskip 2mm} $ descends to the action  ${\hskip 2mm} m_F {\hskip 2mm}$
of $ {\hskip 2mm} H_*(M)$ on $HF^*(M)$ ;

\vskip .2cm

B) ${\hskip 2mm}$  The induced action $m_F: H_*(M) \rightarrow  End( HF^*(M) )
$ does not depend on the choice of ``the Hamiltonian'' $ {\hskip 2mm} H {\hskip
2mm} $ assuming that this Hamiltonian is ``generic'' in the sense of Lemma 4.4.

\vskip .2cm

For the case when  $M$ is a  {\bf positive} almost-Kahler manifold the
Theorem 5.1 was proved by Floer himself [F1]. ${\hskip 2mm}$   The same proof
works with some modifications for Calabi-Yau and the general semi-positive
case.

\vskip .2cm

 We will reproduce here the main steps of the proof of the Theorem 5.1.
In the next section the techniques which is used in this proof will be applied
to prove equivalence of Floer's and quantum multiplication.

\vskip .2cm

 The main idea behind this proof is to consider ``$\tau$-dependent Hamiltonian
perturbation'' of the equation $(4.5)$. More precisely, let
 $ {\hskip 2mm}H {\hskip 2mm}  $ be some smooth function on
 $ {\hskip 2mm} R \times S^1 \times M .{\hskip 2mm}$ Here, as above,
the real line ${\hskip 2mm} R {\hskip 2mm} $ is equipped with the parameter $
{\hskip 2mm} \tau {\hskip 2mm} $, varying from minus infinity to plus infinity,
and the circle $ {\hskip 2mm} S^1 {\hskip 2mm} $ is equipped with the arclength
parameter $ {\hskip 2mm} \theta$.

\vskip .2cm

Let us restrict ourselves to the functions on
 $ {\hskip 2mm} R \times S^1 \times M {\hskip 2mm} $ which are $ {\hskip 2mm}
\tau$-independent in the region
 $ {\hskip 2mm} - \infty < \tau  < -1 {\hskip 2mm}$ and
in the region  $ {\hskip 2mm} 1 < \tau < + \infty . {\hskip 2mm}$ This
condition means that there exist two functions $H_-$ and $H_+$ on
$S^1 \times M$ such that

$$ H(\tau ; \theta) = H_-(\theta)   {\hskip 3mm}  if {\hskip 3mm}
\tau \leq -1    \eqno         (5.4A) $$

$$ H(\tau ; \theta) = H_+(\theta)    {\hskip 3mm} if {\hskip 3mm}
\tau \geq 1   \eqno         (5.4B) $$

Let us denote the space of all such functions ${\cal G}_{H_-,H_+}$

\vskip .2cm

Then we can study the space of solutions of the following PDE

$${{\partial \gamma_{\tau} (\theta)} \over {\partial \tau}} = J {{\partial
\gamma_{\tau } (\theta)} \over {\partial \theta}} - grad   {\hskip 2mm}
H_{\theta}  \eqno         (5.5) $$

which are $L^2$-bounded in the sense of $(4.6)$

\vskip .2cm

The same argument as in the section 4 shows that for any  $L^2$-bounded
solution of $(5.5)$ there exist a critical point $ \gamma_+$ of
$S_{\omega,H_+}$ and
a critical point $ \gamma_-$ of $S_{\omega,H_-}$ such that

$$ \gamma_{\tau} (\theta) \rightarrow   \gamma_- (\theta)  {\hskip 6mm}
\tau \rightarrow - \infty  \eqno         (5.6A) $$

and

$$ \gamma_{\tau} (\theta) \rightarrow   \gamma_+ (\theta)  {\hskip 6mm}
\tau \rightarrow + \infty \eqno         (5.6B)  $$

In the same way as in the section 4, we consider the space
${\cal M}_H (\gamma_- ,\gamma_+)$ of $\tau$-dependent gradient
flow trajectories.

\vskip .2cm

The  virtual dimension of
 $  {\hskip 2mm} {\cal M}^H (\gamma_- ; \gamma_+)  {\hskip 2mm}$ is again given
by the spectral flow of the appropriate family of
$  {\hskip 2mm} {\bar \partial}$-operators and coincides with the actual
dimension for ``generic'' $  {\hskip 2mm} J  {\hskip 2mm} $ and
$  {\hskip 2mm} H .$

\vskip .2cm

The moduli spaces
$ {\hskip 2mm}  \{ {\cal M}_H (\gamma_- ,\gamma_+) \}  {\hskip 2mm}$ of
solutions of $ {\hskip 2mm}  (4.5)  {\hskip 2mm} $ are compactified
 by adding gradient flow trajectories obtained by ``splitting'' and by
``bubbling off''. The compactified moduli spaces
$ {\hskip 2mm} \{  {\bar {\cal M}}_H (\gamma_- ,\gamma_+) \} {\hskip 2mm} $
have the structure of stratified spaces such that each stratum is a smooth
manifold with boundary.

\vskip .2cm

\proclaim Statement 5.2.
``The compactification divisor''
$ {\hskip 1mm}  {\bar {\cal M}}_H (\gamma_- , \gamma_+)  {\hskip 1mm} -
{\hskip 1mm}  {\cal M}_H (\gamma_- ; \gamma_+) {\hskip 1mm} $
 has codimension at least two.

\vskip .2cm

For the case of {\bf positive} symplectic manifold this statement was proved
in [F1]. In the semi-positive case the proof was given in [HS] when
$ {\hskip 2mm} H {\hskip 2mm} $ was ${\hskip 2mm} \tau$-independent.  The same
arguments as in [Hs] give the proof in the general case.

\vskip .2cm

The analogues of $  {\hskip 2mm}  (4.9) - (4.11)  {\hskip 2mm} $ also hold for
the moduli spaces of

 $\tau$-dependent  gradient flow trajrctories. This implies that we can fix the
additive constant ambiguities in the gradings of the critical
 points of  $ {\hskip 2mm}  \{ S_{\omega,H} \} {\hskip 2mm} $
{\bf for all Hamiltonians simultaneously} such that

$$vdim {\hskip 1mm}  {\cal M}_H (\gamma_- ; \gamma_+) {\hskip 2mm} =
{\hskip 2mm}
deg_{CF^*(M,H_+)} ( \gamma_+ )  {\hskip 2mm}  - {\hskip 2mm}
 deg_{CF^*(M,H_-)} ( \gamma_- )   \eqno         (5.7) $$

\proclaim Theorem 5.3.
For any two ``$\tau$-dependent Hamiltonians''
$ {\hskip 2mm}  H^{(0)} {\hskip 2mm} $ and
$ {\hskip 2mm}  H^{(1)} {\hskip 2mm} $   lying in the space
$ {\hskip 2mm}  {\cal G}_{H_-,H_+} {\hskip 2mm} $
 the manifolds of trajectories
 $ {\hskip 2mm} {\bar {\cal M}}_{ H^{(0)}} (\gamma_- ,\gamma_+) {\hskip 2mm} $
and
$ {\hskip 2mm} {\bar {\cal M}}_{ H^{(1)}} (\gamma_- ,\gamma_+) {\hskip 2mm}  $
are cobordant to each other as stratified spaces.

More precisely, there exists a path
$ {\hskip 2mm} \{ H^{(t)} \}  {\hskip 2mm} (0 \leq t \leq 1) {\hskip 2mm} $
in $ {\hskip 2mm}  {\cal G}_{H_-,H_+} {\hskip 2mm}  $ joining
 $ {\hskip 2mm}  H^{(0)} {\hskip 2mm}  $ and
 $ {\hskip 2mm}  H^{(1)} {\hskip 2mm} $ such that
  $ {\hskip 2mm}  \bigcup_{0 \leq t \leq 1} {\bar {\cal M}}_{ H^{(t)}}
(\gamma_- ,\gamma_+) {\hskip 2mm} $
   gives us the desired cobordism.

\vskip .2cm

Since Floer boundary operator ${\hskip 2mm}  \delta {\hskip 2mm} $ (in general)
acts nontrivially on
 $ {\hskip 2mm}  \gamma_- {\hskip 2mm}  $ and Floer coboundary operator
${\hskip 2mm}  \delta^* {\hskip 2mm} $  acts nontrivially on
 $ {\hskip 2mm}  \gamma_+ {\hskip 2mm} $
then the manifolds
 $ {\hskip 2mm}  \{ {\cal M}_{ H^{(t)}} (\gamma_- ,\gamma_+) \} {\hskip 2mm} $
are {\bf the manifolds with boundary} and the cobordism
$  {\hskip 2mm} \bigcup_{0 \leq t \leq 1} {\bar {\cal M}}_{ H^{(t)}}
(\gamma_-, \gamma_+)  {\hskip 2mm} $
will have {\bf two extra boundary components}

$ {\hskip 2mm} \bigcup_{0 \leq t \leq 1} {\bar {\cal M}}_{ H^{(t)}}
( \delta \gamma_- ,\gamma_+) {\hskip 2mm} $
and
$ {\hskip 2mm} \bigcup_{0 \leq t \leq 1} {\bar {\cal M}}_{ H^{(t)}}
(\gamma_-,\delta^* \gamma_+)$

\vskip .2cm

These extra components appear due to the presence of intermediate critical
points of $S_{\omega,H_-}$ and
$S_{\omega,H_+}$

\vskip .2cm

The lemma 2.6 implies
 that such a  path
$ \{ H^{(t)} \}  {\hskip 2mm} (0 \leq t \leq 1)$
in ${\cal G}_{H_-,H_+}$ joining $H^{(0)}$ and $H^{(1)}$ exists.
The Statement 5.2. implies that the corresponding smooth cobordism can be
compactified and the compactification divisor will have codimension at least
two.

\vskip .2cm

Let $  {\hskip 2mm} {\cal J}_M , {\hskip 2mm} $ be the space of all
almost-complex structures on $M$ compatible with symplectic forms
$  {\hskip 2mm} \{ \omega_1,..., \omega_s    \}  {\hskip 2mm} $ and with some
differential form representative of $  {\hskip 2mm} c_1 (TM) . {\hskip 2mm} $
By the theorem of Gromov,
$  {\hskip 2mm} {\cal J}_M  {\hskip 2mm} $ is an open contractible set
containing $  {\hskip 2mm} J_0 .$

\vskip .2cm

We should consider the space
$  {\hskip 2mm} Map (\gamma_- ,\gamma_+)  {\hskip 2mm} $ of all $  {\hskip 2mm}
W^2_{5n}$-Sobolev maps from $  {\hskip 2mm} R \times S^1  {\hskip 2mm} $ to   $
 {\hskip 2mm} M  {\hskip 2mm} $ with the assymptotics
 $  {\hskip 2mm} (5.6)  {\hskip 2mm}$ as
$  {\hskip 2mm} \tau \rightarrow \pm \infty  {\hskip 2mm} $ and the
infinite-dimensional Hilbert bundle
$  {\hskip 2mm} {\cal H}  {\hskip 2mm} $ over
 $  {\hskip 2mm} Map (\gamma_- ,\gamma_+)  \times  {\cal J}_M  {\hskip 2mm}$
The fibre of the  bundle
$  {\hskip 2mm} {\cal H}  {\hskip 2mm} $ over the point
 $  {\hskip 2mm} (\gamma ; J) {\hskip 2mm} $  in
$ {\hskip 2mm} Map (\gamma_- ,\gamma_+)  \times  {\cal J}_M $
will be the space of all  $  {\hskip 2mm} W^2_{5n-1}$-Sobolev
$ {\hskip 2mm} (0,1)$-forms on $  {\hskip 2mm} R \times S^1  {\hskip 2mm}$ with
the coefficients in $  {\hskip 2mm} \gamma^*(TM)  {\hskip 2mm} $ which tend to
zero as
 $  {\hskip 2mm}\tau \rightarrow \pm \infty .$

\vskip .2cm

We can consider the pull-back of this Hilbert bundle  to

$Map (\gamma_- ,\gamma_+)  \times  {\cal J}_M   \times$
$\times  {\cal G}_{H_-,H_+}    $
and construct a (canonical) section $\Phi$ as follows:

$$\Phi (\gamma) = {{\partial \gamma} \over {\partial \tau}} - J {{\partial
\gamma} \over {\partial \theta}} - grad   {\hskip 2mm} H (\tau, \theta){\hskip
2mm}
 d {\bar z}
 \eqno (4.8)  $$

Here $  {\hskip 2mm} d {\bar z}$ is a canonical $  {\hskip 2mm}(0,1)$-form on
$  {\hskip 2mm} R \times S^1   {\hskip 2mm} = {\hskip 2mm} C^* . {\hskip 2mm}$
The identification between   $  {\hskip 2mm} R \times S^1   {\hskip 2mm}$   and
 $  {\hskip 2mm} C^*  {\hskip 2mm}$ is given by the map $  {\hskip 2mm} (4.8)$.

\vskip .2cm

The arguments of McDuff [McD] show that if the function $H$ does not admit any
holomorphic symmetries with respect to parameters on $R \times S^1$ then the
 section $\Phi$  is regular over
$Map (\gamma_- ,\gamma_+)  \times  {\cal J}_M   \times  \{ H   \}    $

\vskip .2cm

Since the space of the functions $\{ H \}$ with this property is open end dense
in    $  {\hskip 2mm} {\cal G}_{H_-,H_+}  {\hskip 2mm}$   this means that the
section $  {\hskip 2mm} \Phi  {\hskip 2mm} $  is regular over

$Map (\gamma_- ,\gamma_+)  \times  {\cal J}_M   \times  {\cal G}_{H_-,H_+} $.
This allows us to apply Lemmas 2.4 and 2.6 to prove existence of the above
cobordism.
The theorem 5.2. is proved.

\vskip .2cm

Now let us remember that the manifolds
$  {\hskip 2mm} \{  {\cal M} (q^{d^1} x ; q^{d^2} y) \}  {\hskip 2mm} $
of gradient flow trajectories can be thought either as submanifolds in the
loop-space or as submanifolds in the  space
$  {\hskip 2mm} Map (q^{d^1} x ; q^{d^2} y)  {\hskip 2mm}$ of maps from the
cylinder $  {\hskip 2mm} R \times S^1  {\hskip 2mm} $ to
$  {\hskip 2mm} M  {\hskip 2mm}$ with the fixed ``boundary values'' at
 $  {\hskip 2mm} \tau \rightarrow \pm \infty .  {\hskip 2mm}$ Thus, we have a
commutative diagram

$$\begin{array}{ccc}

{\cal M} (q^{d^1} x ; q^{d^2} y)  &  \stackrel{{\tilde ev}_1}  \longrightarrow
  &          {\cal L}M  \\
 \downarrow  &                                          &  \downarrow \\
 Map (R \times S^1 ; M)      &  \stackrel{ ev_{(0;1)}}  \longrightarrow  &
M

\end{array}$$

\vskip .2cm

Having this commutative diagram in mind, we can rewrite the definition $(5.3)$
for the matrix element of the Floer multiplication as:

$$ <  q^{d^2} y | m_F( {\widehat C} ) | q^{d^1} x  >  {\hskip 2mm} =
 {\hskip 2mm}
{\cal M} (q^{d^1} x ; q^{d^2} y) \bigcap   ev_{(0;1)}^{-1}({\widehat C} )
  \eqno  (5.9)     $$

Here $  {\hskip 2mm} ev_{0,1}  {\hskip 2mm}$ is the ``evaluation at the point
(0;1)'' map from
$Map (R \times S^1 ; M)  {\hskip 2mm} $ to $  {\hskip 2mm} M$.

\vskip .2cm

The formula $(5.9)$ for the matrix element
$ {\hskip 2mm} <  q^{d^2} y | m_F( {\widehat C} ) | q^{d^1} x  > {\hskip 2mm} $
of the Floer multiplication admits the following generalization:

\vskip .2cm

Let $H_+ , H_- , \gamma_+ , \gamma_-$ and $H$  are defined as above. Then let
us put

$$ <  q^{d^2} \gamma_+  | m_F( {\widehat C} ) | q^{d^1} \gamma_-  >
  {\hskip 2mm} =  {\hskip 2mm}
{\cal M}_H (q^{d^1} \gamma_+ ; q^{d^2} \gamma_-) \bigcap
ev_{(0;1)}^{-1}({\widehat C} )
  \eqno  (5.10)     $$

where the r.h.s., as usual, means the intersection index

\vskip .2cm

Any cycle $ {\hskip 2mm} x  {\hskip 2mm}$ in the Floer Chain complex $  {\hskip
2mm} CF_*(M;H_-)  {\hskip 2mm} $ can be written as
a sum $  {\hskip 2mm} \sum_k n_k x_k  {\hskip 2mm} $ where
 $  {\hskip 2mm} x_k  {\hskip 2mm} $ are (possibly coinciding) critical points
of
$  {\hskip 2mm} S_{\omega,H_-}  {\hskip 2mm} $ and
$  {\hskip 2mm} n_k = \pm 1 .  {\hskip 2mm}$ The same is true for the cycle
$  {\hskip 2mm} y = \sum_l m_l y_l  {\hskip 2mm} $ in
 $  {\hskip 2mm} CF_*(M;H_+) .$

\vskip .2cm

We can consider the manifolds

$${\cal M}_H (q^{d^1} x ; q^{d^2} y)  {\hskip 2mm}  =  {\hskip 2mm}
\bigcup_{k,l} n_k m_l  {\hskip 1mm}
{\cal M}_H (q^{d^1} x_k ; q^{d^2} y_l) $$

Here the factor $  {\hskip 2mm} n_k m_l  = \pm 1  {\hskip 2mm} $ in front means
that the component

${\cal M}_H (q^{d^1} x_k ; q^{d^2} y_l)  {\hskip 2mm} $ should be taken with
the appropriate orientation.

\vskip .2cm

If we glue all the components of
$  {\hskip 2mm} {\cal M}_H (q^{d^1} x ; q^{d^2} y)   {\hskip 2mm}$
together, we will obtain a smooth $   deg ( q^{d^2} y) - deg (q^{d^1} x)
$-dimensional pseudo-manifold {\bf without boundary} (or pseudo-cycle).

\proclaim Theorem 5.4..
For any two $\tau$-dependent Hamiltonians $H^{(0)}$ and $H^{(1)}$ from
${\cal G}_{H_-,H_+}$ we have

$${\cal M}_{H^{(0)}} (q^{d^1} x ; q^{d^2} y) \bigcap
ev_{(0;1)}^{-1}({\widehat C} )  {\hskip 2mm} =  {\hskip 2mm} {\cal M}_{H^{(1)}}
(q^{d^1} x ; q^{d^2} y) \bigcap   ev_{(0;1)}^{-1}({\widehat C} )  \eqno (5.11)
$$

The Theorem 5.3 provides us with a cobordism
$  {\hskip 2mm} {\cal M}^t  {\hskip 2mm} $ between

${\cal M}_{H^{(0)}} (q^{d^1} x ; q^{d^2} y)  {\hskip 2mm} $ and
$  {\hskip 2mm} {\cal M}_{H^{(1)}} (q^{d^1} x ; q^{d^2} y) .  {\hskip 2mm}$
The fact that both $  {\hskip 2mm} x  {\hskip 2mm} $
and $  {\hskip 2mm} y  {\hskip 2mm}$ are cycles in the appropriate Floer
complexes means that the  cobordism $  {\hskip 2mm} {\cal M}^t  {\hskip 2mm} $
between them does not have other boundary components. (All the ``extra boundary
components'' of cobordisms
$  {\hskip 2mm} \{ \bigcup_{0 \leq t \leq 1} {\cal M}_{ H^{(t)}}
(q^{d^1} x_k ; q^{d^2} y_l)  \} {\hskip 2mm}  $  for different $k$ and $l$
will cancel each other after we glue  them together ).

\vskip .2cm

The theorem 2.1 of [McD S] which claims that the map
$ {\hskip 2mm}   ev_{(0;1)}  {\hskip 2mm} $ from

 $ {\cal M}_{ H^{(t)}}
(q^{d^1} x_k ; q^{d^2} y_l)   \times  {\cal J}_M  \times  {\cal G}_{H_-,H_+}
{\hskip 2mm} $ to
$ {\hskip 2mm}   M  {\hskip 2mm} $ is surjective, allows us to apply the Lemma
2.5. By applying this lemma to the evaluation map
$ev_{(0;1)}$ taken as ``projection operator'' we have that the cobordism
 ${\cal M}^t$ intersects transversally with
 $  {\hskip 2mm} ev_{(0;1)}^{-1}({\widehat C} )  {\hskip 2mm}  $
and the corresponding intersection gives us smooth one-dimensional submanifold
(with boundary).

\vskip .2cm

This  submanifold does not intersect the ``compactification divisor''
 ${\bar {\cal M}}^t  - {\cal M}^t$ since the later has codimension $\geq 2$ and
we have in our hands the freedom of putting everything ``in general position''.

\vskip .2cm

Thus, ${\cal M}^t \bigcap   ev_{0,1}^{-1}({\widehat C} )$ gives us the desired
compact one-dimensional cobordism between
${\cal M}_{H^{(0)}} (q^{d^1} x ; q^{d^2} y)
\bigcap   ev_{0,1}^{-1}({\widehat C} )$
and
${\cal M}_{H^{(1)}} (q^{d^1} x ; q^{d^2} y)
\bigcap   ev_{0,1}^{-1}({\widehat C} )$
The statement of the Theorem 5.4. follows.

\vskip .2cm

The same cobordism and transversality arguments  proves the following

\proclaim Lemma 5.5..
If ${\widehat C_1}$ and ${\widehat C_2}$ be two pseudo-manifolds in $M$
homologous to each other (which implies that they are actually cobordant to
each other in the class of pseudo-manifolds) and if $(\tau_0 , \theta_0)$ and
$(\tau_1 , \theta_1)$ be any two points on the cylinder $R \times S^1$ then

$${\cal M}_{H^{(0)}} (q^{d^1} x ; q^{d^2} y) \bigcap   ev_{\tau_0 ,
\theta_0}^{-1}({\widehat C} ) = {\cal M}_{H^{(1)}} (q^{d^1} x ; q^{d^2} y)
\bigcap
 ev_{\tau_1 , \theta_1}^{-1}({\widehat C} )  \eqno (5.12) $$

Now we are ready to prove the Theorem 5.1.
In order to prove it, we should (following Floer):

\vskip .2cm

A) ${\hskip 2mm}$ Construct a chain homotopy
$ {\hskip 2mm} h_H : CF^*(M,H_-) \rightarrow CF^*(M,H_+) {\hskip 2mm} $ (which
depends on the choice of the function $ {\hskip 2mm}H \in {\cal G}_{H_-,H_+} .$

\vskip .2cm

B)${\hskip 2mm}$  Prove that the chain homotopy $ {\hskip 2mm} h_H  {\hskip
2mm} $ gives  a well-defined homomorphism
$h_{H_-,H_+} : HF^*(M,H_-) \rightarrow HF^*(M,H_+)$ on the level of homology,
and this homomorphism  is independent on the choice of $ {\hskip 2mm} H .$

\vskip .2cm

C) ${\hskip 2mm}$ Prove that
 $ {\hskip 2mm} h_{H_1,H_3} = h_{H_2,H_3}  h_{H_1,H_2} {\hskip 2mm} $ for any
triple of ``generic'' Hamiltonians $ {\hskip 2mm} H_1, H_2, H_3 {\hskip 2mm}$
defined as functions from  $ {\hskip 2mm} S^1 \times M {\hskip 2mm}$
to ${\hskip 2mm}  R {\hskip 2mm} $

\vskip .2cm

D) ${\hskip 2mm}$   Prove that for any singular homology class
$ {\hskip 2mm}{\widehat C} {\hskip 2mm} $ in $ {\hskip 2mm} M$

$$h_{H_-,H_+} ( m_F^{H_-}( {\widehat C} )) {\hskip 2mm} =
{\hskip 2mm}    m_F^{H_+}( {\widehat C} )
  h_{H_-,H_+}      \eqno (5.13)   $$

Here  $m_F^{H_-}$ and  $m_F^{H_+}$ are operators of the action of
$H_*(M)$ on the Floer cohomology $HF^*(M,H_-)$ and  $HF^*(M,H_+)$
respectively.

\vskip .2cm

Let $\{ x_1,x_2,...\}$  and $\{ y_1,y_2,...\}$   be the bases
(over $Z_{H_2(M)}$)  of critical points of $S_{\omega,H_-}$   and
$S_{\omega,H_+}$ respectively.

\vskip .2cm

Then the matrix element
$ {\hskip 2mm}  <  q^{d^2} y  | h_H | q^{d^1} x  > {\hskip 2mm}$  of the
desired chain homotopy $ {\hskip 2mm} h_H {\hskip 2mm} $
is by definition the number of zero-dimensional components of

${\cal M}_H (q^{d^1} x ; q^{d^2} y) {\hskip 2mm} $ taken with appropriate
orientation.
This number is non-zero only if
$ {\hskip 2mm} deg (q^{d^1} x) = deg ( q^{d^2} y) . {\hskip 2mm}$
by our convention, the difference
${\hskip 2mm}  deg ( q^{d^2} y) - deg (q^{d^1} x) {\hskip 2mm} $
is given by the spectral flow.

\vskip .2cm

The Theorem 5.4 and Lemma 5.5 imply that above defined
${\hskip 2mm} h_H {\hskip 2mm} $ is really a chain homotopy such that the
statements   A)  and  B)  above hold

\vskip .2cm

The statement    D)  above  is equivalent to the fact that

$${\cal M}_H (q^{d^1} x ; q^{d^2} y) \bigcap   ev_{(2;1)}^{-1}({\widehat C} )
{\hskip 2mm} = {\hskip 2mm} {\cal M}_H (q^{d^1} x ; q^{d^2} y) \bigcap
 ev_{(-2 ;1)}^{-1}({\widehat C} )  \eqno (5.14) $$

The l.h.s. of $ {\hskip 2mm} (5.14) {\hskip 2mm} $ conncide with the l.h.s. of
$ {\hskip 2mm} (5.13) {\hskip 2mm} $ because of
$ {\hskip 2mm} H(2, \theta) = H_+(\theta) . {\hskip 2mm} $ The r.h.s. of
$(5.14)$ conncide with the r.h.s. of $(5.13)$ because of
${\hskip 2mm} H(- 2, \theta) = H_-(\theta) . {\hskip 2mm}$
Thus, we reduced the statement    D) to the particular case of the Lemma 5.5.

\vskip .2cm

The statement   C)  above  is a consequence of the procedure of ``gluing
trajectories'' [AuBr]. Namely, let us glue two half-cylinders
 $ {\hskip 2mm} S^1 \times (- \infty ; T] {\hskip 2mm} $ and
 $ {\hskip 2mm} S^1 \times [-  T ; + \infty ) {\hskip 2mm}  $ along their
boundaries. Sioce we have a  ``$\tau$-dependent Hamiltonian'' $H_{12}$  on the
first half-cylinder and  a  ``$\tau$-dependent Hamiltonian''
$H_{23}$ on the second half-cylinder such that

$$ H_{12}(\tau ; \theta) = H_1(\theta)   {\hskip 3mm}  if {\hskip 3mm}
\tau \leq -T-1   $$

$$ H_{12}(\tau ; \theta) = H_2(\theta)    {\hskip 3mm} if {\hskip 3mm}
\tau \geq -T+1  $$

$$ H_{23}(\tau ; \theta) = H_2(\theta)   {\hskip 3mm}  if {\hskip 3mm}
\tau \leq T-1   $$

$$ H_{23}(\tau ; \theta) = H_3(\theta)    {\hskip 3mm} if {\hskip 3mm}
\tau \geq T+1  $$

then we can glue them together to obtain a new $\tau$-dependent Hamiltonian
 $H_{13}^T$ which is defined as

$$H_{13}^T (\tau ; \theta) = H_{12}(\tau ; \theta) {\hskip 3mm}  if {\hskip
3mm}
\tau \leq 0 $$

$$H_{13}^T (\tau ; \theta) = H_{23}(\tau ; \theta) {\hskip 3mm}  if {\hskip
3mm}
\tau \geq 0 $$

If $x_1$ and $x_3$ are any two cycles in $CF_*(M,H_1)$ and in $CF_*(M,H_3)$
respectively  of relative Morse index zero then the lemma 5.5. implies that

$$ <  x_3  | h_{H_1,H_3} |  x_1  >  {\hskip 2mm}   =
 {\hskip 2mm}   \chi ( {\cal M}_{H_{13}^T} ( x_1 ; x_3 ) )     \eqno (4.15) $$

for {\bf any} value of the gluing parameter $T$. Here $ \chi$ means the Euler
characteristics of the zero-dimensional manifold.

\vskip .2cm

Now if we tend $  {\hskip 2mm} T  {\hskip 2mm}  $ to infinity then any
trajectory in $  {\hskip 2mm}  {\cal M}_{H_{13}^T}  {\hskip 2mm} $ will
``split'' into connected sum of a gradient flow trajectory of
$ {\hskip 2mm}  H_{12} {\hskip 2mm} $ and
a gradient flow trajectory of $ {\hskip 2mm}  H_{23} . {\hskip 2mm} $ These two
trajectories are glued together in some point
 $  {\hskip 2mm} x_2  \in \widehat{{\cal L}M}  {\hskip 2mm} $ which has to be a
critical point of $  {\hskip 2mm} S_{\omega,H_2}  {\hskip 2mm} $ due to the
$L^2$-boundedness condition.

\vskip .2cm

This observation implies that C) holds which proves the Theorem 4.1.

\vskip .2cm

Thus, we have a well-defined map

$$m_F : H^*(M)  \otimes   HF^*(M)  \rightarrow  HF^*(M)   $$

Since the Floer cohomology $ HF^*(M)$ are isomorphic to the classical
cohomology $ H^*(M)   \otimes  N $ then this ``Floer multiplication'' gives us
some bilinear operation

$$m_F : H^*(M)  \otimes   H^*(M)  \rightarrow  H^*(M)  \otimes  N $$

in classical cohomology.

\vskip .2cm

In order to calculate this  bilinear operation
 and prove that it coincides with the quantum cup-product, we should examine
more closely how the isomorphism between  $ HF^*(M)$ and $ H^*(M)   \otimes  N
$
is constructed.
We will do this in the next section.

$$ $$
 \centerline {\bf       {6. THE PROOF OF THE MAIN THEOREM}}
\centerline {  }

\vskip .2cm

	For each cohomology class $C \in H^*(M) $ two linear operators

$$m_Q (C ): H^*(M)  \rightarrow  H^*(M) \otimes  N
 \eqno         (6.1A) $$

and

$$m_F (C): H^*(M)  \rightarrow  H^*(M) \otimes  N
 \eqno         (6.1B) $$

 were defined in the previous three sections.
The map  $m_Q (C )$ was called {\bf quantum multiplication} (from the left) on
the  cohomology class $C $ . The  map $m_F (C )$  was called {\bf Floer
multiplication} (from the left) on the  cohomology class $C$.

\proclaim The Main Theorem 6.1.
Quantum  multiplication coincides with the  Floer multiplication.

	Let us fix  be some (homogenous) basis
$  {\hskip 2mm} \{A,B,...\}  {\hskip 2mm}$ in the total cohomology of
 $  {\hskip 2mm} M  {\hskip 2mm} $. To prove that the homomorphisms
$  {\hskip 2mm} (6.1A)  {\hskip 2mm} $ and
$  {\hskip 2mm} (6.1B)  {\hskip 2mm}$ are in
fact equal, it is sufficient to prove that all their ($N$-valued) matrix
elements

$$ < B | m_Q(C) |  A >  {\hskip 2mm}  =  {\hskip 2mm}
\sum_d  {\hskip 2mm} q^d  {\hskip 2mm}< q^d  B | m_Q(\mu) | A >
    \eqno  (6.2A)     $$

and

$$ < B | m_F(C) |  A >  {\hskip 2mm} =  {\hskip 2mm}
\sum_d   {\hskip 2mm} q^d  {\hskip 2mm} < q^d  B | m_F(\mu) | A >
   \eqno  (6.2B)     $$

are the same.

\vskip .2cm

Let $H$ be some smooth function on $S^1 \times R \times M$ such that

\vskip .2cm

A)  $  {\hskip 2mm}H$ vanishes in the region $|| \tau || > 1 $

\vskip .2cm

B) $  {\hskip 2mm} H {\hskip 2mm}  $ is not invariant under any holomorphic
automorphism of $ {\hskip 2mm}  S^1 \times R {\hskip 2mm} $
(which is identified with $ {\hskip 2mm} C^*$)

\vskip .2cm

Following the logic of the previous section, we can consider the space of
$L^2$-bounded trajectories
$ {\hskip 2mm}  {\cal M}_H ( \widehat{A} ; q^d \eta (\widehat{B})){\hskip 2mm}$
and  compactify it as a stratified space. We assume taht ``the statement 5.2''
holds in this case also. The proof of this generalization of
the statement 5.2 repeats the proof of the original statement

\proclaim Theorem 6.2.
$$  {\bar {\cal M}}_H ( \widehat{A} ; q^d \eta (\widehat{B})) {\hskip 2mm} =
 {\hskip 2mm}
{\bar {\cal M}}_{J_{grad H d {\bar z} } , d} {\hskip 2mm} \bigcap  ev_0^{-1}
(\widehat{A})   \bigcap   ev_{\infty}^{-1} (\widehat{\eta(B)})  \eqno (6.3)$$

Let
 $  {\hskip 2mm}  \gamma {\hskip 2mm} = {\hskip 2mm} \gamma (\tau, \theta)
{\hskip 2mm} $ be any
$  {\hskip 2mm} L^2$-bounded  solution of   {\hskip 2mm} $(5.5)  {\hskip 2mm}$
with $  {\hskip 2mm} H = 0  {\hskip 2mm} $ in the region
 $  {\hskip 2mm} || \tau || > 1  . {\hskip 2mm} $
Then  $  {\hskip 2mm} \gamma  {\hskip 2mm} $ (considered as a  map from
the cylinder $  {\hskip 2mm} S^1 \times R  {\hskip 2mm} $ to
$  {\hskip 2mm} M  ){\hskip 2mm} $ can be continiously extended from
the cylinder  $  {\hskip 2mm} S^1 \times R  {\hskip 2mm} $ to the 2-sphere
 $ {\hskip 2mm} S^2 {\hskip 2mm} $ since the limit value of
 $  {\hskip 2mm} \gamma  {\hskip 2mm} $
 at $ {\hskip 2mm} \tau \rightarrow \pm \infty {\hskip 2mm} $ should be
constant loops.

\vskip .2cm

 Ellipticity of the gradient flow equation $(5.5)$ with the
prescribed boundary conditions at $\tau \rightarrow \pm \infty$ implies that
this extension is actually smooth. Now the statement of the  Theorem 6.2.
follows directly from the definitions of the l.h.s and the r.h.s. of $(6.3)$.

\vskip .2cm

 The fact that ${\hskip 2mm} (6.3) {\hskip 2mm} $ is an isomorphism at the
level of {\bf compactifications} (as stratified spaces) can be observed by
comparing the explicit description of these compactifications that we have.

\vskip .2cm

The previous theorem means that the matrix element of quantum multiplication
can be written as

$$ < B | m_Q(C) |  A > {\hskip 3mm} = {\hskip 3mm}\sum_d  q^d {\hskip 2mm}
 {\cal M}_H ( \widehat{A} ; q^d \eta (\widehat{B}))
\bigcap   ev_{0,1}^{-1}({\widehat C} )
 \eqno (6.4)$$

where the number in the r.h.s., as usual, is understood as intersection index.

\vskip .2cm

\proclaim Remark. The r.h.s. of $ (6.4)$ can be thought as a of Floer
multiplication operation defined for unperturbed symplectic action.

\vskip .2cm

This remark implies that in order to prove the Main Theorem, it is enough to
generalize the program implemented in the previous section as  follows:

\vskip .2cm

A) Construct the chain homotopies
$ {\hskip 2mm} \{  h_{H_i, 0} : CF^*(M,H_i) \rightarrow CF^*(M, 0) \} {\hskip
2mm} (i=1;2) {\hskip 2mm} $ from the Floer complexes
$ {\hskip 2mm} \{ CF^*(M,H_i) \} {\hskip 2mm} $ of perturbed symplectic action
$S_{\omega,H_i}$ to the Floer complex
$ {\hskip 2mm} CF^*(M, 0) {\hskip 2mm} $ of unperturbed symplectic action $
{\hskip 2mm} S_{\omega} {\hskip 2mm}$
and the chain  homotopies
$ {\hskip 2mm} \{ h_{0 , H_i } \} {\hskip 2mm} :  CF^*(M, 0) \rightarrow
CF^*(M,H_i){\hskip 2mm} $ going in the opposite direction

\vskip .2cm

B)${\hskip 2mm}$  Prove that these chain homotopies gives us  well-defined
homomorphisms
on the level of cohomology
 which is independent on the choice of $ {\hskip 2mm} \tau$-dependent
hamiltonians by means of which they are constructed

\vskip .2cm

C) ${\hskip 2mm}$  Prove that for any pair of ``generic'' Hamiltonians
$ {\hskip 2mm} H_1{\hskip 2mm}$ and ${\hskip 2mm}  H_2 {\hskip 2mm}$
considered as functions on  $ {\hskip 2mm} S^1 \times M {\hskip 2mm}$ we have
functoriality property
$ {\hskip 2mm} h_{H_1,H_2} = h_{0,H_2}  h_{H_1, 0} ' {\hskip 2mm}$
	and  prove that  $ {\hskip 2mm} h_{H_1, 0}   h_{0,H_1  }$ gives us identity
map in the Floer cohomology group (defined by unperturbed symplectic action).

\vskip .2cm

D) ${\hskip 2mm}$  Prove that for any singular homology class
 $ {\hskip 2mm} {\widehat C} {\hskip 2mm} $ in
 $ {\hskip 2mm} M {\hskip 2mm}$ we have

$$h_{H_1, 0 } ( m_F^{H_1}( {\widehat C} )) {\hskip 2mm} =
{\hskip 2mm}     m_F^{0}( {\widehat C} )            h_{H_1, 0 }   \eqno (6.5A)
   $$

and

$$h_{0 , H_1} ( m_F^{0}( {\widehat C} )) {\hskip 2mm} =
{\hskip 2mm}    m_F^{H_1}( {\widehat C} )            h_{0 , H_1 }   \eqno
(6.5B)      $$

\vskip .2cm

The proof of the Statements A - D above goes exactly the same way as the proof
of analogous statements in the section five.

\vskip .2cm

Thus, $h_{0, H_1}$ gives us an isomorphism between classical and  Floer
cohomology which maps quantum multiplication  $m_Q(C)$ to the Floer
multiplication
$ m_F(C)$ for any cohomology class $C \in H^*(M)$.

\vskip .2cm

This proves our Main Theorem.

$$ $$

 \centerline {\bf       {7. FLOER COHOMOLOGY OF COMPLEX GRASSMANIANS}}
\centerline {  }
	As an example of applications of our Main theorem, let us give a rigorous
proof of the formula for Floer cohomology ring of the complex Grassmanian
$G(k,N)$ ov $k$-planes in complex $N$-dimensional vector space $V$. The formula
for the quantum cohomology ring $HQ^* (G(k,N))$ was assumptiond long ago by
Vafa [Va1]. More detailed analysis of quantum cohomology of Grassmanians was
worked out by
Intrilligator [I] and recently by Witten [Wi5] in relation with the Verlinde
algebra. Witten also mentioned that  Floer cohomology ring of the  Grassmanian
should be given by the same formula.

\vskip .2cm

	Now we need to discuss the cohomology of $G(k,N)$.  We begin with the
classical cohomology.  Over $G(k,N)$ there is a ``tautological''
$k$-plane bundle $E$ (whose fiber over $x\in G(k,N)$ is the $k$
plane in $V$ labeled by $x$) and a complementary bundle $F$
(of rank $N-k$):
$$0 \rightarrow E \rightarrow  V^* =C^N \rightarrow  F \rightarrow  0 $$
Obvious cohomology classes of $G(k,N)$ come from Chern classes.
We set

$$ x_i=c_i(E^*) $$

where $*$ denotes the dual.  (It is conventional to use $E^*$ rather
than $E$, because $\det E^*$ is ample.)
This is practically where Chern classes come from, as $G(k,N)$ for
$N\to \infty$ is the classifying space of the group $U(k)$.
It is known that the $x_i$ generate $H^*(G(k,N))$ with certain relations.
The relations come naturally from the existence of the complementary
bundle $F$ in   Let $y_j=c_j(F^*)$, and let
$c_t(\cdot)=1+tc_1(\cdot)+t^2c_2(\cdot)+\dots$.  Then
 $H^*(G(k,N))$ is generated by the $\{ x_i,y_j \}$ with relations

$$ c_t(E^*)c_t(F^*)=1     \eqno     (7.1)$$

	Since the left hand side of $(7.1)$ is  {\it a priori}
a polynomial in $t$ of degree $N$) the classical relations are of degree
$2,4,\dots, 2N$. The first $N-k$ of these relations (uniquely)  express the $\{
y_j \} $ in terms
of the $\{ x_i \} $ . This means that the classical cohomology ring of
$H^*(G(k,N))$ is generated by the $k$ generators $\{ x_i \}$ with $k$ relations
 of degree
$2N-2k+2,2N-2k+4,\dots, 2N$.

\vskip .2cm

	Let us now  work out the quantum cohomology ring $HQ^*(G(k,N))$ of the
Grassmannian. We can consider a subring in $HQ^*(G(k,N))$ generated by  $\{
x_i,y_j \}$

\proclaim Conjecture (Vafa).  quantum cohomology ring $HQ^*(G(k,N))$ of the
Grassmannian is generated  by  $\{ x_i,y_j \}$ with ``deformed relations''

$$c_t(E^*)c_t(F^*)=1+ q (-1)^{N-k}t^N  \eqno     (7.2)$$

where $q$ is (the unique) Kahler class in $H^2 (G(k,N),Z)$.

\vskip .2cm

	To prove this Vafa's conjecture it is sufficient to prove that

\vskip .2cm

A) $\{ x_i \}$ generate the whole quantum cohomology ring

\vskip .2cm

B) $\{y_j \}$ are expressed in terms
of the $\{ x_i \} $ by the same formulas as in the classical cohomology ring

\vskip .2cm

C) The relations  on  $\{ x_i \}$ in our  quantum cohomology ring form an ideal

\vskip .2cm

D) This ideal of relations is generated by  $k$ relations
 of degree

$2N-2k+2,2N-2k+4,\dots, 2N$ coming from expansion of the l.h.s. of $(7.2)$ in
powers of $t$ and taking coefficients of degree $2N-2k+2,2N-2k+4,\dots, 2N$
{\bf without any extra relations}

\vskip .2cm

	The fact that $\{y_j \}$ are expressed in terms
of the $\{ x_i \} $ by the same formulas as in the classical cohomology ring
and the fact that these $k$ Vafa's relations indeed take place was proved
(rigorously) by Witten [Wi5] by examining the fact that

\vskip .2cm

a) The classical relations of degree $2,4,\dots, 2N-2$ cannot deform since
$deg [q] = 2N$  , and

\vskip .2cm

b) There is a ``quantum correction'' to the the ``top'' relation

$c_k(E^*)c_{N-k}(F^*)=0$ of degree $2N$ in the classical cohomology. This
``deformed relation'' has the form
$ c_k(E^*)c_{N-k}(F^*)= a$
for some number $a$ which can be  computed by examining degree-one rational
curves in the
Grassmannian. The value of this unknown number $a$ was (rigorously)
computed by Witten and was shown to be equal to  $(-1)^{N-k}$.

\vskip .2cm

	The statement C) that the relations  on  $\{ x_i \}$ in our  quantum
cohomology ring form an ideal will follow from the associativity of the
quantum cohomology ring (which was proved rigorously after [Wi5] was
finished).

\vskip .2cm

	Thus, the only things we need to prove after Witten are

\vskip .2cm

A) $\{ x_i \}$ generate the whole quantum cohomology ring,

 and

\vskip .2cm

D) that there are no extra relations (in degree hihger than $2N$ ) on these
generators.

\vskip .2cm

	The statement  A) can be proved inductively by the degree {\bf deg}.
Let us assume that all the elements in $HQ^*(G(k,N))$ of degree less than $m$
can be expressed as polynomials in $\{ x_i \}$. Let us prove that this also
holds for all the elements in $HQ^*(G(k,N))$ of derree $m$.

\vskip .2cm

	 Let $A \in  H^m(G(k,N),Z) \subset   HQ^*(G(k,N))$ be some homogenous element
of derree $m$. Then we know that in the classical cohomology ring we
have
 $$A = P_m(x_1,...,x_k)$$

 for some polynomial $P_m$ of degree $m$. The fact that $deg [q] = 2N$ is
positive means that in the quantum cohomology ring we
have

 $$A = P_m(x_1,...,x_k) + \sum_d q^d A_d$$

for some (unknown) cohomology classes $ A_d \in  H^{m-2Nd}(G(k,N),Z) $  of
degree $m-Nd$.

	But by our induction hypothesis we know that all $\{ A_d \}$ can be expressed
as some polynomials in  $\{ x_i \}$. This simple observation proves the
statement  A).

\centerline {  }

	To prove the last remaining statement  D) let us note that the rank (over the
ring $Z_{<q>}$) of the quantum cohomology of the Grassmanian $HQ^*(G(k,N))$
should be equal to the rank (over $Z$) of  the classical cohomology
$H^*(G(k,N))$.

\vskip .2cm

	If there were some extra relations among the generators $\{ x_i \}$
this would mean that the rank (over the ring $Z_{<q>}$) of the free polynomial
ring in  $\{ x_i \}$
moded out by the ideal generated by the coefficients
of the l.h.s. of $(6.2)$ would be strictly greater than  the rank of
$H^*(G(k,N))$.

\vskip .2cm

	But we know that any two $Z$-graded
rings generated by  $k$ homogenous generators $\{ x_1,...,x_k \}$ of degrees
$\{ 2,4,...,2k \}$ with $k$ homogenous relations
 of degrees
$2N-2k+2,2N-2k+4,\dots, 2N$ should have the same rank.

\vskip .2cm

	This proves the statement D) and the Vafa's conjecture.

\centerline {  }

	The arguments presented here together with the results of Ruan and Tian  [RT]
who proved ``the handle-gluing formula'' of Witten [Wi1] give a complete proof
to a more refined formula of Intrilligator [I] for the certain intersection
numbers (known as Gromov-Witten invariants) on the moduli space of holomorphic
maps of {\bf higher genus curves} to the Grassmanian. The proof of this formula
was previously known only for the special case $(G(2,N))$ of the Grassmanians
of 2-planes and is due to Bertram,Daskaloupulos and Wentworth [BDW],[Be]. Our
arguments prove this formula in the full generality.

$$ $$

 \centerline {\bf       {8.  QUANTUM  COHOMOLOGY  REVISITED     }}
\centerline {  }

	The significant drawback of the Definition B of quantum cup-product
(which is the only completely justified definition available at the moment) is
that
the Definition B invokes the moduli spaces of $J_g$-holomorphic
spheres instead of the moduli spaces of $J_0$-holomorphic
spheres when $J_0$ is an actual complex structure (which is much more
interesting object from an algebro-geometric point of view).

\vskip .2cm

It is hard to prove that some particular choice of $g$ is ``regular'' and to
make any calculations using this definition.

\vskip .2cm

	In order to give a definition of quantum cup-product  which uses only the
moduli spaces  $\{  {\cal M}_{J_0,d} \}$  of holomorphic
curves we need to introduce some more notations.

\vskip .2cm

	Let us suppose that:

\proclaim Statement 8.1.
 For any ``generalized degree'' $d$ the
moduli space ${\hskip 2mm} {\cal M}_d  = {\cal M}_{J_0,d } {\hskip 2mm} $
will be a finite union of smooth strata (of possibly different dimensions) such
that each stratum is a smooth almost-complex manifold

\proclaim Statement 8.2.
 Each manifold $ {\cal M}_d  $  can be compactified (by adding ``degenerate
$J$-holomorphic curves'') as  a stratified space $\bar{ {\cal M}}_d $ such that
each stratum is a smooth almost-complex manifold

\proclaim Statement 8.3.
``The compactification divisor'' $\bar{ {\cal M}}_d  -  {\cal M}_d $ and ``the
singular strata'' have codimension at least two (or ``complex codimension
one'') in  each irreducible component of the compactified moduli space
 $\bar{ {\cal M}}_d$.

\proclaim Theorem 8.4 (Gromov [Gr2]).
 The statements $(8.1) - (8.3)$ always hold if $M$ is Kahler manifold with its
actual complex structure.

If the almost-complex structure $J$ on $M$ is non-integrable, we will state
$(8.1) - (8.3)$ as assumptions.

\proclaim Note.
 If $M$ is algebraic the Theorem 8.4 follows from the fact that in the case
$\bar{ {\cal M}}_{J,d}$ can be defined in algebraic  terms as a Hilbert scheme.

\vskip .2cm

	Let us note that the formal tangent space to the moduli space           $
{\cal M}_{J,d}$  at the point $\varphi \in {\cal M}_{J,d}$ is equal to
the kernel of the linearised ${\bar \partial}$-operator, acting from the space
of $W_{5n}^2$-sections of (holomorphic) vector bundle  ${\varphi}^*(TM)$ over
$CP^1$ to the space of $W_{5n-1}^2$-one-forms with the coefficients in this
vector bundle.

\vskip .2cm

 Equivalently, the formal tangent space to $ {\cal M}_{J,d}$ at the point
$\varphi$ is isomorphic to  $H^0 [{\varphi}^*(TM)]$.

\vskip .2cm

The cokernel of the same  ${\bar \partial}$-operator  is isomorphic to
$H^1 [{\varphi}^*(TM)]$.

\vskip .2cm

In general, the formal tangent space $H^0 [{\varphi}^*(TM)]$ to the moduli
space of

$J$-holomorphic spheres is not necessarily equal to the actual tangent space to
this moduli space at the point $\varphi$.

\vskip .2cm

There may be ``an obstruction'' to integration of the formal tangent vector to
the local deformation of  the moduli space in the direction of this formal
tangent vector. This obstruction is a non-linear map from $H^0
[{\varphi}^*(TM)]$ to $H^1 [{\varphi}^*(TM)]$ with vanishing first derivative.

\vskip .2cm

The existence of a non-trivial obstruction corresponds to the singularity
of our moduli space  $ {\cal M}_{J,d}$  at the point $\varphi$

\vskip .2cm

According to the theorem 8.4  (or according to the assumption 8.3 if $J$ is
 non-integrable) the space of   $\varphi \in {\cal M}_d$  with obstructed
deformations  has codimension at least two. This means that on the complement
to the lower-dimensional singular startum there is no obstruction and the
moduli space
 $ {\cal M}_d$ is a smooth manifold of dimension
$2 dim (H^0 [{\varphi}^*(TM)])$.

\vskip .2cm

On the  complement to this  singular startum the cokernel of  ${\bar
\partial}$-operator has constant dimension   dim$H^1 [{\varphi}^*(TM)]$

\vskip .2cm

By applying Riemann-Roch theorem to the vector bundle
${\varphi}^*(TM)$ over $CP^1$
we see that

$$2 dim (H^0 [{\varphi}^*(TM)]) - 2 dim (H^1 [{\varphi}^*(TM)]) {\hskip 2mm} =
{\hskip 2mm} dim M {\hskip 2mm} + {\hskip 2mm} \sum_{i=1}^{s} {\hskip 2mm} d_i
{\hskip 2mm} deg [q_i]   \eqno (8.1)   $$

which is equal to the r.h.s of $(2.7)$

\vskip .2cm

	So, the r.h.s. of $(8.1)$ reproduces us ``the virtual dimension''
$vdim [{\cal M}_d]$ of
the moduli space  $ {\cal M}_d$.  This ``virtual dimension'' $vdim$ is equal to
the actual dimension of this moduli space if and only if the first cohomology $
H^1 [{\varphi}^*(TM)]$
is zero-dimensional (or, equivalently, if the   ${\bar \partial}$-operator is
surjective).

\centerline {  }

Unfortunately, this situation almost never takes place if we do not consider
$g$-perturbed ${\bar \partial}$-operator. Usually the different irreducible
components of
the moduli space ${\cal M}_d$ have different dimensions. Unlike the case of
 moduli space of $J_g$-holomorphic spheres, these components may intersect each
other. This why we call them ``irreducible components'' instead of ``connected
components''

\vskip .2cm

 This difference between the actual and the virtual dimension of any
irreducible component  ${\cal M}_c$
of  the moduli space ${\cal M}_d$   is always non-negative and is given by the
number $2b_c = 2 dim [ H^1 [{\varphi}^*(TM)]]$  computed at the complement to
the singular locus of ${\cal M}_c$.  Physicists would call
the number $b_c $
``the number of the fermion zero-modes''.

\vskip .2cm

The dimension $  dim [ H^1 [{\varphi}^*(TM)]]  $ usually is not  constant as
$\varphi$ varies over ${\cal M}_c$. The ``jumping divisor''
 coincides with the singular locus $sing_c$ of  ${\cal M}_c$.

\vskip .2cm

	If `the number of the fermion zero-modes $b_c$ is positive, then let us
introduce
$b_c$-dimensional complex vector bundle $ {\cal F}_c$
 over ${\cal M}_c-sing_c$  which assigns to each holomorphic map $\varphi \in
{\cal M}_c$ the vector space $ H^1 [{\varphi}^*(TM)]$. This vector space varies
 holomorphically as $\varphi$ varies. Physicists would call this bundle ``the
bundle of the fermion zero-modes''.

\vskip .2cm

	Let $\chi ({\cal F}_c)$ be the Euler class of the  bundle $ {\cal F}_c$
(formally) considered as a cohomology class in $H^{b_c}({\cal M}_c)$
The precise meaning of this Euler class  will be specified later
 in this section.

\vskip .2cm

 In the case when $M$ is a projective algebraic manifold, we can think about
${\bar {\cal M}}_{J,d}$ as a Hilbert scheme and about ${\bar {\cal M}}_c$ as
its irreducible component. Then we have a coherent algebraic sheaf
${\cal F} = H^1 [{\varphi}^*(TM)]$ over ${\bar {\cal M}}_{J,d}$. This sheaf is
not locally free. Its restriction on any irreducible component
${\bar {\cal M}}_c$ (compactified algebraically) of our Hilbert scheme is also
 not locally free.

\vskip .2cm

But if we restrict our sheaf ${\cal F}$ on
${\cal M}_c-sing_c$ we will obtain a locally free sheaf of rank $b_c$. This
sheaf over
${\cal M}_c-sing_c$ will be the sheaf of sections of the bundle
${\cal F}_c$ of fermion zero-modes (considered as an algebraic vector bundle).

\vskip .2cm

Thus, in algebraic situation we can think about $\chi ({\cal F}_c) $ as  an
algebraic Euler class of the sheaf on ${\bar {\cal M}}_c$
 which is a well-defined mathematical object.

\vskip .2cm

Now let us give two more definitions of the quantum tri-linear pairings

$<A;B;C>_q$ using only the moduli spaces  $\{  {\cal M}_d \} $

\proclaim Definition C (Witten).

Let $A,B,C  \in H^* (M,Z) \otimes Z_{<C>}$  Then

$$ <A ; B ; C>_q  =  \sum_d q^d  \sum_c
\int_{   {\cal M}_c   }  ev_0^* (A) \bigwedge      ev_{\infty}^* (B)  \bigwedge
    ev_1^* (C)  \bigwedge    \chi ({\cal F}_c)    \eqno   (8.2)$$

Here the second sum is over the irreducible components  $\{ {\cal M}_c \}$
of the moduli space $  {\cal M}_d  $

\vskip .2cm

Strictly speaking, the r.h.s. of $(8.2)$ does not make sence by the same reason
as the r.h.s. of $(3.2)$ does not make sence (we should integrate over
non-compact moduli spaces and extend $\chi ({\cal F}_c) $ to the singular part
of  $ {\cal M}_c $ )

\vskip .2cm

	It is again possible, following [SeSi] and [Wi4], to choose differential form
representative of $ \chi ({\cal F}_c) $ computed through the curvature and
try to extend it over the compactofication divisor and over the singularities.

\vskip .2cm

But in order to show that  the integral  $(8.2)$ over the  compactified moduli
space is well-defined, we should prove  its independence of the choice of
 differential form representatives of cohomology classes $A , B$ and $C$
and $\chi ({\cal F}_c) $

\vskip .2cm

Since the structure of the compactification (and singularities) of
 $  {\cal M}_d  $ is enormously complicated and not well-understood, the proof
of ``independence of the choices'' is not yet available

\vskip .2cm

To get rid of this problem, we will give another definition in the spirit of
Vafa, using the Poincare dual language of intersection of cycles.

\vskip .2cm

But even in this approach, we need to choose some model ${\widehat {\cal M}}_c$
 for
desingularization  of ${\bar {\cal M}}_c$.
 Our strategy will be to give a formula using some choice of desingularization
and then prove that the answer is actually independent of this choice. More
precisely, we need the following:

\vskip .2cm

\proclaim Theorem 8.5. The following five statements hold:

1) There exists a (non-unique) desingularization ${\widehat {\cal M}}_c$
 of  ${\bar {\cal M}}_c$  which coincides with ${\cal M}_c$ on the complement
to
the small tubular neighborhood $B^{\epsilon}_c$ of the singular locus.
 This
 desingularization is
constructed analytically through Kuranishi obstruction theory.

\vskip .2cm

2) The  vector bundle  $ {\cal F}_c$ extends canonically to any of these
desingularizations as a smooth $b_c$-dimensional  vector bundle

\vskip .2cm

3) Different irreducible components
 $\{ {\widehat {\cal M}}_c \}$ of desingularized moduli space
 $ {\widehat {\cal M}}_d$  do not intersect each other

\vskip .2cm

4) Any two  ``desingularizations  constructed a la Kuranishi'' are cobordant to
each other

\vskip .2cm

5) The bundle of the fermion zero-modes
$ {\cal F}_c$ can be extended as the vector bundle over the cobordism between
two different desingularizations of  ${\bar {\cal M}}_c$.

\vskip .2cm

\proclaim Proof.

Following analogous constructions by Taubes [Ta1] , [Ta2] for the moduli space
of anti-self-dual connections on a four manifold, we can argue sa follows:

\vskip .2cm

 For any  $\varphi  \in Map_d $ we have two linear
operators: ${\bar \partial}$-operator acting from $T_{\varphi} (Map_d )$ to
$ {\cal H}_{\varphi}$  and the adjoint ${\bar \partial}^*$- operator acting
from
$ {\cal H}_{\varphi}$ to  $T_{\varphi} (Map_d )$ (since we choose our elliptic
differential operators
to act in Hilbert spaces).

\vskip .2cm

 The non-zero spectrum of the ``Laplace operators''

${\bar \partial}^* {\bar \partial} :  T_{\varphi} (Map_d )  \rightarrow
 T_{\varphi} (Map_d )   $  and
${\bar \partial} {\bar \partial}^*  :   {\cal H}_{\varphi}   \rightarrow
 {\cal H}_{\varphi}   $                is the same.

\vskip .2cm

The kernel of ${\bar \partial}^* {\bar \partial}$ is isomorphic to the kernel
of ${\bar \partial}$ and the   kernel of ${\bar \partial} {\bar \partial}^*$ is
isomorphic to the cokernel of ${\bar \partial}$

\vskip .2cm

For any positive number $\lambda$ let  $  {\cal M}^{\lambda}_d  $ be the
topological subspace in $Map_d $ consisting of those  $\varphi  \in Map_d $
for which the operator ${\bar \partial}^* {\bar \partial}$ has eigenvalues less
than $\lambda$. In particular, $  {\cal M}_d \subset {\cal M}^{\lambda}_d  $

\vskip .2cm

Since the spectrum of any of  ``Laplace operators''
(papemetrized by the points  in $Map_d $ ) is discrete and changes smoothly as
$  {\hskip 2mm} \varphi  {\hskip 2mm} $ varies then for any particular choice
of $\varphi  \in Map_d $ such that
$\lambda$ is not in the spectrum the following statements hold:

\vskip .2cm

1)  The intersection of
$  {\cal M}^{\lambda}_d$ with the small ball  $B^{\epsilon}(\varphi)$  of
radius ${\epsilon}$ with the center $ {\hskip 2mm} \varphi  {\hskip 2mm}$
  is a smooth
finite-dimensional submanifold in  $ {\hskip 2mm} Map_d  {\hskip 2mm} $ and

\vskip .2cm

2) $  {\cal M}^{\lambda_1}_d \bigcap B^{\epsilon}(\varphi)$ is a smooth
submanifold of $  {\cal M}^{\lambda_2}_d \bigcap B^{\epsilon}(\varphi)$
if  ${\lambda_1} < {\lambda_2}$ and both ${\lambda_1}$ and ${\lambda_2}$ are
not in the spectrum

\vskip .2cm

3) If restricted to the smooth part of $  {\cal M}^{\lambda}_d$, the
infinite-dimensional Hilbert bundle ${\cal H}$ splits into direct orthogonal
sum
${\cal H}_{ <  \lambda} \oplus      {\cal H}_{ > \lambda}$.

\vskip .2cm

Here
${\cal H}_{ <  \lambda}$ is a finite-dimensional subbundle in  ${\cal H}$
spanned by the eigenvectors of the Laplacian ${\bar \partial} {\bar
\partial}^*$
with the eigenvalues less than
${\hskip 2mm} \lambda$.

\vskip .2cm

  ${\cal H}_{ >  \lambda}$ is an infinite-dimensional subbundle  in  ${\cal H}$
spanned by the eigenvectors of
 ${\hskip 2mm} {\bar \partial} {\bar \partial}^* {\hskip 2mm}$
with the eigenvalues greater than
${\hskip 2mm} \lambda$

\vskip .2cm

Moreover, Kuranishi-type techniques [Ku],[Ta1] gives us that:

\vskip .2cm

4) If $\varphi$ lies in ${\cal M}_d$ then there exists a preferred section
$\Psi_{\lambda}$ of the bundle
${\cal H}_{ <  \lambda}$ over
$  {\cal M}^{\lambda}_d \bigcap B^{\epsilon}(\varphi)$ such that

$$  {\cal M}_d \bigcap B^{\epsilon}(\varphi) = \Psi^{-1}_{\lambda}(0) \bigcap
 {\cal M}^{\lambda}_d \bigcap B^{\epsilon}(\varphi)  \eqno (8.3) $$

This  preferred section $\Psi_{\lambda}$ is called ``Kuranishi map'' [Ku],[Ta1]

\vskip .2cm

5)  If the point  $\varphi \in {\cal M}_c$ does not lie in the singular locus
$sing_c$ then the derivative of the  Kuranishi map $\Psi_{\lambda}$ has the
constant
rank. The corank of this derivative is equal to the number of fermion
zero-modes ${\hskip 2mm} b_c$.

If  the point  ${\hskip 2mm} \varphi \in {\cal M}_c {\hskip 2mm}$ lies in the
singular locus than the corank
 of the derivative of  ${\hskip 2mm} \Psi_{\lambda} {\hskip 2mm}$ jumps near
the singular locus ${\hskip 2mm} sing_c {\hskip 2mm}$
and is equal to ${\hskip 2mm} b_c {\hskip 2mm} $ outside  the singular locus.

\vskip .2cm

6)  If ${\lambda_1} < {\lambda_2}$ then the corresponding ``Kuranishi maps''
are related as

 $$\Psi_{\lambda_2}  = \Psi_{\lambda_1}
\oplus \Pi^{\lambda_2}_{\lambda_1}  {\bar \partial}   \eqno (8.4) $$

where $ \Pi^{\lambda_2}_{\lambda_1}$ is the orthogonal projector on the
subbundle in ${\cal H}$ generated by eigenvectors of Laplacian
 ${\hskip 2mm} {\bar \partial} {\bar \partial}^* {\hskip 2mm}$
 with eigenvalues between
${\lambda_1}$ and ${\lambda_2}$

\vskip .2cm

Since the Gromov compactification of the moduli space ${\cal M}_d$ consists of
degenerate curves and we know from the work of Seeley and Singer [Se Si] that
${\bar \partial}$-operator can be continiously extended to such degenerate
curves, then the above described Kuranishi machinery is also applicable to the
case when
${\hskip 2mm} \varphi$ lies in the compactification of ${\bar {\cal M}}_c$

\vskip .2cm

We can cover the neighborhood of the compactification divisor by a system of
charts $\{  {\bar {\cal M}}^{\lambda}_d \bigcap B^{\epsilon}(\varphi)  \}$.
 We also have a  Kuranishi map on each chart such that the piece of
$ {\bar {\cal M}}_c$  covered by this  chart coincides with the zero locus of
the corresponding Kuranishi map

\vskip .2cm

Now we can construct the desired desingularization.

\vskip .2cm

For any point
$\varphi \in sing_c$  we can perturb slightly the Kuranishi map
$\Psi_{\lambda}$
on
$  {\cal M}^{\lambda}_d \bigcap B^{\epsilon}(\varphi)$  without changing it
outside the ball ${\hskip 2mm} B^{\epsilon / 2}(\varphi) {\hskip 2mm}$.  We
make the perturbation such that the new
 map
will have constant rank
$dim ({\cal H}_{ <  \lambda} ) - b_c$. Let us denote this ``perturbed
Kuranishi map'' $\Psi^,_{\lambda}$

\vskip .2cm

 Moreover, we can do these perturbations on all charts
$\{  {\cal M}^{\lambda}_d \bigcap B^{\epsilon}(\varphi)  \}$  covering the
singular locus   $sing_c$  simultaneously. We can make
 these perturbations on different charts consistent with each other in the
sense of $(8.4)$.

\vskip .2cm

Thus, we will have a new smooth compact manifold
 ${\widehat {\cal M}}_c$  which we will call {\bf a desingularization} of
${\bar {\cal M}}_c$

\vskip .2cm

The extension of the vector bundle $ {\cal F}_c$ on ${\widehat {\cal M}}_c$
is defined as follows:

\vskip .2cm

let us define the restriction of $ {\cal F}_c$
on the chart
$(\Psi^,_{\lambda})^{-1}  (0) \bigcap
 {\cal M}^{\lambda}_d \bigcap B^{\epsilon}(\varphi) {\hskip 2mm}$
as cokernel of derivative of the perturbed Kuranishi map
 ${\hskip 2mm} \Psi^,_{\lambda}$

\vskip .2cm

This definition is consistent with the definition of the bundle $ {\cal F}_c$
over the region   ${\hskip 2mm} {\widehat {\cal M}}_c  {\bigcap  {\cal M}}_c
{\hskip 2mm} $ where the Kuranishi map is not perturbed.

\vskip .2cm

Moreover, the bundle  $ {\cal F}_c$ on ${\widehat {\cal M}}_c$ is by its
construction
a factor-bundle of the infinite-dimensional Hilbert bundle  ${\cal H}$
restricted to  ${\widehat {\cal M}}_c$. Since  ${\cal H}$ is a  Hilbert bundle
then we can also think of $ {\cal F}_c$ as a subbundle of
  ${\cal H}_{|{\widehat {\cal M}}_c}$

\vskip .2cm

Thus, we have already proved parts one and two of the Theorem 8.5.
Part three follows from the transversality arguments. What remains
to be proved is parts  four and five.

\vskip .2cm

Let  ${\widehat {\cal M}}^,_c$ and  ${\widehat {\cal M}}^{,,}_c$ be two
different desingularizations corresponding to two different perturbations
$\Psi^,_{\lambda}$ and $\Psi^{,,}_{\lambda}$ of the Kuranishi map.

\vskip .2cm

Although we denote  the (perturbed) Kuranishi map by one symbol
$\Psi^,_{\lambda}$, we understand that actually we have a system of
perturbed  Kuranishi maps on different charts
 which match on their intersection.

\vskip .2cm

Then by a version of Sard Lemma we can construct a function
${\hskip 2mm}  \Psi_{\lambda}(t){\hskip 2mm}  $ on
$  {\hskip 2mm}$ [ the union of charts] $\times {\hskip 2mm}  [0;1]$
which interpolates between
$\Psi^,_{\lambda}$ at $t=0$ and $\Psi^{,,}_{\lambda}$ at $t=1$.
Here $t$ is the parameter on  ${\hskip 2mm}  [0;1]$

\vskip .2cm

We can choose the interpolating function  $\Psi_{\lambda}(t)$ to  coincides
with the unperturbed Kuranishi map outside the tubular neighborhood
$ {\hskip 2mm} B^{\epsilon / 2}(sing_c)  {\hskip 2mm} $ of the singular locus
and to be
 independent of
 $ {\hskip 2mm} t  {\hskip 2mm}$ in that region.
We can also choose
  $\Psi_{\lambda}(t)$ to have constant corank
$b_c$

\vskip .2cm

This proves the existence of the required cobordism and the part four of the
theorem.

\vskip .2cm

The above construction actually proves the part five as well since the cokernel
of derivative of  $\Psi_{\lambda}(t)$ is the desired bundle extension.

\vskip .2cm

At this point it is time to state the following

\proclaim Conjecture.
 If $M$ is a projective algebraic manifold then any of algebraic
resolutions of singularities of ${\bar {\cal M}}_d$ considered as a Hilbert
Scheme   can be obtained by the above
described  Kuranishi-type construction.

\vskip .2cm

We are not going to rely on this conjecture anywhere in the present paper.

\vskip .2cm

The proof of the above conjecture will after certain work give us equivalence
between the definition $(8.2)$  of the quantum
cup-product interpreted algebraically (or sheaf-theoretically) and the
Gromov-type Ruan's definition $(3.3)$ which has symplectic origin.

\vskip .2cm

We are going to work out these matters in a separate publication.

\vskip .2cm

Let us choose some  desingularization
${\widehat {\cal M}}_c$ of
${\bar {\cal M}}_c$ and some
smooth section $f_c$ of the vector bundle $ {\cal F}_c$
over this  desingularization

\vskip .2cm

We choose this section to be ``regular'' in the sense that its zero-set is a
smooth $dim({\cal M}_c)-2b_c$-dimensional submanifold
in ${\widehat {\cal M}}_c$

\proclaim Assumption 8.6. We  assume that ``regularity at infinity'' holds,
i.e., ${\widehat {\cal M}}_c   \bigcap  f_c^{-1}(0) - Map_d$ has codimension
two
in  ${\widehat {\cal M}}_c   \bigcap  f_c^{-1}(0) $

\vskip .2cm

\proclaim Definition D.
$ {\hskip 2mm} <A ; B ; C>_q {\hskip 2mm} = $

 $$ =  {\hskip 2mm}        \sum_d q^d {\hskip 2mm} \sum_c
 \sum_{ [\varphi \in {\widehat {\cal M}}_c
 \bigcap  ev_0^{-1} ({\widehat A}) \bigcap      ev_{\infty}^{-1} ({\widehat B})
\bigcap ev_1^{-1} ({\widehat C})  \bigcap
f_c^{-1}(0)] } \pm{1}   \eqno   (8.5)$$

\vskip .2cm

One can prove by the standard cobordism methods that the Definition D  is
independent of the choices of  the pseudo-manifold representatives of
${\widehat A} , {\widehat B}$ , ${\widehat C}$  and the choices of the sections
    $\{ f_c \}$.

The part 4) of the theorem 2.14 implies that the r.h.s. of $(8.5)$ is also
independent on the choice of desingularization  ${\widehat {\cal M}}_c$

\vskip .2cm
What requires a careful proof is

\proclaim Theorem 8.7.
For any choice of the desingularization   ${\widehat {\cal M}}_c$
    and the sections
    $\{ f_c \}$ such that assumption 2.15 holds we have
${\hskip 2mm} <A ; B ; C>_q = <A ; B ; C>_q^{VR}$

\vskip .2cm

To prove  the Theorem 8.7 it is enough to prove that:

\vskip .2cm

A) there exists a smooth cobordism ${\cal M}^t$ inside $Map_d$ between
the manifolds  ${\hskip 2mm} {\cal M}_{J,g,d} {\hskip 2mm} $ and
 $ {\hskip 2mm} {\bigcup}_c { {\widehat {\cal M}}_c   \bigcap  f_c^{-1}(0) }
\bigcap   Map_d {\hskip 2mm}$

\vskip .2cm

B) The compactification of $ {\cal M}_{J,g,d}$ and of
 $ {\hskip 2mm} {\bigcup}_c { {\widehat {\cal M}}_c   \bigcap  f_c^{-1}(0) }
\bigcap  Map_d {\hskip 2mm} $
can be extended to the compactification
  ${\hskip 2mm} {\bar {\cal M}}^t {\hskip 2mm}$ of the cobordism
 ${\hskip 2mm} {\cal M}^t {\hskip 2mm}$ (considered as a stratified space) such
that ``the compactification divisor''
${\bar {\cal M}}^t - {\cal M}^t$ has codimension at least two.

\vskip .2cm

C) The finite-codimensional  cycles  ${\hskip 2mm} ev_0^{-1} ({\widehat
A}){\hskip 2mm}$  ;    $ev_{\infty}^{-1} ({\widehat B})$ and  $ev_1^{-1}
({\widehat C}){\hskip 2mm} $
 in ${\hskip 2mm} Map_d {\hskip 2mm}$ intersect the
cobordism ${\hskip 2mm} {\cal M}^t {\hskip 2mm}$ transversally

\vskip .2cm

D)
${\hskip 2mm}  {\cal M}^t \bigcap ev_0^{-1} ({\widehat A})  \bigcap
ev_{\infty}^{-1} ({\widehat B}) \bigcap  ev_1^{-1} ({\widehat C}) {\hskip 2mm}$
is a compact one-dimensional cobordism between
$ {\hskip 2mm} {\cal M}_{J,g,d}  \bigcap ev_0^{-1} ({\widehat A})  \bigcap
ev_{\infty}^{-1} ({\widehat B}) \bigcap  ev_1^{-1} ({\widehat C}) {\hskip 2mm}
$
and

 ${\bigcup}_c  \{   {\widehat {\cal M}}_c   \bigcap f_c^{-1}(0) \} \bigcap
ev_0^{-1}
({\widehat A})  \bigcap   ev_{\infty}^{-1} ({\widehat B}) \bigcap  ev_1^{-1}
 ({\widehat C})   $

\vskip .2cm

This cobordism lies
inside ${\hskip 2mm} {\cal M}^t {\hskip 2mm}$
and does not touch the compactification divisor
${\hskip 2mm} {\bar {\cal M}}^t - {\cal M}^t$

\vskip .2cm

If we prove A) - D), we will be done since the statement D) already implies
the Theorem 8.5.

\vskip .2cm

Let us consider the Hilbert manifold
 ${\hskip 2mm} Map_d \times {\cal G}_0     \times {\cal G}_0 \times  [0 ; 1] $

\vskip .2cm

The general element of this space has the form
${\hskip 2mm} (\varphi ; g_1 ; g_2 ; s )$

\vskip .2cm

This space contains two submanifolds ${\hskip 2mm} Map_d \times {\cal G}_0
\times \{ 0  \}  \times \{ 0  \} {\hskip 2mm} $  and

 $Map_d \times \{ 0  \}  \times {\cal G}_0 \times \{ 1  \} $. Both of them are
(canonically) diffeomorphic to $Map_d \times {\cal G}_0 $

\vskip .2cm

We have a Hilbert bundle ${\cal H} $ over $Map_d \times {\cal G}_0     \times
{\cal G}_0 \times  [0 ; 1]$ and a canonical section
$ {\bar \partial}  =  {\bar \partial}_{J_0} - g$ of the bundle
${\cal H}$  restricted to
$Map_d \times {\cal G}_0     \times \{ 0  \}   \times \{ 0  \} $.

\vskip .2cm

The zero set of this  section ${\bar \partial}$ restricted to
$Map_d \times \{ g_1 \}   \times \{ 0  \}  \times \{ 0  \} $ will be the moduli
space
 ${\cal M}_{J_{g_1},d}$

\vskip .2cm

The section ${\bar \partial}$ is ``regular'' which means that the image of the
derivative of ${\bar \partial}$ on its zero set is surjective linear operator.

\vskip .2cm

Our strategy of constructing a cobordism will be to extend the section
${\bar \partial}$   to some ``regular'' (in the sense defined above) section
$\Phi$ over $Map_d \times {\cal G}_0     \times {\cal G}_0  \times  [0 ; 1] $
such that:

\vskip .2cm

A) The zero set of $\Phi$ restricted to $Map_d      \times \{ 0  \} \times \{
g_2  \}  \times \{ 1  \}   $ will be a union of smooth submanifolds inside some
desingularization of ${\cal M}_d$ (in the sense defined above).

\vskip .2cm

B) For each irreducible component ${\cal M}_c$ of ${\cal M}_d$ there exists a
section $f_c$  of the ``bundle of the fermion zero-modes '' ${\cal F}_c$ over
${\widehat {\cal M}_c}$ such that

$${\Phi}^{-1}(0) \bigcap  [ Map_d   \times \{ 0  \} \times \{ g_2 \}  \times \{
1  \} ] =
{\bigcup}_c  \{ {\widehat {\cal M}_c }  \bigcap f_c^{-1}(0) \}  \times \{ 0  \}
\times \{ g_2  \}  \times \{ 1  \}  $$

The sections $\{ f_c \}$  will depend on the choice of $g_2$ and vary smoothly
as $g_2 \in {\cal G}_0 $    varies.

\vskip .2cm

We already have a canonical section
$ {\bar \partial} =  {\bar \partial}_{J_0} - g$ of the bundle
${\cal H}$  considered as a bundle over $Map_d \times {\cal G}_0$.

\vskip .2cm

We can take a pull-back ${\Phi}_1$  of this section to  $Map_d \times {\cal
G}_0 \times {\cal G}_0 \times  [0 ; 1]l$ with respect to the projection
$Map_d \times {\cal G}_0  \times {\cal G}_0 \times  [0 ; 1]
 \rightarrow Map_d \times {\cal G}_0 $

on the product of the first two factors.

\vskip .2cm

By construction of the section ${\Phi}_1$ (which is the ordinary $ {\bar
\partial}$-operator  if restricted to $Map_d      \times \{ 0  \} \times {\cal
G}_0  \times \{ 0  \} $ ) we have

$$ {\Phi}_1^{-1}(0) \bigcap \{ Map_d      \times \{ 0  \} \times {\cal G}_0
\times \{ 0  \} \} {\hskip 2mm} = {\hskip 2mm}
{\cal M}_d \times \{ 0  \} \times {\cal G}_0   \times \{ 0  \} $$

\vskip .2cm

For any irreducible component ${\widehat {\cal M}_c }$  of desingularized
moduli space
 ${\widehat {\cal M}_d}$ we have a finite-dimensional bundle ${\cal F}_c$ (of
fermion zero-modes) over it. This finite-dimensional bundle is constructed as a
quotient of an infinite-dimensional Hilbert bundle ${\cal H}$ restricted to
${\widehat {\cal M}_c }$. Using the inner product in the fibers of
${\cal H}|_{\widehat {\cal M}_c }$ we can think of  ${\cal F}_c$ as a subbundle
of ${\cal H}$.

\vskip .2cm

 Let us consider a tubular neighborhood of the finite-dimensional manifold
${\cal M}_c$  inside the
 infinite-dimensional manifold $Map_d$. Let us denote  this tubular
neighborhood  ${\cal M}^{\epsilon}_c$. Inside this tubular neighborhood let us
consider some smaller  tubular neighborhood of the desingularized component
${\widehat {\cal M}}_c $. Let us denote  this smaller tubular neighborhood
 ${\widehat {\cal M}}^{\delta}_c$

\vskip .2cm

 We assume that the tubular neighborhoods
$\{ {\widehat {\cal M}}^{\delta}_c \} $ of different
irreducible components ${\widehat {\cal M}}_c $ do not intersect each other.
(We can make this assumption by the part three of the theorem 8.5).

\vskip .2cm

By definition of ``perturbed Kuranishi maps'' the zero locus of which defines
``the finite part of our   desingularization''
 ${\hskip 2mm} {\widehat {\cal M}_c }  \bigcap  Map_d {\hskip 2mm}$  we can
match together different ``perturbed Kuranishi maps'' $\{ \Psi^,_{\lambda} \}$
defined as a sections of finite-dimensional bundles ${\cal H}_{ < \lambda}$ on
the corresponding charts. The result will be a section ${\Psi}^,$ of the
infinite-dimensional Hilbert bundle ${\hskip 2mm}  {\cal H} {\hskip 2mm} $
defined over $ {\hskip 2mm}  {\cal M}^{\epsilon}_c {\hskip 2mm} $.

\vskip .2cm

 The finite part of the desingularization
${\hskip 2mm} {\widehat {\cal M}}_c  \bigcap  Map_d {\hskip 2mm} $
 is (by its definition) a zero-locus of this ``perturbed section''  ${\Psi}^,$
 defined over ${\cal M}^{\epsilon}_c$.

\vskip .2cm

 Since the normal bundle to  ${\widehat {\cal M}_c }  \bigcap  Map_d $ in
$Map_d$
(as any other infinite-dimensional Hilbert bundle) is trivial,
 the tubular neighborhood
 ${\hskip 2mm} {\widehat {\cal M}}^{\delta}_c {\hskip 2mm}$   is diffeomorphic
to a product of
${\hskip 2mm} {\widehat {\cal M}_c }  \bigcap  Map_d {\hskip 2mm}$ and an
infinite-dimensional ball. Let
${\hskip 2mm} \pi: {\widehat {\cal M}}^{\delta}_c \rightarrow
{\widehat {\cal M}_c } {\hskip 2mm} $
be the
corresponding projection.

\vskip .2cm

\proclaim Lemma 8.8.
We can extend the subbundle
${\hskip 2mm} {\cal F}_c \subset {\cal H} {\hskip 2mm}$ from
${\widehat {\cal M}_c }$ to  ${\widehat {\cal M}}^{\delta}_c$
 such that:

A) ${\cal F}_c$ will be a smooth $b_c$-dimensional subbundle in
 ${\widehat {\cal M}}^{\delta}_c$

\vskip .2cm

B) For any point
${\hskip 2mm} \varphi \in  {\widehat {\cal M}}^{\delta}_c {\hskip 2mm}$
the fiber ${\hskip 2mm} {\cal F}_c|_{\varphi} {\hskip 2mm}$ will be a
$b_c$-dimensional subspace in
${\hskip 2mm} {\cal H}|_{\varphi} {\hskip 2mm}$ which is orthogonal to the
value of the section
${\Psi}^,$
 at the  point ${\hskip 2mm} \varphi$.

\vskip .2cm

Since any two infinite-dimensional Hilbert bundles over any topological space
are isomorphic to each other (and trivial), let us trivialize the bundle
$ {\hskip 2mm}{\cal H} {\hskip 2mm}$ over
${\hskip 2mm} {\widehat {\cal M}}^{\delta}_c {\hskip 2mm}$.
The bundle
${\hskip 2mm} {\cal H}|_{ {\widehat {\cal M}}^{\delta}_c  }  {\hskip 2mm}$
 will thus be isomorphic to  the bundle
 ${\hskip 2mm} \pi^*( {\cal H}|_{ {\widehat {\cal M}}_c } )$.

\vskip .2cm

 Let us fix some isomorphism between these two Hilbert bundles and let ${\cal
F}^`_c$ be the image of
 ${\hskip 2mm} \pi^*( {\cal F}_c|_{ {\widehat {\cal M}}_c } ) {\hskip 2mm}$
under this isomorphism.

\vskip .2cm

So, we have constructed some $b_c$-dimensional subbundle ${\cal F}^`_c$ over
 ${\widehat {\cal M}}^{\delta}_c$

\vskip .2cm

Then over ${\widehat {\cal M}}^{\delta}_c $
 we have a codimension-one-subbundle
${\cal H}^` \subset {\cal H}$ whose fiber at the point $\varphi$ is defined as
orthogonal complement  to the one-dimensional subspace generated by the value
of the section
${\Psi}^,$ at the  point $\varphi$.
We can also define an operator ${\hskip 2mm} \pi^, {\hskip 2mm}$ of orthogonal
projection onto ${\hskip 2mm} {\cal H}^` \subset {\cal H} {\hskip 2mm}$.

\vskip .2cm

Let us put ${\tilde {\cal F}}_c =
\pi^, ({\cal F}^`_c)$ to be the subbundle in ${\cal H}$ defined over
 ${\widehat {\cal M}}^{\delta}_c - {\widehat {\cal M}}_c  $.

\vskip .2cm

We claim that the subbundle ${\tilde {\cal F}}_c$ can be smoothly extended to
$ {\widehat {\cal M}}_c  $ if we put it equal to
 ${\hskip 2mm} {\cal F}_c {\hskip 2mm}$ there.

\vskip .2cm
Now let us check
that this bundle   has the properties required by the Lemma 8.8.

\vskip .2cm

The property B of the Lemma 8.8 is obvious by construction.

\vskip .2cm

Since $ {\hskip 2mm} {\cal F}_c {\hskip 2mm}$ is orthogonal to the image of the
derivative of the section $ {\hskip 2mm} {\Psi}^, {\hskip 2mm}$, the
neighboring to $ {\hskip 2mm} {\hskip 2mm}{\widehat {\cal M}}_c  $ fibers of
the bundle $ {\hskip 2mm} {\cal F}^`_c {\hskip 2mm} $ are ``closed to be
orthogonal'' to the image of $ {\hskip 2mm} {\Psi}^, {\hskip 2mm} $. This
implies that $ {\hskip 2mm} {\tilde {\cal F}}_c {\hskip 2mm}$  smoothly extends
to    ${\hskip 2mm} {\widehat {\cal M}}_c {\hskip 2mm} $ (which is the zero-set
of   $ {\hskip 2mm} {\Psi}^, {\hskip 2mm} $ in
 $ {\hskip 2mm}{\cal M}^{\epsilon}_c {\hskip 2mm} $)

\vskip .2cm

The property A) and
the Lemma 8.8 follows.

\vskip .2cm

Now it is time to define ``a perturbed section'' $ \Phi$ over
$ {\hskip 2mm} Map_d \times \{ 0 \}  \times {\cal G}_0  \times \{ 1 \} $

\vskip .2cm

For each irreducible component  ${\hskip 2mm} {\cal M}_c {\hskip 2mm} $ of
 $ {\hskip 2mm}{\cal M}_d {\hskip 2mm} $   we have a
$b_c$-dimensional subbundle   $ {\hskip 2mm} {\tilde {\cal F}}_c {\hskip 2mm}$
in
${\hskip 2mm} {\cal H}|_{ {\widehat   {\cal M}}^{\delta}_c}$

\vskip .2cm

Let us take some ``cut-off function'' $h$ on the positive real axis  which
is identically one on $[0 ; \delta / 2]$ and identically zero on
$[\delta ; + \infty )$

\vskip .2cm

Then for any ${\hskip 2mm} g_2 \in  {\cal G}_0 {\hskip 2mm}$ we can:

\vskip .2cm

1) $ {\hskip 2mm}$ Take a section $ {\hskip 2mm} g_2 {\hskip 2mm}$ of
$ {\hskip 2mm} {\cal H}|_{ {\widehat   {\cal M}}^{\delta}_c} {\hskip 2mm}$

\vskip .2cm

2) $  {\hskip 2mm} $  Take a projection $f_c(g_2)$ of this section to the
subbundle
${\tilde {\cal F}}_c   \subset {\cal H}  $ to have a smooth section of
${\tilde {\cal F}}_c$

\vskip .2cm

3) $  {\hskip 2mm} $  Multiply the above section of  ${\tilde {\cal F}}_c$
 by a cut-off function

$ {\hskip 2mm} h(||\varphi - {\cal M}_c||) {\hskip 2mm} $ to have another
smooth section of a subbundle $ {\hskip 2mm} {\tilde {\cal F}}_c {\hskip 2mm} $
  in ${\cal H}  $  which can be extended (by zero) from
${\widehat   {\cal M}}^{\delta}_c$
 to $Map_d$ as a some smooth section of ${\cal H}$. Let us denote this section
$(g_2)_c$

\vskip .2cm

4) Take the sum $ {\hskip 2mm} { \tilde g_2} = \sum_c (g_2)_c {\hskip 2mm} $
over different components of
${\cal M}_d$

\vskip .2cm

If we put $  {\hskip 2mm} \Phi =   {\Psi}^, - { \tilde g_2}  {\hskip 2mm} $ as
a section of
$ {\hskip 2mm } {\cal H}$ over $Map_d
\times \{ 0 \}  \times \{ g_2 \} \times \{ 1 \}  {\hskip 2mm}$
 and consider it as a section  of
$  {\hskip 2mm} {\cal H}$ over $Map_d  \times \{ 0 \}
 \times {\cal G}_0  \times \{ 1 \} {\hskip 2mm} $ (as $g_2 \in {\cal G}_0$
varies) then this section will be ``regular'' by construction.

\vskip .2cm

Here, as usual, ``regular'' means that the image of derivative of the  section
$ \Phi$ is surjective over its zero-set.

\vskip .2cm

This section $\Phi$  can be extended from $Map_d \times  \{ 0 \} \times {\cal
G}_0  \times \{ 1 \} $ to
$Map_d \times  {\cal G}_0 \times {\cal G}_0  \times [ 0 ; 1]  $ by the formula

$$ \Phi =  (1-s) {\bar \partial}_{J_0} + s {\Psi}^,  -  g_1 - { \tilde g_2}
\eqno  (8.6)  $$

\proclaim Lemma 8.9. The section $ \Phi$ is regular

\proclaim Lemma 8.10.
 The zero-set of $ \Phi$ restricted to
$Map_d \times  \{ 0 \} \times {\cal G}_0  \times  \{ 1 \}$ lies inside

 ${\widehat {\cal M}}_d \times  \{ 0 \} \times {\cal G}_0 $

\proclaim Lemma 8.11.
$ \Phi^{-1}(0) \bigcap {\cal M}_c \times  \{ 0 \}  \times   \{ g_2 \}
 \times  \{ 1 \}  $
is the zero-set of the section  $f_c(g_2)$ of the bundle ${\cal F}_c$

\proclaim Lemma 8.12.
For a generic choice of $g_2 \in {\cal G}_0$ all the sections $\{ f_c(g_2) \}$
of the bundles $\{ {\cal F}_c \}$ are ``regular'' and their zero-sets are
smooth manifolds of dimension equal to $ {\hskip 2mm}  vdim [{\cal M}_d]$.

\proclaim Assumption 8.13.
If $ {\hskip 2mm}  g_1 + g_2  {\hskip 2mm} $ lies in
 $ {\hskip 2mm}  {\cal G}_0  {\hskip 2mm} $
then the zero-submanifolds
 $  {\hskip 2mm} \Phi^{-1}(0) \bigcap Map_d \times  \{ g_1 \}
 \times   \{ g_2 \} \times  \{ s \}  {\hskip 2mm}  $
 can be compactified as  stratified spaces such that their
 ``compactification divisors'' have codimension at least two.

This ``assumption'' can be proved using estimates on norms of our ``perturbed
${\bar \partial}$-operators''

\vskip .2cm

 Using Lemmas 2.4 and 2.6 we have that there exists a smooth cobordism
 $ {\hskip 2mm} {\cal M}^t  {\hskip 2mm}$ inside
 $ {\hskip 2mm} Map_d  \times {\cal G}_0 \times {\cal G}_0  \times [0 ; 1]
{\hskip 2mm} $
between
the manifolds

 $ {\hskip 2mm} {\cal M}_{J,g_1,d} \times  \{ g_1 \} \times  \{ 0 \} \times  \{
0 \}  {\hskip 2mm} $
 and
${\hskip 2mm} \{ {\cal M}_d \bigcap f^{-1}(0) \} \times  \{ 0 \}   \times  \{
g_2 \}
 \times  \{ 1 \} {\hskip 2mm} $

\vskip .2cm

Assuming that 8.13 holds we have that this cobordism can be compactified and
``the compactification divisor''
${\bar {\cal M}}^t - {\cal M}^t$ has codimension at least two in the total
space of the cobordism.

\vskip .2cm

Thus, we have established the statements A and B after the theorem 8.7.
Now let us observe that the statement D (and the Theorem 8.7  itself) will
follow from the statement C. But the statement C follows from the
Lemma 2.5.

\vskip .2cm

The Theorem 8.7 is proved.

\vskip .2cm

The above proof of the Theorem 8.7 is very similiar to the proof of Ruan [Ru2]
of the following fact (we state it using our notation ):

\vskip .2cm

Let us take a non-regular almost-complex structure ${\hskip 2mm} J {\hskip
2mm}$ on $ {\hskip 2mm} M  {\hskip 2mm}$ satisfying
Assumptions 8.1 - 8.3.
Let us assume that  we can be perturbe ${\hskip 2mm} J {\hskip 2mm}$ inside the
space of almost-complex structures on ${\hskip 2mm} M {\hskip 2mm} $   (without
going to $CP^1 \times M$ ) such that ``generic'' perturbation is regular, then
the r.h.s. of $(8.5)$ and the r.h.s. of $(3.3)$
are equal as formal power series in $\{ q_i \}$.

\vskip .2cm

Here we take the expression $(8.5)$ for a a non-regular almost-complex
structure $J$. The expression
$(3.3)$ is taken for a regular generic perturbation of  $J$ inside the space of
almost-complex structures on $M {\hskip 2mm}$ (if such a perturbation exists).

\vskip .2cm

The advantage of our approach is that it handles the problem what to do if
there are no regular perturbations of the almost-comple structure $J$. Such
situation almost always happens when $J$-holomorphic curve $\varphi$ is a
multiple branched cover of some other $J$-holomorphic curve of lower degree
(see [McD1 ] for the best exposition of this analytic problem).

\vskip .2cm

\vskip .2cm

\vskip .2cm

\vskip .2cm

$$ $$

 \centerline {\bf       {9. DISCUSSION }}
\centerline {  }

The Main Theorem 6.1, proved in the present paper can be thought as
mathematical implementation of the program of Vafa [Va]  $ ${\hskip 2mm} $  of
understanding quantum cohomology through geometry of the loop space. The notion
of $${\hskip 2mm} $ ``BRST-quantization on the loop space'' considered by
string theorists (see [Wi1] for the best treatment), can be put in the
mathematically rigorous framework of symplectic Floer homology.

\vskip .2cm

If we are studying geometry of Kahler manifold $M$ from the point of view of
the string propagating on it, we can extract more algebrogeometrical
information on $M$ than is contained in its quantum cohomology ring $HQ^*(M)$

\vskip .2cm

The String Theory on $M$ also provides us with

\vskip .2cm

A) Deformation of the classical cohomology ring $H^*(M)$ with respect to all
(and not just two-dimensional) cohomology classes

\vskip .2cm

B) Some explicitely constructed cohomology classes of the moduli spaces of
punctured curves known as Gromov-Witten classes

\vskip .2cm

We recommend the reader  a very interesting recent paper by Kontsevich and
Manin [KM] with new developments in ``quantum cohomology'' (in this broader
sense) and applications of these new invariants to classical problems in
algebraic geometry

\vskip .2cm

During the preparation of the present paper we also received two very
interesting papers by Fukaya [Fu1],[Fu2]  with new developments in Floer
homology. What Fukaya did is that he constructed analogues of the classical
Massey products in Floer homology of Lagrangian intersections. In order to
construct these ``quantized Massey products'', Fukaya used the loop space
generalization of a   finite-dimensional Morse-theoretic construction, which
was not known before.

\vskip .2cm

These new results together with the work of Cohen-Jones-Segal [CJS2] and
Betz-Cohen [BK] give a hope to understand what is ``quantum homotopy type''
and ``Floer homotopy type'' of a  semi-positive almost-Kahler manifold.

\vskip .2cm

If our Kahler ``manifold'' $M$ is the moduli space of flat connections on a
two-dimensional surface  (which is actually  a stratified space and not a
manifold), symplectic Floer homology of this ``manifold'' is conjectured [A] to
be isomorphic to  instanton Floer homology of a circle bundle over this surface
( see [DS],[Y] for the proof of this conjecture and [Ta2] for further
developements).

\vskip .2cm

The multiplicative structure in
symplectic Floer homology coresponds under this isomorphism to relative
Donaldson invariants of some  4-dimensional manifolds with boundary. Thinking
about these  relative Donaldson invariants as some matrix elements of quantum
multiplication on the moduli space of flat connections we can interpret gluing
formulas
[BrD] and recursion relations [KrM] for  Donaldson invariants  as recursion
relations coming from associativity of quantum multiplication.

\vskip .2cm

\vskip .2cm

 {\hskip 6mm}
\centerline {  }
\centerline {\bf  ACKNOWLEDGEMENTS}
\centerline {  }

	I am greatly indebted to V.Sadov and to I.M.Singer for introducing me to
the problem and for many helpful remarks without which this work would not be
possible and to
 E.Witten for explaining me his work on quantum cohomology of grassmanians
which inspired me to work out this example rigorously.

\vskip .2cm

 M.Gromov, D.Mcduff and C.Taubes taught me about analysis relevant to
Floer-Gromov theory of pseudo-holomorphic curves.

\vskip .2cm

 C.Taubes helped me with the literature on this subject.

\vskip .2cm

S.Axelrod, D.Kazhdan, M.Kontsevich and expecially  S-T.Yau read carefully the
early versions of my manuscript and made many suggestions towards its
improvement.

\vskip .2cm

     I also benefited very much from discussions with  V.I.Arnold,
 A.Astashkevich, R.Bott, S.Chang, R.Constantinescu, R.Dijkgraaf, K.Fukaya,
A.Givental, A.Goncharov, K.Intrilligator, A-K.Liu, Yu.I.Manin, D.Morrison,
 S.P.Novikov, G.Segal,
G.Tian, C.Vafa and R.Wentworth, to whom I express my sincere gratitude.

\clearpage


\begin{thebibliography}{WW}

\bibitem[A]{8} M.Atiyah, {\em  New invariants of three- and four-dimensional
manifolds.}
Proc. Symp. Pure Math. {\bf 48} (1988), 38-48.

\bibitem[AB]{9} M.Atiyah, R.Bott, {\em The moment map and
equivariant cohomology.} Topology {\bf 23} (1984), 1--28.

\bibitem[AuBr]{9} D.Austin, P.Braam, {\em Morse-Bott Theory and
equivariant cohomology.} Preprint, 1993


\bibitem[AJ]{9} M.Atiyah, L.Jeffrey, {\em Topological Lagrangians and
cohomology.} J. Geom. Phys. {\bf 7} (1990), 119--136.



\bibitem[AM]{9} P.Aspinwal, D.Morrison, {\em Topological Field Theory and
Rational Curves.}  Commun. Math. Phys. {\bf 151} (1993), 245--262.



\bibitem[AS1]{9} M.Atiyah, I.Singer, {\em The index of elliptic operators.}
Ann.Math. {\bf 87} (1968), 546--604 ; {\bf 93} (1971), 119--138.

\bibitem[AS]{9} A.Astashkevich, V.Sadov, {\em Quantum Cohomology of Partial
Flag Manifolds $F_{n_1...n_k}$.} hep-th 9401103.


\bibitem[BS]{12} L.Baulieu, I.Singer ,{\em  Topologucal sigma-models.}
Commun. Math. Phys {\bf 125} (1989), 227--237





\bibitem[Ba]{18}  V.Batyrev, {\em Quantum cohomology ring of Toric Manifolds.}
 ${\hskip 2mm}$ alg-geom  9310004


\bibitem[BDW]{18} A.Bertram, G.Daskalopoulos, R. Wentworth,
 {\em Gromov Invariants for Holomorphic Maps from Riemann Surfaces to
Grassmannians.}

alg-geom 9306005

\bibitem[Be]{18} A.Bertram,  {\em Towards a Schubert calculus for maps from a
Riemann surface to a Grassmannian.}
 ${\hskip 2mm}$ alg-geom 9403007




\bibitem[BC]{18}  M.Betz, R.Cohen, {\em Graph Moduli Spaces and Cohomology
Operations.} ${\hskip 2mm}$ Preprint, 1993.



\bibitem[BrD]{18} P.Braam, S.Donaldson,  {\em Fukaya-Floer Homology and Gluing
Formulas for Polynomial Invariants.}  ${\hskip 2mm}$ Preprint, 1






\bibitem[BF]{12} J-M. Bismut, D.Freed, {\em  Analysis of elliptic families.}
Commun. Math. Phys {\bf 106} (1986), 159--176  and  {\bf 107} (1986), 103--163


\bibitem[COGP]{5} P.Candelas, X.de la Ossa, P.Green, L.Parkes, {\em A Pair of
Calabi-Yau manifolds as an exactly soluble superconformal field theory.}
Nucl. Phys. {\bf 359} (1991) 21--74

\bibitem[CJS1]{18}  R.Cohen, J.D.S.Jones, G.Segal {\em  Morse Theory and
Classifying Spaces.}
Preprint 1992.



\bibitem[CJS2]{18}  R.Cohen, J.D.S.Jones, G.Segal {\em Floer's
Infinite-Dimensional Morse Theory and Homotopy Theory.}
Preprint 1993.



\bibitem[CKM]{3} H.Clemens, J.Kollar, S.Mori, {\em Higher dimensional complex
geometry.}     Asterusque {\bf 166} (1988).


\bibitem[DK]{3} S.Donaldson, P.Kronheimer, {\em The geometry of
Four-Manifolds.}    (Oxford Math. Monogr.) Oxford, Clarendon, 1990.


\bibitem[DoSa]{5} S.Dostoglou, D.Salamon, {\em Self-Dual Instantons and
Holomorphic Curves.} Preprint, 1992



\bibitem[DSWW]{5} M.Dine, N.Seiberg, X-S.Wen, E.Witten, {\em Non-perturbative
effects on string worldsheet.}
Nucl. Phys. {\bf 278} (1986) 769--789  and    {\bf 289} (1987)   319--363




\bibitem[F1]{12} A.Floer,{\em Symplectic fixed points and holomorphic spheres.}
Commun.Math.Phys. {\bf 120} (1989), 575--611.



\bibitem[F2]{11} A.Floer, {\em Cuplength estimates on lagrangian
intersections.}
 Commun. Pure. Appl. Math.  {\bf 42} (1989), 335--356.



\bibitem[F3]{11} A.Floer, {\em Morse theory and lagrangian intersections.}
 J. Diff. Geom.  {\bf 28} (1988), 513--547.


\bibitem[F4]{11} A.Floer, {\em Witten's complex and Infinite-dimensional Morse
theory.}
 J. Diff. Geom.  {\bf 30} (1989), 207--221.




\bibitem[F5]{11} A.Floer, {\em A relative Morse index for the symplectic
action.}
 Commun. Pure. Appl. Math.  {\bf 41} (1988), 393--407.


\bibitem[F6]{11} A.Floer, {\em The unfegularized gradient flow for the
symplectic action.}
 Commun. Pure. Appl. Math.  {\bf 41} (1988), 775--813.




\bibitem[F7]{11} A.Floer, {\em A refinement of the Conley index and the
application to the stability of hyperbolic invariant sets.}
 Ergod. Theory and Dynam. Systems  {\bf 7} (1987), 93--103.




\bibitem[F8]{11} A.Floer, {\em A proof of the Arnold Assumption for surfaces
and generalizations to certain Kahler manifolds.}
 Duke Math. J.  {\bf 53} (1986), 1--32.



\bibitem[FH]{11} A.Floer, H.Hofer, {\em Coherent orientations for periodic
orbit problems in symplectic geometry.}
 Mathematische Zeichrift  {\bf 212: 1} (1993), 13--38.

\bibitem[Fu1]{11} K.Fukaya, {\em Morse homotopy, $A_{\infty}$-category and
Floer homologies.}  Preprint, 1993

\bibitem[Fu2]{11} K.Fukaya, {\em Morse homotopy and its quantization.}
Preprint, 1994



\bibitem[FU]{11} D.Freed, K.Uhlenbeck, {\em Instantons and 4-Manifolds.}
 Second Edition, New York, Springer-Verlag  (1991)



\bibitem[Ge]{22} D.Gepner  {\em Exactly solvable String compactifications fon
manifolds with $SU(N)$-holonomy.}
Phys. Let.  {\bf 199B} (1987), 380--388.


\bibitem[G]{16} V.A.Ginsburg {\em Equivariant cohomology and Kahler geometry.}
Funct. Anal. Appl. {\bf 21:4} (1987), 271--283.





\bibitem[GK]{20} A.Givental, B.Kim, {\em Quantum Cohomology of Flag Manifolds
and Toda Lattices.} hep-th 9312096.


\bibitem[Gr1]{20} M.Gromov, {\em Pseudo-holomorphic curves in almost complex
manifolds.} Invent. Math. {\bf 82:2} (1985), 307--347.

\bibitem[Gr2]{20} M.Gromov, Private communication.


\bibitem[GP]{20} V.Guillemin, A.Pollack, {\em Differential Topology.}
Prentice Hall, 1974.


\bibitem[HS]{18}  H.Hofer, D.Salamon, {\em Floer homology and Novikov rings.}
Preprint, 1992.



\bibitem[Ho]{18}  H.Hofer, {\em   }
Annales de l`Institut Poincare {\bf 5:5} (1988), 495--499.





\bibitem[I]{22} K.Intrilligator  {\em Fusion Residues.}
Mod. Phys. Let.  {\bf A6} (1987), 3543--



\bibitem[K]{24} M.Kontsevich, {\em $A^{\infty }$-algebras in mirror symmetry.}
Preprint, 1993.

\bibitem[KM]{24} M.Kontsevich, Yu.I.Manin,  {\em  Gromov-Witten	Classes,
Quantum Cohomology and Enumerative Geometry.}
hep-th 9402147

\bibitem[KrM]{24} P.Kronheimer, T.Mrowka,  {\em  Recurrence Relations and
Assymptotics for 4-manifold invariants.}
 Bull.AMS. {\bf 30} (1994), 215-221.









\bibitem[McD1]{20} D.McDuff, {\em Examples of symplectic structures.} Invent.
Math. {\bf 89:1} (1987), 13-36 ; {\em Elliptic methods in symplectic geometry.}
Bull.AMS. {\bf 23} (1990), 311-358.

\bibitem[McD S]{20} D.McDuff, D.Salamon, {\em  $J$-holomorphic
curves and Quantum Cohomology.} preprint, 1994


\bibitem[Mi]{20} J.Milnor, {\em Topology from the differential viewpoint.}
The Univ. Press of Virginia ,1965

\bibitem[Q]{16} D.Quillen {\em Determinants of Cauchy-Riemann operators over a
Riemann surface.}
Funct. Anal. Appl. {\bf 19} (1987), 31--34

\bibitem[No]{16} S.P.Novikov   {\em The Hamiltonian formalism and a
multi-valued analogue of Morse Theory.}   Russ. Math. Surv.   {\bf 37:5}
(1982), 1--56


\bibitem[Ru1]{21} Y.Ruan, {\em Topological sigma model and Donaldson type
invariants in Gromov theory.} Preprint.


\bibitem[Ru2]{21} Y.Ruan, {\em Symplectic Topology and Extremal Rays.}
Geometric And Funct. Anal. {\bf 3:4} (1993), 395--430



\bibitem[RT]{21} Y.Ruan, G.Tian {\em Mathematical theory of Quantum
cihomology.} Pre-preprint, 1994


\bibitem[RRW]{18} M.S.Ravi, J. Rosenthal, X.Wang, {\em Degree of the
generalized Pl\"ucker embedding of a Quot scheme and Quantum
cohomology.}

alg-geom/9402011



\bibitem[S]{7} V.Sadov, {\em On equivalence of Floer's and quantum cohomology.}
Preprint HUTP-93/A027.


\bibitem[Sa]{5} D.Salamon, {\em Morse Theory, the Conley Index and Floer
homology.} Bulletin of London Math. Soc.  {\bf 22} (1990) 113 - 140





\bibitem[SeSi]{5} R.Seeley, I.Singer, {\em Extending  ${\bar \partial}$ to
Singular Riemann Surfaces.}
J. Geom. Phys.  {\bf 5:1} (1988)



\bibitem[SU]{5} J.Sacks, K.Ulenbeck, {\em The existence of minimal immersions
of two-spheres.}
Ann. Math. {\bf 13} (1981), 1--24


\bibitem[SaZ]{5} D.Salamon, E.Zendler {\em Morse Theory
for periodic solutions of Hamiltonian Systems and the Maslov Index.}
Comm. Pure Appl. Math.   {\bf 45} (1992) 1303 - 1360



\bibitem[ST]{5}   B.Siebert, G.Tian, {\em On Quantum Cohomology Rings of Fano
Manifolds and a Formula of Vafa and
Intriligator.}    alg-geom/9403010







\bibitem[SW]{5} A.Strominger, E.Witten, {\em New Manifolds for Superstring
Compactification.}
Commun. Math. Phys {\bf 101} (1985), 341--361



\bibitem[Ta]{5} C.Taubes, {\em Self-Dual Connections on 4-manifolds with
Indefinite Intersection Matrix.}
Journal of Diff. Geom,  {\bf 19:2} (1984), 517--560





\bibitem[Ta2]{5} C.Taubes,  in preparation




\bibitem[Va1]{1} C.Vafa,{\em Topological mirrors and quantum rings.}
in: S.-T. Yau ({\em Ed.}), {\em  Essays on mirror manifolds.}
    International Press Co., Hong Kong, 1992.



\bibitem[Vt]{17} C.Viterbo, {\em The cup-product on the Thom--Smale--Witten
complex, and Floer cohomology.} To appear in: Progress in Math., v. 93,
Birkhauser, Basel.




\bibitem[Wi1]{5} E.Witten, {\em Topologucal sigma-models.}
Commun. Math. Phys {\bf 118} (1988), 353--386

\bibitem[Wi2]{5} E.Witten, {\em Two-dimensional gravity and intersection theory
on moduli space.} Surveys in Diff. Geom. {\bf 1} (1991), 243--310.

\bibitem[Wi3]{1} E.Witten,{\em  Mirror manifilds and topological field
theories.}
in: S.-T. Yau ({\em Ed.}), {\em  Essays on mirror manifolds.}
    International Press Co., Hong Kong, 1992.


\bibitem[Wi4]{5} E.Witten, {\em The N-matrix Model and gauged WZW  Models.}
Nucl. Phys.  {\bf B371} (1992), 191--245


\bibitem[Wi5]{5} E.Witten, {\em The Verlinde algebra and the cohomology of the
grassmanian.}
hep-th 9312104


\bibitem[Wi6]{5} E.Witten, {\em On the structure of the topological phase of
two-dimensional gravity.}
Nucl. Phys. {\bf B340} (1990) 281--332







\bibitem[Wi7]{10} E.Witten, {\em Supersymmetry and Morse theory.} J. Diff.
Geom.
 17 (1982), 661--692.




\bibitem[Y]{5} T.Yoshida, {\em Floer homology and Holomorphic Curves - the
Atiyah Conjecture.}    Preprint, 1993.



\end{thebibliography}
\end{document}